\documentclass[twocolumn,aps,pra,amsmath,superscriptaddress,notitlepage,longbibliography]{revtex4-1}
\usepackage{amsmath,amssymb,amsfonts,bm}
\usepackage{graphicx}
\usepackage{dcolumn}
\usepackage{mathrsfs}
\usepackage{bbold}
\usepackage{dsfont}
\usepackage{dcolumn}
\usepackage[colorlinks=true,linkcolor=blue,citecolor=blue, urlcolor=blue,bookmarks=false]{hyperref}
\usepackage{changes}

\begin{document}


\title{Anomalous thermoelectric and thermal Hall effects in irradiated altermagnets}

\author{Fang Qin}
\email{qinfang@just.edu.cn}
\affiliation{School of Science, Jiangsu University of Science and Technology, Zhenjiang, Jiangsu 212100, China}

\author{Xiao-Bin Qiang}
\email{12331032@mail.sustech.edu.cn}
\affiliation{State Key Laboratory of Quantum Functional Materials, Department of Physics, and Guangdong Basic Research Center of Excellence for Quantum Science, Southern University of Science and Technology (SUSTech), Shenzhen 518055, China}
\affiliation{Division of Physics and Applied Physics, School of Physical and Mathematical Sciences, Nanyang Technological University, 21 Nanyang Link, Singapore 637371, Singapore}

\begin{abstract}
We show that a $d$-wave altermagnet can be transformed into a Chern insulator by irradiating it with elliptically polarized light from a high-frequency photon beam. We further explore the intrinsic anomalous thermoelectric and thermal Hall effects in light-irradiated altermagnets. At low temperatures, the thermoelectric Hall coefficient exhibits a linear temperature dependence but vanishes within the energy gap between the conduction and valence bands near the $M$ point. However, it displays pronounced peaks and dips at the gap boundaries near both the $M$ and $\Gamma$ points, suggesting that thermoelectric Hall conductivity is a sensitive probe for these gapped regions. Similarly, the low-temperature thermal Hall coefficient, which also shows a linear temperature dependence, becomes quantized across the bandwidth, reflecting the underlying topological character of the light-induced Chern insulating phase. These results establish thermoelectric and thermal Hall transports as powerful signatures of topology in driven altermagnetic systems.
\end{abstract}
\maketitle

\section{Introduction}

Temperature gradients are ubiquitous in condensed-matter systems and form the basis of thermoelectric phenomena that enable energy conversion between heat and electricity. A central example is the Nernst effect, where a transverse electric response is induced by a temperature gradient. While a finite Nernst signal conventionally requires an external magnetic field, in magnetic and topological materials the Berry curvature of electronic bands can generate an intrinsic transverse thermoelectric response even in zero field, known as the anomalous Nernst effect~\cite{xiao2006berry,yokoyama2011transverse,qiang2023topological}. Recently, thermoelectric and thermal transports have emerged as powerful probes of topological phases of matter~\cite{xiao2006berry,yokoyama2011transverse,qiang2023topological,hirschberger2016chiral,fu2020topological,xu2020finite,zhang2021topological,chen2021anomalous,pan2022giant} and within linear response theory, universal relations such as the Mott relation and Wiedemann-Franz law establish fundamental connections between charge and heat transports in systems with broken time-reversal symmetry~\cite{xiao2006berry,yokoyama2011transverse,qiang2023topological}.

Thermoelectric and thermal transport phenomena have also been actively explored in altermagnetic materials~\cite{badura2025observation,han2025nonvolatile,zhou2024crystal,weissenhofer2026magnon,debnath2026spin}. In particular, the anomalous Nernst effect has been observed in the altermagnetic Mn$_5$Si$_3$~\cite{badura2025observation}, an anomalous Nernst effect associated with a collinear N\'{e}el vector has been reported in altermagnets~\cite{han2025nonvolatile}, and thermal transport has been investigated in RuO$_2$~\cite{zhou2024crystal,weissenhofer2026magnon}, CrSb~\cite{weissenhofer2026magnon}, and MnTe~\cite{debnath2026spin}.
Motivated by these developments, we investigate the anomalous thermoelectric and thermal Hall effects in Floquet-engineered altermagnets.

Altermagnets constitute a recently identified class of collinear magnetic materials that is distinct from both conventional ferromagnets and antiferromagnets. From a symmetry perspective, altermagnets are protected by spin-group symmetries, which permit anisotropic spin-split electronic band structures while enforcing vanishing macroscopic magnetization~\cite{wu2004dynamic,wu2007fermi,lee2009theory,yuan2026unconventional,hayami2019momentum,hayami2020bottom,ma2021multifunctional,song2025altermagnets,xu2026chemical,hu2025catalog,smejkal2022giant,smejkal2022beyond,smejkal2022emerging,smejkal2022anomalous}.
As a result, altermagnets combine key features of ferromagnetic and antiferromagnetic systems, enabling sizable spin-dependent responses without stray magnetic fields. A rapidly growing set of candidate materials, including  KV$_2$Se$_2$O~\cite{jiang2025metallic,wang2025atomic,yang2026visualizing}, RbV$_2$Te$_2$O~\cite{zhang2025crystal,hu2026observation},  RuO$_2$~\cite{li2025exploration,ahn2019antiferromagnetism,vsmejkal2020crystal,shao2021spin,rafael2021efficient,bose2022tilted,bai2022observation,karube2022observation,guo2024direct,he2025evidence},  MnF$_2$~\cite{bhowal2024ferroically,li2024creation}, MnTe~\cite{mazin2023altermagnetism,krempasky2024altermagnetic,lee2024broken,osumi2024observation,orlova2025magnetocaloric,belashchenko2025giant}, CrSb~\cite{reimers2024direct,ding2024large,peng2025scaling,zhou2025manipulation},  FeSb$_2$~\cite{mazin2021prediction,attias2024intrinsic,phillips2025electronic},  Mn$_5$Si$_3$~\cite{leiviska2024anisotropy,reichlova2024observation,rial2024altermagnetic,han2024electrical}, and BiFeO$_3$~\cite{urru2025g,george2026topological,sajid2026signatures,gui2026electric}, has been proposed and experimentally investigated.

The unconventional symmetry and band structure of altermagnets give rise to a broad range of emergent electronic and transport phenomena. Recent theoretical studies have predicted anomalous Josephson effects~\cite{ouassou2023dc,zhang2024finite,cheng2024orientation,beenakker2023phase,lu2024varphi,sun2025tunable,fukaya2025josephson,pal2025josephson}, unconventional Andreev reflection~\cite{sun2023andreev,papaj2023andreev}, nonlinear charge transport~\cite{ezawa2024intrinsic,fang2024quantum,liu2025enhancement}, magnetoresistance effects~\cite{sun2025tunneling}, parity anomalies~\cite{wan2025altermagnetism}, bulk photovoltaic effects~\cite{ezawa2025bulk}, altermagnetic proximity effect~\cite{zhu2026altermagnetic}, and layer Hall effect~\cite{qin2026layer}.  Extensions to quasicrystalline lattices~\cite{chen2026quasicrystalline,shao2025classification,li2026unconventional}, Coulomb drag~\cite{lin2025coulomb}, spin-orbital coupled responses~\cite{wang2025spin}, momentum-space alternating spin polarization~\cite{chen2025probing}, spin triplet states~\cite{fu2026floquet,fu2025light,fukaya2025superconducting,maeda2025classification}, and spintronics~\cite{fu2025all} have further broadened the scope of altermagnetic phenomena.
Moreover, altermagnets provide a natural platform for realizing diverse topological phases. Recent studies have predicted altermagnet-induced topological states~\cite{ezawa2024detecting,ma2024altermagnetic,antonenko2025mirror,qu2025altermagnetic,parshukov2025topological,fernandes2024topological}, higher-order topological phases~\cite{li2024creation}, and light-induced odd- and even-parity altermagnetism~\cite{zhuang2025odd,huang2025light,zhu2025floquet,liu2026light,zhu2026light}, as well as Floquet-engineered topological phases under periodic driving~\cite{zhu2025floquet_Chen,ganguli2026tunable,ghorashi2025dynamical,cheraghchi2026floquet,yarmohammadi2026floquet}.
These developments establish altermagnetism as a versatile platform for exploring symmetry-driven transport and topological effects, providing a natural motivation for the present study.

Recent studies have demonstrated that altermagnetic materials can host intrinsic anomalous Nernst effects despite possessing compensated magnetic order and vanishing net magnetization. In particular, the observation of the anomalous Nernst effect in the altermagnetic candidate Mn$_5$Si$_3$~\cite{badura2025observation,han2025nonvolatile} reveals that unconventional spin-split electronic structures in altermagnets can generate sizable transverse thermoelectric responses through intrinsic Berry-curvature effects. These equilibrium anomalous Nernst phenomena originate from material-specific electronic structures determined by the altermagnetic symmetry, magnetic ordering, and spin splitting. Indeed, a variety of altermagnetic compounds, including RuO$_2$~\cite{zhou2024crystal,weissenhofer2026magnon}, CrSb~\cite{weissenhofer2026magnon}, and  MnTe~\cite{debnath2026spin}, have been identified as promising platforms owing to their unconventional spin-split band structures without net magnetization.

Compared with these previously reported intrinsic anomalous Nernst effects, the Floquet-induced mechanism proposed in this work provides a fundamentally different route for controlling thermoelectric transport. Instead of relying solely on the intrinsic electronic structures of equilibrium altermagnetic materials, periodic optical driving enables dynamic reconstruction of the band dispersion and Berry-curvature distribution, thereby offering an external knob to manipulate the anomalous Nernst response. Such light-driven control allows the magnitude, sign, and topological nature of the anomalous Nernst response to be tuned by varying the driving parameters, including the frequency, amplitude, and polarization of the incident optical field. Therefore, Floquet engineering offers a flexible platform for realizing light-controlled anomalous Nernst effects and accessing optically induced thermoelectric phases beyond those available in equilibrium altermagnetic systems.

\begin{figure}[htpb]
\centering
\includegraphics[width=\columnwidth]{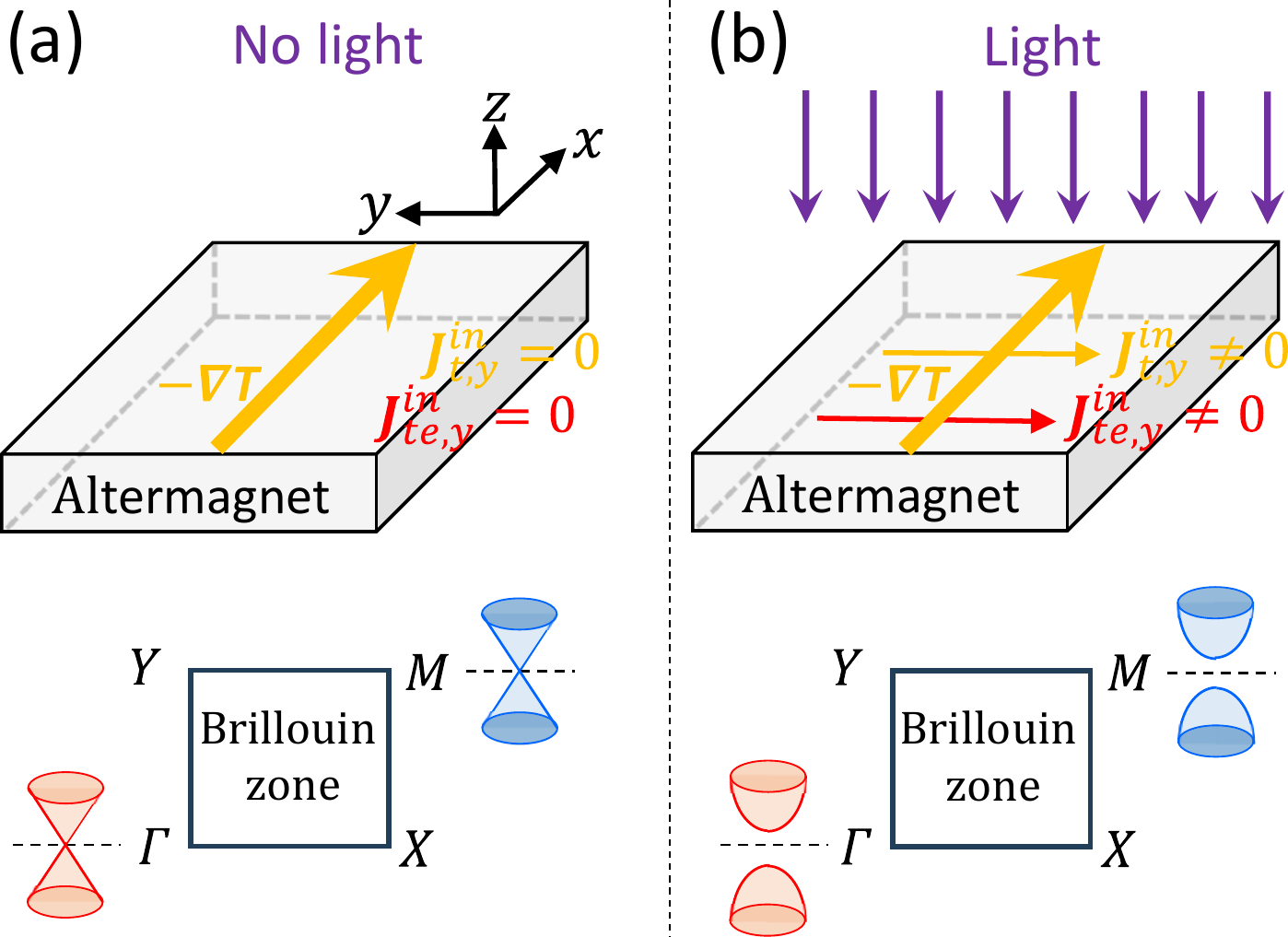}
\caption{Schematic illustration of intrinsic thermoelectric and thermal Hall effects in a $d$-wave altermagnet, with and without light irradiation. 
Here, $J_{te,y}^{\rm in}$ and $J_{t,y}^{\rm in}$ denote the intrinsic thermoelectric and thermal Hall current densities, respectively. $-\nabla T$ is the longitudinal temperature gradient and $T$ is the system temperature.
(a) Without light irradiation. All intrinsic Hall responses vanish, i.e., $J_{te,y}^{\rm in}\!=\!J_{t,y}^{\rm in}\!=\!0$, and the Dirac cones around the $\Gamma$ and $M$ points in the Brillouin zone remain gapless. 
(b) With light irradiation. Finite intrinsic Hall thermoelectric and thermal currents emerge, $J_{te,y}^{\rm in}\!\neq\!0$ and $J_{t,y}^{\rm in}\!\neq\!0$, accompanied by the opening of gaps at the Dirac cones around the $\Gamma$ and $M$ points in the Brillouin zone.} \label{Fig:Schematic_Altermagnet}
\end{figure}

In this work, we demonstrate that a $d$-wave altermagnet can be driven into a Chern-insulating phase by irradiation with elliptically polarized, high-frequency light. As schematically illustrated in Fig.~\ref{Fig:Schematic_Altermagnet}, light irradiation plays a crucial role in activating intrinsic thermoelectric and thermal Hall responses in the system. In the absence of irradiation, symmetry-protected Dirac cones at the 
$\Gamma$ and $M$ points of the Brillouin zone remain gapless, leading to the complete suppression of intrinsic thermoelectric and thermal Hall currents. Upon exposure to elliptically polarized light, these Dirac points acquire finite gaps through Floquet-induced symmetry breaking, generating a nonzero Berry curvature in momentum space. Consequently, intrinsic thermoelectric and thermal Hall currents are simultaneously induced under the temperature gradient.
Beyond the topological phase transition, we further investigate the intrinsic anomalous thermoelectric and thermal Hall effects in the light-driven altermagnetic phase. At extremely low temperatures, the thermoelectric Hall coefficient exhibits a linear temperature dependence and vanishes inside the bulk energy gap between the conduction and valence bands. In contrast, pronounced peaks and dips emerge at the gap edges near both the $M$ and $\Gamma$ points, indicating that the thermoelectric Hall response serves as a sensitive probe of the Floquet-induced band gaps. In parallel, the low-temperature thermal Hall coefficient also displays linear temperature scaling, while the thermal Hall conductivity becomes quantized across the bandwidth. This quantization underscores the potential of thermal Hall transport as a robust probe of the system's underlying topological character.

The remainder of this paper is organized as follows. In Section~\ref{2}, we present the general formalism for calculating the intrinsic anomalous thermoelectric and thermal Hall conductivities. In Section~\ref{3}, we derive the low-temperature expansions of the corresponding Hall transport coefficients using the Sommerfeld expansion. Section~\ref{4} introduces the Floquet Hamiltonian for a $d$-wave altermagnet. In Section~\ref{5}, we analytically derive the Berry curvature within a lattice regularization and analyze the associated Chern number. Numerical results are presented in Section~\ref{6}, followed by a discussion of the heating rate in Section~\ref{7}. In Section~\ref{8}, we provide the quantitative analysis of the anomalous thermoelectric and thermal Hall effects. Finally, Section~\ref{9} summarizes the main conclusions.

\section{Transport Coefficients}\label{2}

We start from the semiclassical equations of motion for electronic wavepackets under the temperature gradient $\nabla_{\bf r}T$~\cite{chang1995berry,sundaram1999wave,xiao2010berry,shen2017topological,qiang2023topological,qin2024light}
\begin{eqnarray}
&&\dot{\bf r}\!=\!\frac{1}{\hbar}\nabla_{\bf k}\varepsilon_{n,\bf k}^{} \!-\! \dot{\bf k}\times{\bf\Omega}_{n,\bf k},\label{eq:r}\\
&&\dot{\bf k}\!=\!-\frac{1}{\hbar}\mathbf{F}_{\nu,\bf k}(\nabla_{\bf r}T,\mu),\label{eq:k}
\end{eqnarray} where $\mathbf{r}$ and $\mathbf{k}$ denote the wavepacket's position and momentum, respectively, 
$\varepsilon_{n,\bf k}^{}$ is the band dispersion for $n$th band,
${\bf\Omega}_{n,\mathbf{k}}$ is the Berry curvature vector~\cite{li2016negative,lu2017quantum,qin2025emergent}, $\hbar\!=\! h/(2\pi)$ is the reduced Planck constant with Planck constant $h$, $-e$ is the electron charge, $k_B$ is the Boltzmann constant, $T$ is the system temperature, and $\mu$ is the chemical potential. The thermal driving term $\mathbf{F}_{\nu,\bf k}(\nabla_{\bf r}T,\mu)$ is given by
\begin{eqnarray}
\textbf{F}_{\nu,\bf k}(\nabla_{\bf r}T,\mu)\!=\!\int_{\varepsilon_{n,\bf k}^{}\!-\!\mu}^{\infty}\varepsilon^{\nu\!-\!1}\nabla_{\bf r}f_{0}(\varepsilon)d\varepsilon, \label{eq:F_0}
\end{eqnarray} 
where $\nu\!=\!1,2$, and $f_{0}(\varepsilon)\!=\!1/[e^{\varepsilon/(k_{B}T)}\!+\!1]$ is the Fermi-Dirac distribution function.
Here, we set $\varepsilon_{n,\bf k}^{}\!<\!\mu$ in the lower limit of Eq.~\eqref{eq:F_0}.

The non-equilibrium distribution function $f(\mathbf{r},\mathbf{k},t)$ satisfies the Boltzmann equation~\cite{chang1995berry,sundaram1999wave,xiao2010berry,shen2017topological,qiang2023topological,qin2024light,qin2025nonlinear,qin2014thermal}
\begin{eqnarray}\label{eq:Boltzmann_1}
\dot{\bf r}\cdot\frac{\partial f}{\partial{\bf r}} \!+\! \dot{\bf k}\cdot\frac{\partial f}{\partial{\bf k}} \!+\! \frac{\partial f}{\partial t} \!=\! I\{f\}.
\end{eqnarray}
Within the relaxation time approximation~\cite{qin2024light,qin2025nonlinear,qin2014thermal}, the collision term is  
\begin{eqnarray}\label{eq:coll}
I\{f\} \!=\! -\frac{f-f_{0}}{\tau},
\end{eqnarray} where $\tau$ is the relaxation time.
Substituting Eqs.~\eqref{eq:r}, \eqref{eq:k}, and \eqref{eq:coll} into Eq.~\eqref{eq:Boltzmann_1} allows one to expand the distribution function in powers of the thermal field, 
\begin{eqnarray}\label{eq:Boltzmann_3}
f(\varepsilon_{n,\bf k}^{}\!-\!\mu) 
\!\simeq\!\sum_{m=0}^{\infty} f_{m}(\varepsilon_{n,\bf k}^{}\!-\!\mu),
\end{eqnarray} where $f_{m}(\varepsilon_{n,\bf k}^{}\!-\!\mu)$ denotes the $m$th-order contribution in $\tau |\mathbf{F}_{\nu,\mathbf{k}}| / \hbar$, i.e., in the regime of small temperature gradients. 

In the following, we focus on the intrinsic contributions, which are independent of the relaxation time.
The intrinsic anomalous thermoelectric current density is then expressed as (see Sec. SI of the Supplemental Material~\cite{SuppMat}) 
\begin{eqnarray}\label{eq:Jte_in}
\textbf{J}^{\rm in}_{te}(\nabla_{\bf r}T,\mu) 
\!=\! -\frac{e}{V}\sum_{n,\bf k}\dot{\textbf{R}}_{1}f_{0}(\varepsilon_{n,\bf k}^{}\!-\!\mu),
\end{eqnarray} and the intrinsic anomalous thermal current density can be obtained as
\begin{eqnarray}\label{eq:Jt_in}
\textbf{J}_{t}^{\rm in}(\nabla_{\bf r}T,\mu) 
\!=\! \frac{1}{V}\sum_{n,\bf k}\dot{\textbf{R}}_{2}f_{0}(\varepsilon_{n,\bf k}^{}\!-\!\mu).
\end{eqnarray} where 
\begin{eqnarray}
\dot{\textbf{R}}_{\nu}\!=\!\frac{1}{\hbar}\mathbf{F}_{\nu,\bf k}(\nabla_{\bf r}T,\mu)\times{\bf\Omega}_{n,\bf k}.
\end{eqnarray}
From Eqs.~\eqref{eq:Jte_in} and \eqref{eq:Jt_in}, the intrinsic anomalous thermoelectric (Nernst or Peltier), and thermal Hall conductivities can be written as (see Sec. SII of the Supplemental Material~\cite{SuppMat}) 
\begin{eqnarray}
\alpha_{xy}^{\rm in} 
&\!=\!& \frac{ek_{B}}{h}{\cal C}_{1}^{\rm in},\label{eq:alpha_xy_in}\\
\kappa_{xy}^{\rm in} 
&\!=\!& -\frac{k_{B}^{2}T}{h}{\cal C}_{2}^{\rm in},\label{eq:kappa_xy_in}
\end{eqnarray} where
\begin{eqnarray}
{\cal C}_{\nu}^{\rm in}\!=\!\chi_{xy}^{\rm in}(\mu)\int_{\varepsilon_{n,\bf k}^{}\!-\!\mu}^{\infty}d\varepsilon\left[-\frac{\partial f_{0}(\varepsilon)}{\partial\varepsilon}\right]\left(\frac{\varepsilon}{k_{B}T}\right)^{\nu},\label{eq:C_n_in}
\end{eqnarray}
with $\nu\!=\!1,2$, and
\begin{eqnarray}
\chi_{xy}^{\rm in}(\mu)\!=\!\frac{2\pi}{V}\sum_{n,{\bf k}}\Omega_{n,\bf k}^{xy}f_{0}(\varepsilon_{n,\bf k}\!-\!\mu).\label{eq:chi_0}
\end{eqnarray}

\section{Low-temperature limit}\label{3}

At extremely low temperatures $T\!\to\!0$, the quantity $\chi_{xy}^{\rm in}(\mu)$ [Eq.~\eqref{eq:chi_0}] reduces to
\begin{eqnarray}
\chi_{xy}^{\rm in}(\mu)&\!\to\!&\frac{2\pi}{V}\sum_{n,{\bf k}}\Omega_{n,\bf k}^{xy}\Theta(\mu\!-\!\varepsilon_{n,\bf k}),\label{eq:chi_T0}
\end{eqnarray} where $\Theta(x)$ is the Heaviside step function~\cite{qin2015three,qin2016universal,qin2018high}.

Applying the Sommerfeld expansion~\cite{pathria1996statistical,qin2009comparative,qin2010finite,qin2011joule,qin2012adiabatic,qin2012pauli,qin2014thermal,qiang2023topological} (see Sec. SIII of the Supplemental Material~\cite{SuppMat}), the intrinsic anomalous thermoelectric and thermal Hall conductivities simplify to
\begin{eqnarray}
&&\alpha_{xy}^{\rm in} 
\xrightarrow{T\!\to\!0} \frac{\pi^2}{3}\frac{ek_{B}^{2}}{h}T\frac{2\pi}{V}\sum_{n,{\bf k}}\Omega_{n,\bf k}^{xy}\delta(\mu\!-\!\varepsilon_{n,\bf k}),\label{eq:alpha_xy_in_LT}\\
&&\kappa_{xy}^{\rm in} 
\xrightarrow{T\!\to\!0} -\frac{\pi^2}{3}\frac{k_{B}^{2}}{h}T\frac{2\pi}{V}\sum_{n,{\bf k}}\Omega_{n,\bf k}^{xy}\Theta(\mu\!-\!\varepsilon_{n,\bf k}),\label{eq:kappa_xy_in_LT}
\end{eqnarray}  where $\delta(x)$ is the Dirac delta function and we used the relation $\delta(x)\!=\!d\Theta(x)/dx$.
The delta function can be represented in the low-temperature limit as a Lorentzian~\cite{qin2025emergent,qin2024light}
\begin{eqnarray}
\delta(\mu\!-\!\varepsilon_{n,\bf k}) \!=\! \lim_{k_{B}T\to0}\frac{1}{\pi}\frac{k_{B}T}{(\mu\!-\!\varepsilon_{n,\bf k})^{2} \!+\! (k_{B}T)^{2}}.
\end{eqnarray}
Equation~\eqref{eq:alpha_xy_in_LT} shows that the intrinsic thermoelectric coefficient is proportional to the Berry curvature at the Fermi energy, and thus can be used to probe the Berry phase structure by varying the chemical potential $\mu$. In contrast, the intrinsic thermal Hall coefficient [Eq.~\eqref{eq:kappa_xy_in_LT}] sums the Berry curvature over all occupied states, and therefore reflects the topological properties of the filled bands.

From these low-temperature expressions, universal relations naturally emerge. The Mott relation~\cite{qiang2023topological} follows as
\begin{eqnarray}
\alpha_{xy}^{\rm in} \!=\!-eL_{0}T\frac{\partial\sigma_{xy}^{\rm in}}{\partial\mu},
\end{eqnarray} and the Wiedemann-Franz law~\cite{qiang2023topological} is
\begin{eqnarray}
\frac{\kappa_{xy}^{\rm in}}{\sigma_{xy}^{\rm in}} \!=\!L_{0}T,
\end{eqnarray} where $L_{0}\!=\!(\pi k_{B}/e)^{2}/3$ is the Lorentz number and $\sigma_{xy}^{\rm in}$ is the zero-temperature intrinsic anomalous electric Hall conductivity that is given by
\begin{eqnarray}
\sigma_{xy}^{\rm in} 
\!=\!-\frac{e^{2}}{h}\frac{2\pi}{V}\sum_{n,{\bf k}}\Omega_{n,\bf k}^{xy}\Theta(\mu\!-\!\varepsilon_{n,\bf k}).\label{eq:sigma_xy_in_T0}
\end{eqnarray}
These results highlight that at $T \to 0$, the intrinsic thermoelectric and thermal Hall responses are fully determined by the topology of the electronic bands.

\section{Model}\label{4}

We consider an effective low-energy Hamiltonian describing a two-dimensional $d$-wave altermagnet~\cite{smejkal2022beyond,ghorashi2025dynamical},
\begin{eqnarray}\label{eq:H_d}
\hat{\cal H}\!=\! v(k_{y}\sigma_{x} \!-\! k_{x}\sigma_{y}) \!+\! J_{d}(k_{y}^{2} \!-\! k_{x}^{2})\sigma_{z},
\end{eqnarray} where $\sigma_{i}^{}$ ($i\!=\!x,y,z$) are Pauli matrices acting in spin space. The parameters $v$ and $J_{d}$ characterize the strength of spin-orbit coupling and altermagnetic order, respectively. In particular, the second term in Eq.~\eqref{eq:H_d} arises from the proximity-induced $d$-wave spin splitting.

We subject the system to a time-periodic optical field propagating along the $z$ direction, described by the electric field ${\bf E}(t) \!=\! (E_x\cos(\omega t), E_y\cos(\omega t\!+\!\varphi))$. The phase $\varphi$ controls the polarization, with $\varphi\!=\!0$ corresponding to linear polarization and $\varphi\!=\!\mp\pi/2$ to left- and right-handed elliptic polarization. Introducing the vector potential ${\bf A}(t)\!=\!\omega^{-1}(E_x\sin(\omega t),E_y\sin(\omega t\!+\!\varphi))$, which is periodic with period $\tilde{T}\!=\!2\pi/\omega$, the driven Hamiltonian follows from minimal coupling,
\begin{eqnarray}\label{eq:Ht}
\hat{\cal H}({\bf k},t)
\!=\!\hat{\cal H}\left[k_x \!-\! A_{x}\sin(\omega t),k_y\!-\!A_{y}\sin(\omega t \!+\! \varphi)\right],
\end{eqnarray} where we define $A_{x}\!=\!eE_{x}/(\hbar\omega)$ and $A_{y}\!=\!eE_{y}/(\hbar\omega)$.

\begin{figure}[htpb]
\centering
\includegraphics[width=1\columnwidth]{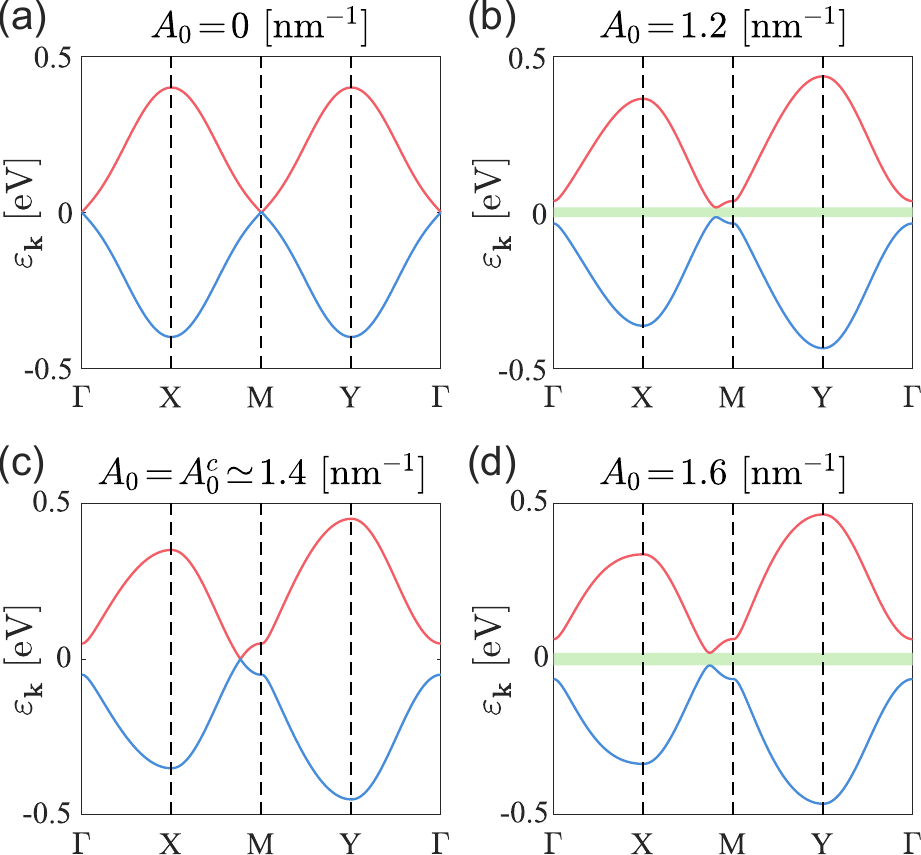}
\caption{Band structures [Eq.~\eqref{eq:Ek}] of altermagnets in the absence ($A_{0}\!=\!0$) and presence ($A_{0}\!\neq\!0$) of optical irradiation. (a) $A_{0}\!=\!0$, (b) $A_{0}\!=\!1.2$ nm$^{-1}$, (c) $A_{0}\!=\!A_{0}^{c}\simeq1.4$ nm$^{-1}$, and (d) $A_{0}\!=\!1.6$ nm$^{-1}$. The conduction and valence bands are shown in red and blue, respectively, while the green shaded regions denote the bandwidths, which remain much smaller than the driving optical frequency. Other parameters are $v\!=\!0.1$ eV$\cdot$nm, $J_{d}\!=\!0.1$ eV$\cdot$nm$^2$, $\alpha\!=\!1$, $\hbar\omega\!=\!0.4$ eV, $\varphi\!=\!\pi/2$, and $a\!=\!1$ nm.} 
\label{Fig:E_Gamma_X_M_Y_TB_together}
\end{figure}

We focus on the off-resonant high-frequency regime, where Floquet replicas are well separated and a Magnus expansion is applicable~\cite{magnus1954exponential,blanes2009magnus,kuwahara2016floquet}. Throughout this work, we choose a representative driving frequency $\hbar\omega\!=\!0.4$ eV~\cite{qin2022phase,qin2023light,qin2022light}, which largely exceeds the bandwidth in our chosen material in the following discussions. The effective static Floquet Hamiltonian is given by~\cite{oka2009photovoltaic,qin2023light,qin2022light,qin2022phase}
\begin{equation}\label{eq:HF0}
\hat{\cal H}^{(F)}({\bf k}) \!=\! \hat{\cal H}_{0} + \sum_{n=1}^{\infty}\frac{[\hat{\cal H}_{-n}, \hat{\cal H}_{n}]}{n\hbar\omega}+{\cal O}(\omega^{-2}),
\end{equation}
where $\hat{\cal H}_{n} \!=\! (1/\tilde{T}) \int_{0}^{\tilde{T}}\hat{\cal H}({\bf k},t) e^{in\omega t}dt$ denotes the Fourier components. For the Hamiltonian in Eq.~\eqref{eq:H_d}, only the $n\!=\!1$ term contributes, yielding (see Section SIV of the Supplemental Material~\cite{SuppMat})
\begin{eqnarray}
\hat{\cal H}^{(F)}({\bf k}) &\!=\!& 
k_{y}\left(v \!-\! v_{0}^{}\right)\sigma_{x} \!-\! k_{x}\left(v \!+\! v_{0}^{}\right)\sigma_{y} \nonumber\\
&&\!+  \left[J_{d}(k_{y}^{2} \!-\! k_{x}^{2}) \!+\! J_{0}\right]\!\sigma_{z},
\label{eq:H_F}
\end{eqnarray} where we set that $A_{y}\!=\!\alpha A_{x}\!=\!\alpha A_{0}$, and 
\begin{eqnarray} 
v_{0}^{}&\!=\!&\frac{2\alpha J_{d}A_{0}^{2}\sin\varphi}{\hbar\omega}v,\\
J_{0}&\!=\!&\frac{1}{2}\left(\alpha^{2} \!-\! 1\right)A_{0}^{2}J_{d} \!+\! \frac{\alpha v^{2}A_{0}^{2}\sin\varphi}{\hbar\omega}.
\end{eqnarray}
The light-induced Dirac mass is therefore
\begin{eqnarray}
M\!=\!J_{0}. \label{eq:M}
\end{eqnarray}
For simplicity, we set $\alpha\!=\!1$ in the following. To assess the validity of the high-frequency expansion, we estimate the maximal instantaneous energy of the driven Hamiltonian at the $\Gamma$ point and require $(1/\tilde{T})\int_{0}^{\tilde{T}}dt~\text{max}\left\{\big|\big|\hat{\cal H}({\bf k},t)\big|\big|\right\}\!<\!\hbar\omega$. This yields the constraint $A_{0}\!<\!{\rm min}\left(\hbar\omega/v,\sqrt{\hbar\omega/J_{d}}\right)$.
Using representative parameters $v\!=\!0.1$ eV$\cdot$nm and $J_{d}\!=\!0.1$ eV$\cdot$nm$^2$, which are comparable to those of the altermagnetic thin film with spin-orbit coupling~\cite{jiang2025metallic,wang2025atomic,yang2026visualizing,grutter2021magnetic,zhang2009topological,chang2013experimental,mogi2022experimental,marfoua2025strain}, we obtain $A_{0}\!\sim\!eE_{x}/(\hbar\omega)\!<\!2$ nm$^{-1}$. 
Furthermore, numerical calculations for the lattice model confirm that the energy gap near the $M$ point is smaller than that at the $\Gamma$ point, as shown in Fig.~\ref{Fig:E_Gamma_X_M_Y_TB_together}. This ensures that the high-frequency expansion remains well controlled throughout the Brillouin zone.

The electric field amplitude is related to the vector potential amplitude through
$E_{0}\!=\!A_{0}\hbar\omega/e$, where $E_{0}\!=\!E_{x}$ denotes the electric field amplitude. For a fixed value of $A_{0}$, the electric field amplitude $E_{0}$ increases linearly with the driving frequency $\omega$. The corresponding light intensity is given by~\cite{qin2024light} $I\!=\!\frac{1}{2}\varepsilon_{0}cn\left|E_{0}\right|^{2}$, where $\varepsilon_{0}$ is the permittivity of vacuum, $c$ is the velocity of light in vacuum, and $n$ is the relative refractive index.

To maintain the off-resonant Floquet regime, where the photon energy $\hbar\omega$ remains significantly larger than the electronic bandwidth, the vector potential amplitude should satisfy $A_{0}\!<\!2$ nm$^{-1}$ for the parameters considered in this work. Under this condition, the required electric field amplitude is below approximately $10^{8}$ V/m, corresponding to peak optical intensities below approximately $10^{15}$ W/m$^2$.

These values are experimentally accessible. For comparison, the upper bound $A_{0}\!\sim\!2$ nm$^{-1}$ corresponds to an electric field amplitude of approximately $10^{8}$ V/m, which is substantially smaller than the peak fields reported in ultrafast optical experiments, including $3.0\!\times\!10^9$ V/m~\cite{wang2018light,bao2022light} and $1.06\!\times\!10^{10}$ V/m~\cite{sentef2015theory}. Therefore, the optical driving strength required in our proposal is within experimentally achievable ranges.

\section{Berry Curvature and Topological Invariant}\label{5}

To obtain the Berry curvature analytically within a lattice regularization, we map the continuum Floquet Hamiltonian in Eq.~\eqref{eq:H_F} onto a square-lattice tight-binding model (see Section SV of the Supplemental Material~\cite{SuppMat}) 
\begin{eqnarray}
\hat{\cal H}_{\rm TB}^{(F)}({\bf k}) \!=\! \sum_{i=x,y,z}h_{i}^{(F)}\sigma_{i},\label{eq:H_F_TB}
\end{eqnarray} where
\begin{eqnarray}
h_{x}^{(F)}&\!=\!&t_{x}\sin(k_{y}a), \label{eq:hx_F}\\
h_{y}^{(F)}&\!=\!&-t_{y}\sin(k_{x}a), \label{eq:hy_F}\\
h_{z}^{(F)}&\!=\!&2J\left[\cos(k_{x}a) \!-\! \cos(k_{y}a)\right] \!+\! J_{0}. \label{eq:hz_F}
\end{eqnarray} Here, $t_{x}\!=\!(v \!-\! v_{0}^{})/a$, $t_{y}\!=\!(v \!+\! v_{0}^{})/a$, $J\!=\!J_{d}/a^{2}$, and $a_{x}\!=\!a_{y}\!=\!a$ is the lattice constant. 
The corresponding Floquet quasienergy spectra are given by
\begin{eqnarray}
\varepsilon_{\pm,\bf k}^{}\!=\! \pm\! \sqrt{\left[h_{x}^{(F)}\right]^{2} \!+\! \left[h_{y}^{(F)}\right]^{2} \!+\! \left[h_{z}^{(F)}\right]^{2}},\label{eq:Ek}
\end{eqnarray} where the upper (lower) sign denotes the conduction (valence) band.

\begin{figure}[htpb]
\centering
\includegraphics[width=1\columnwidth]{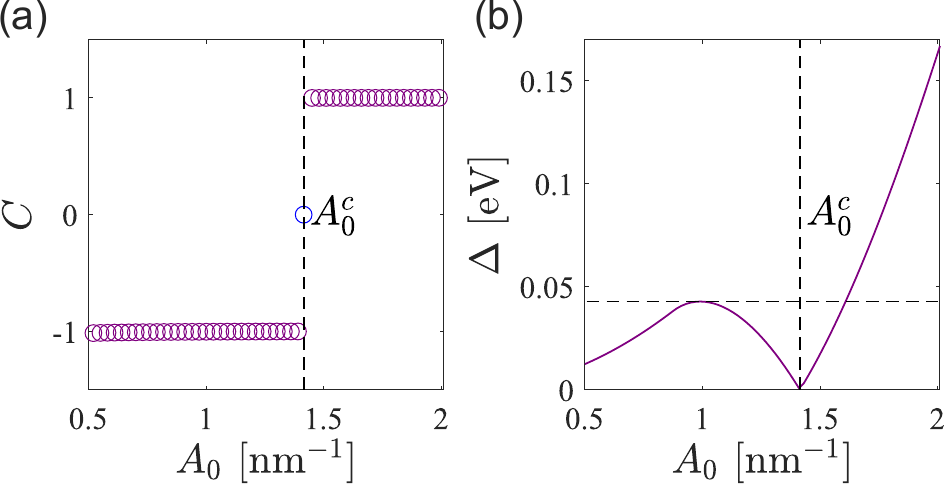}
\caption{(a) Chern number $C$ for the valence band [Eq.~\eqref{eq:C_0}] as a function of the light amplitude $A_{0}$ in the gapped phase at $\mu\!=\!0$. (b) Bandwidth $\Delta$ versus the light amplitude $A_{0}$. All other parameters are the same as those used in Fig.~\ref{Fig:E_Gamma_X_M_Y_TB_together}.} \label{Fig:C_Gap_phi05pi_vs_A0_analytic_TB_together}
\end{figure}

The Berry curvature associated with the Floquet bands is given by~\cite{thouless1982quantized,qin2023light,qin2024kinked,qin2024light,qin2025nonlinear,qin2025emergent} (see Section SVI of the Supplemental Material~\cite{SuppMat})
\begin{eqnarray}
\Omega_{\pm,\bf k}^{xy} 
\!=\!\frac{\pm (v^{2} \!-\! v_{0}^{2})J_{u}}{2\left[ t_{x}^{2}\sin^{2}(k_{x}a) \!+\! t_{y}^{2}\sin^{2}(k_{y}a) \!+\! J_{\text{TB}}^{2} \!\right]^{3/2}},\label{eq:Omega_xy_F} 
\end{eqnarray} where $J_{u}\!=\!2J\left[\cos(k_{x}a) \!-\! \cos(k_{y}a)\right] \!-\! J_{0}\cos(k_{x}a)\cos(k_{y}a)$, $J_{\text{TB}}\!=\!2J\left[\cos(k_{x}a) \!-\! \cos(k_{y}a)\right] \!+\! J_{0}$, and the symbol ``$\pm$'' corresponds to the conduction and valence band states.

With the Berry curvature \eqref{eq:Omega_xy_F}, we can obtain the Chern number $C$ for the valence band (occupied band) as
\begin{eqnarray}
C\!=\!-\frac{2\pi}{V}\sum_{{\bf k}}\Omega_{-,\bf k}^{xy}\!=\!-2\pi\int\frac{d^{2}{\bf k}}{(2\pi)^{2}}\Omega_{-,\bf k}^{xy},\label{eq:C_0}
\end{eqnarray} where we have used $(1/V)\sum_{\bf k}\!\to\!\int d^{2}{\bf k}/(2\pi)^{2}$ for a two-dimensional system~\cite{pathria1996statistical,qin2009comparative,qin2010finite,qin2011joule,qin2012adiabatic,qin2012pauli,qin2014thermal}.

From Eq.~\eqref{eq:Omega_xy_F}, the Berry curvature vanishes when $v\!=\!v_{0}$, corresponding to the critical light amplitude $A_{0}^{c}\!=\!\sqrt{\hbar\omega/(2\alpha J_{d}\sin\varphi)}$, which signals a topological phase transition, as illustrated in Figs.~\ref{Fig:E_Gamma_X_M_Y_TB_together} and \ref{Fig:C_Gap_phi05pi_vs_A0_analytic_TB_together}.

Importantly, spin-orbit coupling plays an essential role in realizing the light-induced topological phase transition. In particular, setting $v\!=\!0$ in Eq.~\eqref{eq:H_d} leads to a vanishing Berry curvature in Eq.~\eqref{eq:Omega_xy_F}, which in turn results in a zero Chern number in Eq.~\eqref{eq:C_0}. Therefore, a finite spin-orbit coupling is indispensable in our proposal.

Comparing Figs.~\ref{Fig:E_Gamma_X_M_Y_TB_together}(a) and \ref{Fig:E_Gamma_X_M_Y_TB_together}(b), we observe that optical irradiation breaks the $\hat{\cal C}_{4}\hat{\cal T}$ symmetry and opens energy gaps at both the $\Gamma$ and $M$ points. Each gapped high-symmetry point contributes a half-integer Chern number, $|C|\!=\!1/2$, yielding a total Chern number $|C|\!=\!1$ for the gapped phase, as shown in Fig.~\ref{Fig:C_Gap_phi05pi_vs_A0_analytic_TB_together}(a).

As the light amplitude increases to the critical value $A_{0}^{c}$, the gap near the $M$ point closes [Fig.~\ref{Fig:E_Gamma_X_M_Y_TB_together}(c)] and subsequently reopens for larger $A_{0}$ [Fig.~\ref{Fig:E_Gamma_X_M_Y_TB_together}(d)], signaling a band inversion across the topological phase transition [Fig.~\ref{Fig:C_Gap_phi05pi_vs_A0_analytic_TB_together}(b)]. Importantly, the light-induced band inversion occurs in the vicinity of the $M$ point rather than exactly at $M$, as illustrated in Fig.~\ref{Fig:E_Gamma_X_M_Y_TB_together}(c). At the transition point, when the gap closes near the $M$ point, the gapped $\Gamma$ and $M$ points each contribute a half-integer Chern number with opposite signs, resulting in a total Chern number $C\!=\!0$ for the gapless phase, as indicated by the blue circle in Fig.~\ref{Fig:C_Gap_phi05pi_vs_A0_analytic_TB_together}(a).

\section{Numerical Results}\label{6}

In the following, we present numerical results for the intrinsic anomalous thermoelectric and thermal Hall conductivities, $\alpha_{xy}^{\rm in}$ and $\kappa_{xy}^{\rm in}$, in the low-temperature limit $T \!\to\! 0$ within topological regions.

Substituting the zero-temperature intrinsic anomalous electric Hall conductivity \eqref{eq:sigma_xy_in_T0} into Eqs.~\eqref{eq:alpha_xy_in_LT} and \eqref{eq:kappa_xy_in_LT}, the low-temperature expressions reduce to 
\begin{eqnarray}
\alpha_{xy}^{\rm in} 
&\!\xrightarrow{T\!\to\!0}\!& -\frac{\pi^2}{3}\frac{ek_{B}^{2}}{h}T\frac{\partial\tilde{\sigma}_{xy}^{\rm in}}{\partial\mu}
\!=\!-\frac{e^3}{h}L_{0}T\frac{\partial\tilde{\sigma}_{xy}^{\rm in}}{\partial\mu},\label{eq:alpha_xy_in_LT_C}\\
\kappa_{xy}^{\rm in} 
&\!\xrightarrow{T\!\to\!0}\!& \frac{\pi^2}{3}\frac{k_{B}^{2}}{h}T\tilde{\sigma}_{xy}^{\rm in}
\!=\!\frac{e^2}{h}L_{0}T\tilde{\sigma}_{xy}^{\rm in},\label{eq:kappa_xy_in_LT_C}
\end{eqnarray} where $L_{0}\!=\!(\pi k_{B}/e)^{2}/3$ is the Lorentz number and 
$\tilde{\sigma}_{xy}^{\rm in}\!=\!\sigma_{xy}^{\rm in}/(e^{2}/h)$ is the dimensionless zero-temperature intrinsic anomalous electric Hall conductivity.

Equations~\eqref{eq:alpha_xy_in_LT_C} and \eqref{eq:kappa_xy_in_LT_C} indicate that, at low temperatures, the intrinsic anomalous thermoelectric and thermal Hall conductivities are directly determined by the zero-temperature intrinsic anomalous electric Hall conductivity \eqref{eq:sigma_xy_in_T0}, reflecting the topological nature of the system.

\begin{figure}[htpb]
\centering
\includegraphics[width=1\columnwidth]{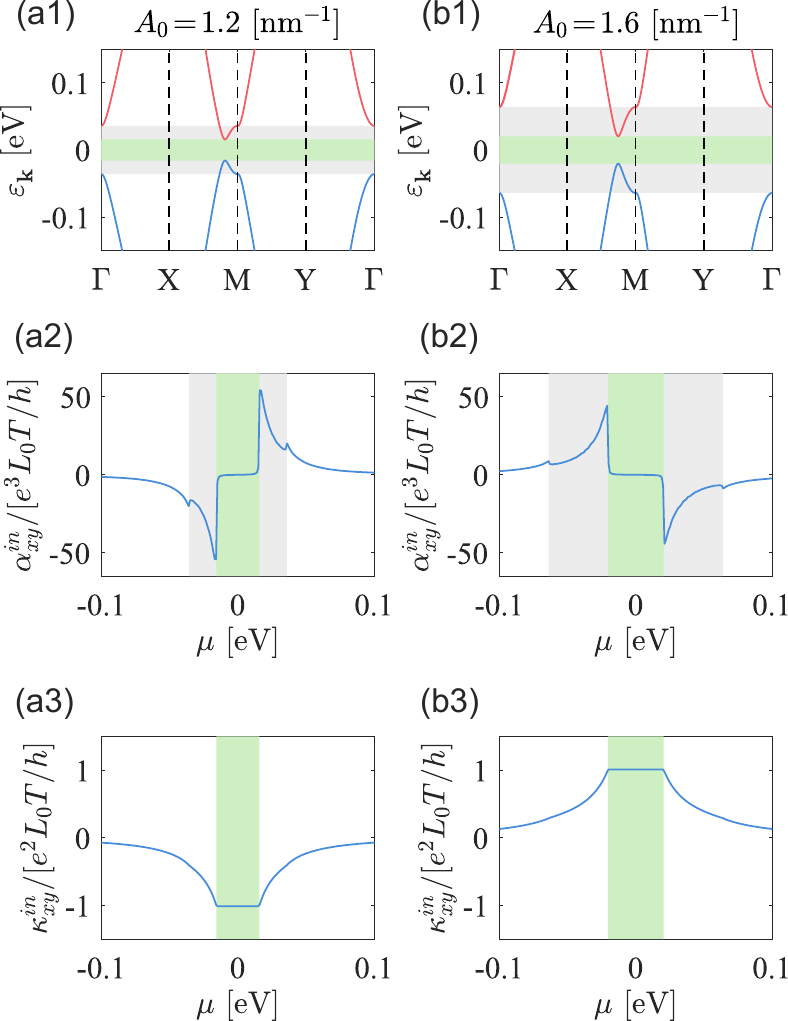}
\caption{Top row: [(a1) and (b1)] Band structures [Eq.~\eqref{eq:Ek}] for (a1) $A_{0}\!=\!1.2$ nm$^{-1}$ and (b1) $A_{0}\!=\!1.6$ nm$^{-1}$. The red and blue curves denote the conduction and valence bands, respectively. The green shaded regions highlight the energy gaps between the conduction and valence bands, while the gray shaded regions indicate the bandwidths near the $\Gamma$ point.
Middle row: [(a2) and (b2)] Reduced thermoelectric Hall coefficient $\alpha_{xy}^{\rm in}/(e^{3}L_{0}T/h)$ [Eq.~\eqref{eq:alpha_xy_in_LT_C}] as a function of the chemical potential $\mu$ at $T\!=\!2$ K for (a2) $A_{0}\!=\!1.2$ nm$^{-1}$ and (b2) $A_{0}\!=\!1.6$ nm$^{-1}$.
Bottom row: [(a3) and (b3)] Reduced thermal Hall coefficient $\kappa_{xy}^{\rm in}/(e^{2}L_{0}T/h)$ [Eq.~\eqref{eq:kappa_xy_in_LT_C}] versus the chemical potential $\mu$ for (a3) $A_{0}\!=\!1.2$ nm$^{-1}$ and (b3) $A_{0}\!=\!1.6$ nm$^{-1}$.
All other parameters are the same as those used in Fig.~\ref{Fig:E_Gamma_X_M_Y_TB_together}.}
\label{Fig:E_alpha_kappa_phi05pi_Alter_TB_T20K_together}
\end{figure}

Figures~\ref{Fig:E_alpha_kappa_phi05pi_Alter_TB_T20K_together}(a1) and \ref{Fig:E_alpha_kappa_phi05pi_Alter_TB_T20K_together}(b1) show the band structures, where the green shaded regions denote the energy gaps between the conduction and valence bands, and the gray shaded regions highlight the bandwidths near the $\Gamma$ point.

Figures~\ref{Fig:E_alpha_kappa_phi05pi_Alter_TB_T20K_together}(a2) and \ref{Fig:E_alpha_kappa_phi05pi_Alter_TB_T20K_together}(b2) show the reduced thermoelectric Hall coefficient $\alpha_{xy}^{\rm in}/(e^3 L_0 T/h)$ as a function of the chemical potential $\mu$ at $T\!=\!2$ K. The coefficient vanishes within the gapped regions (green shaded areas) but exhibits peaks and dips at the edges of the green and gray regions. This behavior indicates that the thermoelectric Hall conductivity is sensitive to both the bandwidth and the gapped regions near the $\Gamma$ point, and can thus serve as a probe of the electronic structure.

Figures~\ref{Fig:E_alpha_kappa_phi05pi_Alter_TB_T20K_together}(a3) and \ref{Fig:E_alpha_kappa_phi05pi_Alter_TB_T20K_together}(b3) show the reduced thermal Hall coefficient $\kappa_{xy}^{\rm in}/(e^2 L_0 T/h)$ versus $\mu$. In contrast to the thermoelectric response, the thermal Hall conductivity is quantized in the gapped regions (green shaded areas), demonstrating that it provides a direct signature of the topological phase.

Together, these results highlight that while $\alpha_{xy}^{\rm in}$ can map both band edges and bandwidths, $\kappa_{xy}^{\rm in}$ serves as a robust probe of the topological regions in irradiated altermagnets.

\section{Discussion}\label{7}

In Floquet-engineered systems, photon absorption may result in heating and thereby obscure the intrinsic topological responses. Therefore, evaluating the influence of heating effects is an important consideration for the present proposal.

In the present work, we intentionally operate in the off-resonant (high-frequency) regime, where the photon energy $\hbar\omega\!=\!0.4$ eV is much larger than the bandwidth of the Floquet bands over the range of driving amplitudes considered, $0\!\leqslant\! A_{0}\!\leqslant\!1.6$ nm$^{-1}$ (see Fig.~\ref{Fig:E_Gamma_X_M_Y_TB_together}). This regime is essential for suppressing resonant photon absorption. When the driving frequency is sufficiently high, photon-induced electronic transitions are strongly suppressed because there are no available electronic states within the photon-energy window required for resonant excitation. Consequently, the dominant effect of the periodic driving is a virtual modification of the electronic band structure, which is accurately captured by the Floquet-Magnus expansion~\cite{magnus1954exponential,blanes2009magnus,kuwahara2016floquet}, rather than direct excitation of charge carriers.

From a theoretical perspective, the heating rate of periodically driven systems has been extensively studied~\cite{kuwahara2016floquet,abanin2015exponentially,mori2016rigorous,mori2018thermalization,bukov2016heating,abanin2017effective}. For clean periodically driven systems at low temperatures, Abanin et al.~\cite{abanin2015exponentially} showed within linear-response theory that the energy absorption rate is exponentially suppressed in the high-frequency regime. This result was subsequently established rigorously and generalized beyond linear response in Refs.~\cite{mori2016rigorous,kuwahara2016floquet}. In particular, Mori \emph{et al}.~\cite{mori2016rigorous} derived an upper bound for the energy absorption rate per lattice site, $\frac{\dot{E}}{N}\!\lesssim\!\frac{N_V}{N}
\exp[-\mathcal{O}(\omega/g)]$, where $E$ is the total energy, $N$ is the total number of lattice sites, $N_V$ is the number of driven sites, $g$ represents the maximum local energy scale per site, and $\omega$ is the driving frequency. This result indicates that the heating time increases exponentially with the driving frequency, leading to a long-lived prethermal Floquet regime~\cite{mori2018thermalization}. These theoretical predictions have been further supported by numerical simulations~\cite{bukov2016heating} and subsequent analytical studies~\cite{abanin2017effective}. 

Based on these results, the heating rate is expected to be strongly suppressed when the driving frequency greatly exceeds the intrinsic electronic energy scales. For the parameters used in our work, $\hbar\omega\!=\!0.4$ eV, while the local maximum Floquet-induced bandwidth is only approximately $\Delta\!\approx\!0.04$ eV over the range $0\!\leqslant\! A_{0}\!\leqslant\!1.6$ nm$^{-1}$ [see Fig.~\ref{Fig:C_Gap_phi05pi_vs_A0_analytic_TB_together}(b)]. Therefore, the condition $\hbar\omega\!\gg\!\Delta$ is well satisfied, ensuring that photon-assisted absorption processes are strongly suppressed and that Floquet heating does not significantly affect the predicted topological transport properties within experimentally relevant time scales.

We also note that the proposed measurements are considered at a low temperature of $T\!=\!2$ K, where efficient heat dissipation can be achieved using standard low-temperature experimental platforms. Moreover, in thermal Hall measurements, the Hall response is determined from the transverse temperature gradient rather than the absolute sample temperature. Therefore, spatially uniform heating would mainly modify the background temperature and would not directly generate or obscure the transverse Hall signal.

\section{Quantitative Analysis}\label{8}

In this section, we provide quantitative estimates of the anomalous thermoelectric and thermal Hall conductivities using the parameters adopted in our model Hamiltonian.

The effective Hamiltonian in Eq.~\eqref{eq:H_d} is constructed to capture the essential low-energy physics of a $d$-wave altermagnet with Rashba spin-orbit coupling. Similar minimal effective models have been widely employed in the altermagnet literature (see, e.g., Refs.~\cite{smejkal2022anomalous,ghorashi2025dynamical,cheraghchi2026floquet,yarmohammadi2026floquet,amundsen2024rkky,rao2024tunable,duan2026neel}). Physically, Rashba spin-orbit coupling can be induced in altermagnetic systems through interface engineering or external electric fields, such as gate voltages~\cite{amundsen2024rkky}. Moreover, the coexistence of Rashba spin-orbit coupling and altermagnetic spin splitting has been demonstrated by first-principles calculations for monolayer V$_2$SeTeO~\cite{marfoua2025strain}.  

We emphasize that our Hamiltonian is a minimal low-energy effective model rather than a complete \emph{ab initio} description of a specific material. The tight-binding regularization in Eq.~\eqref{eq:H_F_TB} is a standard approach for preserving the Berry curvature distribution and topological invariants on a lattice. Therefore, the model provides a direct connection between the symmetry-allowed low-energy theory and material parameters.

For the thermoelectric Hall coefficient $\alpha_{xy}^{\rm in}$ [Eq.~\eqref{eq:alpha_xy_in_LT_C}], the derivative $\partial\tilde{\sigma}_{xy}^{\rm in}/\partial\mu$ vanishes inside the bulk gap, where $\tilde{\sigma}_{xy}^{\rm in}$ remains quantized, resulting in $\alpha_{xy}^{\rm in}\!=\!0$ as shown in Figs.~\ref{Fig:E_alpha_kappa_phi05pi_Alter_TB_T20K_together}(a2) and \ref{Fig:E_alpha_kappa_phi05pi_Alter_TB_T20K_together}(b2). Near the band edges, this derivative becomes finite, leading to pronounced peaks. For the case of $A_{0}\!=\!1.2$ nm$^{-1}$, the magnitude of these peaks can be estimated from Fig.~\ref{Fig:E_alpha_kappa_phi05pi_Alter_TB_T20K_together}(a2) as $\left|\partial\tilde{\sigma}_{xy}^{\rm in}/\partial\mu\right|\!\sim\!P_{\rm max}$ with $P_{\rm max}\!=\!54.1486$ eV$^{-1}$, so that the peak value can be estimated as
\begin{eqnarray}
\left|\tilde{\alpha}_{xy}^{\rm in}\right|\!\sim\!\frac{P_{\rm max}e^{3}L_{0}T}{h}
\!=\!\frac{P_{\rm max}e(\pi k_B)^2T}{3h}.
\end{eqnarray}
Taking $T\!=\!2$ K, $k_B\!\approx\!1.38\!\times\!10^{-23}$ J/K, $h\!\approx\!6.626\!\times\!10^{-34}$ J$\cdot$s, $e\!\approx\!1.602\!\times\!10^{-19}$ C, and $P_{\rm max}\!=\!54.1486$ eV$^{-1}$, we obtain 
\begin{eqnarray}
|\alpha_{xy}^{\rm in}|\!\sim\!10^{-10}~{\rm A/K}. 
\end{eqnarray}
Since experimental measurements are typically performed on thin films with finite thickness, we normalize this value by an effective film thickness $L_z\!\sim\!1$ nm for comparison with experimental measurements. The corresponding three-dimensional coefficient is therefore estimated as
\begin{eqnarray}
\left|\tilde{\alpha}_{xy}^{\rm in}\right|\!=\!
\left|\alpha_{xy}^{\rm in}/L_z\right|
\!\sim\!0.1~{\rm A/(K\cdot m)},
\end{eqnarray}
which is comparable to the anomalous Nernst coefficients experimentally reported in the altermagnetic candidate Mn$_5$Si$_3$~\cite{badura2025observation}.

For the Chern insulating phase with Chern number $\left|C\right|\!=\!1$, $\tilde{\sigma}_{xy}^{\rm in}$ becomes quantized within the bulk gap [Figs.~\ref{Fig:E_alpha_kappa_phi05pi_Alter_TB_T20K_together}(a3) and \ref{Fig:E_alpha_kappa_phi05pi_Alter_TB_T20K_together}(b3), green shaded regions]. Specifically, when the chemical potential lies inside the gap, $\left|\tilde{\sigma}_{xy}^{\rm in}\right|\!=\!\left|C\right|\!=\!1$. Therefore, at $T=2$ K, Eq.~\eqref{eq:kappa_xy_in_LT_C} gives
\begin{eqnarray}
\left|\kappa_{xy}^{\rm in}\right|
&\!=\!&
\frac{e^2L_0T}{h}\left|\tilde{\sigma}_{xy}^{\rm in}\right|
\!=\!
\frac{\pi^2k_B^2}{3h}T .
\label{eq:kappa_xy_in_LT_C_0}
\end{eqnarray}
Substituting $k_B\!\approx\!1.38\!\times\!10^{-23}$ J/K, $h\!\approx\!6.626\!\times\!10^{-34}$ J$\cdot$s, and $T\!=\!2$ K, we obtain
\begin{eqnarray}
\left|\kappa_{xy}^{\rm in}\right|\!\approx\!1.89\!\times\!10^{-12}~{\rm W/K}.
\end{eqnarray}
To compare with experimentally reported thermal Hall conductances, we again normalize by the film thickness $L_z\!\sim\!1$ nm, yielding
\begin{eqnarray}
\left|\tilde{\kappa}_{xy}^{\rm in}\right|
\!=\!
\left|\kappa_{xy}^{\rm in}/L_z\right|
\!\sim\!10^{-3}~{\rm W/(K\cdot m)}.
\end{eqnarray}
This value is comparable to the thermal Hall responses measured in topological materials (e.g., Ref.~\cite{xu2020finite}).

These estimates demonstrate that both the anomalous thermoelectric Hall response and the thermal Hall conductivity predicted in our Floquet-engineered altermagnetic system with Rashba spin-orbit coupling are within the detection capability of current experimental techniques, suggesting that the proposed phenomena are experimentally accessible.

\section{Summary}\label{9}

In summary, this work investigates how elliptically polarized light breaks $\hat{\cal C}_{4}\hat{\cal T}$ symmetry and induces gaps at both the $\Gamma$ and $M$ points in the spectrum of a $d$-wave altermagnet. The gap around the $\Gamma$ point gives rise to Chern bands with Chern number $|C|\!=\!1/2$, while the gap around the $M$ point contributes another Chern number $|C|\!=\!1/2$, leading to a total Chern number of $|C|\!=\!1$. This indicates that the $d$-wave altermagnet can be transformed into a Chern insulator under light irradiation with a high-frequency photon beam.

Particularly, there exists a critical value of the light amplitude at which a topological phase transition occurs. As the light amplitude increases to this critical point, the gap at the $\Gamma$ point remains open, but the gap near the $M$ point closes. Beyond the critical value, the gap near the $M$ point reopens, signifying a band inversion at the critical point.

At extremely low temperatures, the thermoelectric Hall coefficient, which shows a linear temperature dependence, vanishes within the gapped region (bandwidth) between the conduction and valence bands. However, it exhibits peaks and dips at the edges of the gap regions near both the $M$ and $\Gamma$ points, suggesting that thermoelectric Hall conductivity can be used to probe both the bandwidth and the gapped region around the $\Gamma$ point.

Similarly, the low-temperature thermal Hall coefficient, which shows a linear temperature dependence, becomes quantized in the gapped region (bandwidth) between the conduction and valence bands. This observation suggests that thermal Hall conductivity can serve as a probe for the topological properties of the system.

Analogous effects are predicted in $g$- and $i$-wave altermagnets: light irradiation can open gaps at high-symmetry points and induce anisotropic spin-orbit coupling. As a result, the predicted light-driven Chern insulating states, as well as the associated anomalous thermoelectric and thermal Hall effects, can also be realized in these systems.


\begin{acknowledgments}
This work is supported by the Guangdong Provincial Quantum Science Strategic Initiative (Grant No. GDZX2401001).
F.Q. acknowledges support from the Jiangsu Specially-Appointed Professor Program in Jiangsu Province and the Doctoral Research Start-Up Fund of Jiangsu University of Science and Technology. Xiao-Bin Qiang  acknowledges the program of China Scholarship Council (Grant No. 202508440272).
\end{acknowledgments}

\section*{Data Availability}
There are no publicly available research data or software supporting this manuscript. Requests for further information or data should be sent to the authors.

%
%
%
\twocolumngrid
\bibliography{references_Altermagnets_Therm}

\begin{thebibliography}{157}%
\makeatletter
\providecommand \@ifxundefined [1]{%
 \@ifx{#1\undefined}
}%
\providecommand \@ifnum [1]{%
 \ifnum #1\expandafter \@firstoftwo
 \else \expandafter \@secondoftwo
 \fi
}%
\providecommand \@ifx [1]{%
 \ifx #1\expandafter \@firstoftwo
 \else \expandafter \@secondoftwo
 \fi
}%
\providecommand \natexlab [1]{#1}%
\providecommand \enquote  [1]{``#1''}%
\providecommand \bibnamefont  [1]{#1}%
\providecommand \bibfnamefont [1]{#1}%
\providecommand \citenamefont [1]{#1}%
\providecommand \href@noop [0]{\@secondoftwo}%
\providecommand \href [0]{\begingroup \@sanitize@url \@href}%
\providecommand \@href[1]{\@@startlink{#1}\@@href}%
\providecommand \@@href[1]{\endgroup#1\@@endlink}%
\providecommand \@sanitize@url [0]{\catcode `\\12\catcode `\$12\catcode
  `\&12\catcode `\#12\catcode `\^12\catcode `\_12\catcode `\%12\relax}%
\providecommand \@@startlink[1]{}%
\providecommand \@@endlink[0]{}%
\providecommand \url  [0]{\begingroup\@sanitize@url \@url }%
\providecommand \@url [1]{\endgroup\@href {#1}{\urlprefix }}%
\providecommand \urlprefix  [0]{URL }%
\providecommand \Eprint [0]{\href }%
\providecommand \doibase [0]{http://dx.doi.org/}%
\providecommand \selectlanguage [0]{\@gobble}%
\providecommand \bibinfo  [0]{\@secondoftwo}%
\providecommand \bibfield  [0]{\@secondoftwo}%
\providecommand \translation [1]{[#1]}%
\providecommand \BibitemOpen [0]{}%
\providecommand \bibitemStop [0]{}%
\providecommand \bibitemNoStop [0]{.\EOS\space}%
\providecommand \EOS [0]{\spacefactor3000\relax}%
\providecommand \BibitemShut  [1]{\csname bibitem#1\endcsname}%
\let\auto@bib@innerbib\@empty
\bibitem [{\citenamefont {Xiao}\ \emph {et~al.}(2006)\citenamefont {Xiao},
  \citenamefont {Yao}, \citenamefont {Fang},\ and\ \citenamefont
  {Niu}}]{xiao2006berry}%
  \BibitemOpen
  \bibfield  {author} {\bibinfo {author} {\bibfnamefont {Di}~\bibnamefont
  {Xiao}}, \bibinfo {author} {\bibfnamefont {Yugui}\ \bibnamefont {Yao}},
  \bibinfo {author} {\bibfnamefont {Zhong}\ \bibnamefont {Fang}}, \ and\
  \bibinfo {author} {\bibfnamefont {Qian}\ \bibnamefont {Niu}},\ }\bibfield
  {title} {\enquote {\bibinfo {title} {{Berry-Phase Effect in Anomalous
  Thermoelectric Transport}},}\ }\href {\doibase 10.1103/PhysRevLett.97.026603}
  {\bibfield  {journal} {\bibinfo  {journal} {Phys. Rev. Lett.}\ }\textbf
  {\bibinfo {volume} {97}},\ \bibinfo {pages} {026603} (\bibinfo {year}
  {2006})}\BibitemShut {NoStop}%
\bibitem [{\citenamefont {Yokoyama}\ and\ \citenamefont
  {Murakami}(2011)}]{yokoyama2011transverse}%
  \BibitemOpen
  \bibfield  {author} {\bibinfo {author} {\bibfnamefont {Takehito}\
  \bibnamefont {Yokoyama}}\ and\ \bibinfo {author} {\bibfnamefont {Shuichi}\
  \bibnamefont {Murakami}},\ }\bibfield  {title} {\enquote {\bibinfo {title}
  {Transverse magnetic heat transport on the surface of a topological
  insulator},}\ }\href {\doibase 10.1103/PhysRevB.83.161407} {\bibfield
  {journal} {\bibinfo  {journal} {Phys. Rev. B}\ }\textbf {\bibinfo {volume}
  {83}},\ \bibinfo {pages} {161407} (\bibinfo {year} {2011})}\BibitemShut
  {NoStop}%
\bibitem [{\citenamefont {Qiang}\ \emph {et~al.}(2023)\citenamefont {Qiang},
  \citenamefont {Du}, \citenamefont {Lu},\ and\ \citenamefont
  {Xie}}]{qiang2023topological}%
  \BibitemOpen
  \bibfield  {author} {\bibinfo {author} {\bibfnamefont {Xiao-Bin}\
  \bibnamefont {Qiang}}, \bibinfo {author} {\bibfnamefont {Z.~Z.}\ \bibnamefont
  {Du}}, \bibinfo {author} {\bibfnamefont {Hai-Zhou}\ \bibnamefont {Lu}}, \
  and\ \bibinfo {author} {\bibfnamefont {X.~C.}\ \bibnamefont {Xie}},\
  }\bibfield  {title} {\enquote {\bibinfo {title} {{Topological and disorder
  corrections to the transverse Wiedemann-Franz law and Mott relation in kagome
  magnets and Dirac materials}},}\ }\href {\doibase
  10.1103/PhysRevB.107.L161302} {\bibfield  {journal} {\bibinfo  {journal}
  {Phys. Rev. B}\ }\textbf {\bibinfo {volume} {107}},\ \bibinfo {pages}
  {L161302} (\bibinfo {year} {2023})}\BibitemShut {NoStop}%
\bibitem [{\citenamefont {Hirschberger}\ \emph {et~al.}(2016)\citenamefont
  {Hirschberger}, \citenamefont {Kushwaha}, \citenamefont {Wang}, \citenamefont
  {Gibson}, \citenamefont {Liang}, \citenamefont {Belvin}, \citenamefont
  {Bernevig}, \citenamefont {Cava},\ and\ \citenamefont
  {Ong}}]{hirschberger2016chiral}%
  \BibitemOpen
  \bibfield  {author} {\bibinfo {author} {\bibfnamefont {Max}\ \bibnamefont
  {Hirschberger}}, \bibinfo {author} {\bibfnamefont {Satya}\ \bibnamefont
  {Kushwaha}}, \bibinfo {author} {\bibfnamefont {Zhijun}\ \bibnamefont {Wang}},
  \bibinfo {author} {\bibfnamefont {Quinn}\ \bibnamefont {Gibson}}, \bibinfo
  {author} {\bibfnamefont {Sihang}\ \bibnamefont {Liang}}, \bibinfo {author}
  {\bibfnamefont {Carina~A}\ \bibnamefont {Belvin}}, \bibinfo {author}
  {\bibfnamefont {Bogdan~Andrei}\ \bibnamefont {Bernevig}}, \bibinfo {author}
  {\bibfnamefont {Robert~J}\ \bibnamefont {Cava}}, \ and\ \bibinfo {author}
  {\bibfnamefont {Nai~Phuan}\ \bibnamefont {Ong}},\ }\bibfield  {title}
  {\enquote {\bibinfo {title} {{The chiral anomaly and thermopower of Weyl
  fermions in the half-Heusler GdPtBi}},}\ }\href {\doibase 10.1038/nmat4684}
  {\bibfield  {journal} {\bibinfo  {journal} {Nature materials}\ }\textbf
  {\bibinfo {volume} {15}},\ \bibinfo {pages} {1161} (\bibinfo {year}
  {2016})}\BibitemShut {NoStop}%
\bibitem [{\citenamefont {Fu}\ \emph {et~al.}(2020)\citenamefont {Fu},
  \citenamefont {Sun},\ and\ \citenamefont {Felser}}]{fu2020topological}%
  \BibitemOpen
  \bibfield  {author} {\bibinfo {author} {\bibfnamefont {Chenguang}\
  \bibnamefont {Fu}}, \bibinfo {author} {\bibfnamefont {Yan}\ \bibnamefont
  {Sun}}, \ and\ \bibinfo {author} {\bibfnamefont {Claudia}\ \bibnamefont
  {Felser}},\ }\bibfield  {title} {\enquote {\bibinfo {title} {Topological
  thermoelectrics},}\ }\href {\doibase 10.1063/5.0005481} {\bibfield  {journal}
  {\bibinfo  {journal} {APL Materials}\ }\textbf {\bibinfo {volume} {8}},\
  \bibinfo {pages} {040913} (\bibinfo {year} {2020})}\BibitemShut {NoStop}%
\bibitem [{\citenamefont {Xu}\ \emph {et~al.}(2020)\citenamefont {Xu},
  \citenamefont {Li}, \citenamefont {Lu}, \citenamefont {Collignon},
  \citenamefont {Fu}, \citenamefont {Koo}, \citenamefont {Fauqu{\'e}},
  \citenamefont {Yan}, \citenamefont {Zhu},\ and\ \citenamefont
  {Behnia}}]{xu2020finite}%
  \BibitemOpen
  \bibfield  {author} {\bibinfo {author} {\bibfnamefont {Liangcai}\
  \bibnamefont {Xu}}, \bibinfo {author} {\bibfnamefont {Xiaokang}\ \bibnamefont
  {Li}}, \bibinfo {author} {\bibfnamefont {Xiufang}\ \bibnamefont {Lu}},
  \bibinfo {author} {\bibfnamefont {Cl{\'e}ment}\ \bibnamefont {Collignon}},
  \bibinfo {author} {\bibfnamefont {Huixia}\ \bibnamefont {Fu}}, \bibinfo
  {author} {\bibfnamefont {Jahyun}\ \bibnamefont {Koo}}, \bibinfo {author}
  {\bibfnamefont {Beno{\^\i}t}\ \bibnamefont {Fauqu{\'e}}}, \bibinfo {author}
  {\bibfnamefont {Binghai}\ \bibnamefont {Yan}}, \bibinfo {author}
  {\bibfnamefont {Zengwei}\ \bibnamefont {Zhu}}, \ and\ \bibinfo {author}
  {\bibfnamefont {Kamran}\ \bibnamefont {Behnia}},\ }\bibfield  {title}
  {\enquote {\bibinfo {title} {{Finite-temperature violation of the anomalous
  transverse Wiedemann-Franz law}},}\ }\href {\doibase 10.1126/sciadv.aaz3522}
  {\bibfield  {journal} {\bibinfo  {journal} {Science Advances}\ }\textbf
  {\bibinfo {volume} {6}},\ \bibinfo {pages} {eaaz3522} (\bibinfo {year}
  {2020})}\BibitemShut {NoStop}%
\bibitem [{\citenamefont {Zhang}\ \emph {et~al.}(2021)\citenamefont {Zhang},
  \citenamefont {Xu},\ and\ \citenamefont {Ke}}]{zhang2021topological}%
  \BibitemOpen
  \bibfield  {author} {\bibinfo {author} {\bibfnamefont {H.}~\bibnamefont
  {Zhang}}, \bibinfo {author} {\bibfnamefont {C.~Q.}\ \bibnamefont {Xu}}, \
  and\ \bibinfo {author} {\bibfnamefont {X.}~\bibnamefont {Ke}},\ }\bibfield
  {title} {\enquote {\bibinfo {title} {{Topological Nernst effect, anomalous
  Nernst effect, and anomalous thermal Hall effect in the Dirac semimetal
  ${\mathrm{Fe}}_{3}{\mathrm{Sn}}_{2}$}},}\ }\href {\doibase
  10.1103/PhysRevB.103.L201101} {\bibfield  {journal} {\bibinfo  {journal}
  {Phys. Rev. B}\ }\textbf {\bibinfo {volume} {103}},\ \bibinfo {pages}
  {L201101} (\bibinfo {year} {2021})}\BibitemShut {NoStop}%
\bibitem [{\citenamefont {Chen}\ \emph {et~al.}(2021)\citenamefont {Chen},
  \citenamefont {Tomita}, \citenamefont {Minami}, \citenamefont {Fu},
  \citenamefont {Koretsune}, \citenamefont {Kitatani}, \citenamefont
  {Muhammad}, \citenamefont {Nishio-Hamane}, \citenamefont {Ishii},
  \citenamefont {Ishii} \emph {et~al.}}]{chen2021anomalous}%
  \BibitemOpen
  \bibfield  {author} {\bibinfo {author} {\bibfnamefont {Taishi}\ \bibnamefont
  {Chen}}, \bibinfo {author} {\bibfnamefont {Takahiro}\ \bibnamefont {Tomita}},
  \bibinfo {author} {\bibfnamefont {Susumu}\ \bibnamefont {Minami}}, \bibinfo
  {author} {\bibfnamefont {Mingxuan}\ \bibnamefont {Fu}}, \bibinfo {author}
  {\bibfnamefont {Takashi}\ \bibnamefont {Koretsune}}, \bibinfo {author}
  {\bibfnamefont {Motoharu}\ \bibnamefont {Kitatani}}, \bibinfo {author}
  {\bibfnamefont {Ikhlas}\ \bibnamefont {Muhammad}}, \bibinfo {author}
  {\bibfnamefont {Daisuke}\ \bibnamefont {Nishio-Hamane}}, \bibinfo {author}
  {\bibfnamefont {Rieko}\ \bibnamefont {Ishii}}, \bibinfo {author}
  {\bibfnamefont {Fumiyuki}\ \bibnamefont {Ishii}},  \emph {et~al.},\
  }\bibfield  {title} {\enquote {\bibinfo {title} {{Anomalous transport due to
  Weyl fermions in the chiral antiferromagnets Mn$_3$X, X= Sn, Ge}},}\ }\href
  {\doibase 10.1038/s41467-020-20838-1} {\bibfield  {journal} {\bibinfo
  {journal} {Nature communications}\ }\textbf {\bibinfo {volume} {12}},\
  \bibinfo {pages} {572} (\bibinfo {year} {2021})}\BibitemShut {NoStop}%
\bibitem [{\citenamefont {Pan}\ \emph {et~al.}(2022)\citenamefont {Pan},
  \citenamefont {Le}, \citenamefont {He}, \citenamefont {Watzman},
  \citenamefont {Yao}, \citenamefont {Gooth}, \citenamefont {Heremans},
  \citenamefont {Sun},\ and\ \citenamefont {Felser}}]{pan2022giant}%
  \BibitemOpen
  \bibfield  {author} {\bibinfo {author} {\bibfnamefont {Yu}~\bibnamefont
  {Pan}}, \bibinfo {author} {\bibfnamefont {Congcong}\ \bibnamefont {Le}},
  \bibinfo {author} {\bibfnamefont {Bin}\ \bibnamefont {He}}, \bibinfo {author}
  {\bibfnamefont {Sarah~J}\ \bibnamefont {Watzman}}, \bibinfo {author}
  {\bibfnamefont {Mengyu}\ \bibnamefont {Yao}}, \bibinfo {author}
  {\bibfnamefont {Johannes}\ \bibnamefont {Gooth}}, \bibinfo {author}
  {\bibfnamefont {Joseph~P}\ \bibnamefont {Heremans}}, \bibinfo {author}
  {\bibfnamefont {Yan}\ \bibnamefont {Sun}}, \ and\ \bibinfo {author}
  {\bibfnamefont {Claudia}\ \bibnamefont {Felser}},\ }\bibfield  {title}
  {\enquote {\bibinfo {title} {{Giant anomalous Nernst signal in the
  antiferromagnet YbMnBi$_2$}},}\ }\href {\doibase 10.1038/s41563-021-01149-2}
  {\bibfield  {journal} {\bibinfo  {journal} {Nature materials}\ }\textbf
  {\bibinfo {volume} {21}},\ \bibinfo {pages} {203} (\bibinfo {year}
  {2022})}\BibitemShut {NoStop}%
\bibitem [{\citenamefont {Badura}\ \emph {et~al.}(2025)\citenamefont {Badura},
  \citenamefont {Campos}, \citenamefont {Bharadwaj}, \citenamefont {Kounta},
  \citenamefont {Michez}, \citenamefont {Petit}, \citenamefont {Rial},
  \citenamefont {Leivisk{\"a}}, \citenamefont {Baltz}, \citenamefont {Krizek}
  \emph {et~al.}}]{badura2025observation}%
  \BibitemOpen
  \bibfield  {author} {\bibinfo {author} {\bibfnamefont {Anton{\'\i}n}\
  \bibnamefont {Badura}}, \bibinfo {author} {\bibfnamefont {Warlley~H}\
  \bibnamefont {Campos}}, \bibinfo {author} {\bibfnamefont {Venkata~K}\
  \bibnamefont {Bharadwaj}}, \bibinfo {author} {\bibfnamefont {Isma{\"\i}la}\
  \bibnamefont {Kounta}}, \bibinfo {author} {\bibfnamefont {Lisa}\ \bibnamefont
  {Michez}}, \bibinfo {author} {\bibfnamefont {Matthieu}\ \bibnamefont
  {Petit}}, \bibinfo {author} {\bibfnamefont {Javier}\ \bibnamefont {Rial}},
  \bibinfo {author} {\bibfnamefont {Miina}\ \bibnamefont {Leivisk{\"a}}},
  \bibinfo {author} {\bibfnamefont {Vincent}\ \bibnamefont {Baltz}}, \bibinfo
  {author} {\bibfnamefont {Filip}\ \bibnamefont {Krizek}},  \emph {et~al.},\
  }\bibfield  {title} {\enquote {\bibinfo {title} {{Observation of the
  anomalous Nernst effect in altermagnetic candidate Mn$_5$Si$_3$}},}\ }\href
  {\doibase 10.1038/s41467-025-62331-7} {\bibfield  {journal} {\bibinfo
  {journal} {Nature Communications}\ }\textbf {\bibinfo {volume} {16}},\
  \bibinfo {pages} {7111} (\bibinfo {year} {2025})}\BibitemShut {NoStop}%
\bibitem [{\citenamefont {Han}\ \emph {et~al.}(2025)\citenamefont {Han},
  \citenamefont {Fu}, \citenamefont {He}, \citenamefont {Dai}, \citenamefont
  {Zhu}, \citenamefont {Yang}, \citenamefont {Chen}, \citenamefont {Zhang},
  \citenamefont {Zhu}, \citenamefont {Bai}, \citenamefont {Chen}, \citenamefont
  {Hou}, \citenamefont {Wan}, \citenamefont {Han}, \citenamefont {Song},
  \citenamefont {Liu},\ and\ \citenamefont {Pan}}]{han2025nonvolatile}%
  \BibitemOpen
  \bibfield  {author} {\bibinfo {author} {\bibfnamefont {L.}~\bibnamefont
  {Han}}, \bibinfo {author} {\bibfnamefont {X.~Z.}\ \bibnamefont {Fu}},
  \bibinfo {author} {\bibfnamefont {W.~Q.}\ \bibnamefont {He}}, \bibinfo
  {author} {\bibfnamefont {J.~K.}\ \bibnamefont {Dai}}, \bibinfo {author}
  {\bibfnamefont {Y.~X.}\ \bibnamefont {Zhu}}, \bibinfo {author} {\bibfnamefont
  {W.~F.}\ \bibnamefont {Yang}}, \bibinfo {author} {\bibfnamefont {Y.~L.}\
  \bibnamefont {Chen}}, \bibinfo {author} {\bibfnamefont {J.~C.}\ \bibnamefont
  {Zhang}}, \bibinfo {author} {\bibfnamefont {W.~X.}\ \bibnamefont {Zhu}},
  \bibinfo {author} {\bibfnamefont {H.}~\bibnamefont {Bai}}, \bibinfo {author}
  {\bibfnamefont {C.}~\bibnamefont {Chen}}, \bibinfo {author} {\bibfnamefont
  {D.~Z.}\ \bibnamefont {Hou}}, \bibinfo {author} {\bibfnamefont {C.~H.}\
  \bibnamefont {Wan}}, \bibinfo {author} {\bibfnamefont {X.~F.}\ \bibnamefont
  {Han}}, \bibinfo {author} {\bibfnamefont {C.}~\bibnamefont {Song}}, \bibinfo
  {author} {\bibfnamefont {J.~W.}\ \bibnamefont {Liu}}, \ and\ \bibinfo
  {author} {\bibfnamefont {F.}~\bibnamefont {Pan}},\ }\bibfield  {title}
  {\enquote {\bibinfo {title} {{Nonvolatile anomalous Nernst effect in
  ${\mathrm{Mn}}_{5}{\mathrm{Si}}_{3}$ with a collinear N\'{e}el vector}},}\
  }\href {\doibase 10.1103/PhysRevApplied.23.044066} {\bibfield  {journal}
  {\bibinfo  {journal} {Phys. Rev. Appl.}\ }\textbf {\bibinfo {volume} {23}},\
  \bibinfo {pages} {044066} (\bibinfo {year} {2025})}\BibitemShut {NoStop}%
\bibitem [{\citenamefont {Zhou}\ \emph {et~al.}(2024)\citenamefont {Zhou},
  \citenamefont {Feng}, \citenamefont {Zhang}, \citenamefont
  {\ifmmode~\check{S}\else \v{S}\fi{}mejkal}, \citenamefont {Sinova},
  \citenamefont {Mokrousov},\ and\ \citenamefont {Yao}}]{zhou2024crystal}%
  \BibitemOpen
  \bibfield  {author} {\bibinfo {author} {\bibfnamefont {Xiaodong}\
  \bibnamefont {Zhou}}, \bibinfo {author} {\bibfnamefont {Wanxiang}\
  \bibnamefont {Feng}}, \bibinfo {author} {\bibfnamefont {Run-Wu}\ \bibnamefont
  {Zhang}}, \bibinfo {author} {\bibfnamefont {Libor}\ \bibnamefont
  {\ifmmode~\check{S}\else \v{S}\fi{}mejkal}}, \bibinfo {author} {\bibfnamefont
  {Jairo}\ \bibnamefont {Sinova}}, \bibinfo {author} {\bibfnamefont {Yuriy}\
  \bibnamefont {Mokrousov}}, \ and\ \bibinfo {author} {\bibfnamefont {Yugui}\
  \bibnamefont {Yao}},\ }\bibfield  {title} {\enquote {\bibinfo {title}
  {{Crystal Thermal Transport in Altermagnetic ${\mathrm{RuO}}_{2}$}},}\ }\href
  {\doibase 10.1103/PhysRevLett.132.056701} {\bibfield  {journal} {\bibinfo
  {journal} {Phys. Rev. Lett.}\ }\textbf {\bibinfo {volume} {132}},\ \bibinfo
  {pages} {056701} (\bibinfo {year} {2024})}\BibitemShut {NoStop}%
\bibitem [{\citenamefont {Wei{\ss}enhofer}\ \emph {et~al.}(2026)\citenamefont
  {Wei{\ss}enhofer}, \citenamefont {Mrudul}, \citenamefont {Mankovsky},\ and\
  \citenamefont {Oppeneer}}]{weissenhofer2026magnon}%
  \BibitemOpen
  \bibfield  {author} {\bibinfo {author} {\bibfnamefont {Markus}\ \bibnamefont
  {Wei{\ss}enhofer}}, \bibinfo {author} {\bibfnamefont {M.~S.}\ \bibnamefont
  {Mrudul}}, \bibinfo {author} {\bibfnamefont {Sergiy}\ \bibnamefont
  {Mankovsky}}, \ and\ \bibinfo {author} {\bibfnamefont {Peter~M.}\
  \bibnamefont {Oppeneer}},\ }\bibfield  {title} {\enquote {\bibinfo {title}
  {{Magnon orbital Nernst effect in altermagnets}},}\ }\href {\doibase
  10.1038/s41535-026-00853-z} {\bibfield  {journal} {\bibinfo  {journal} {npj
  Quantum Materials}\ }\textbf {\bibinfo {volume} {11}},\ \bibinfo {pages} {25}
  (\bibinfo {year} {2026})}\BibitemShut {NoStop}%
\bibitem [{\citenamefont {Debnath}\ \emph {et~al.}(2026)\citenamefont
  {Debnath}, \citenamefont {Saha},\ and\ \citenamefont
  {Dutta}}]{debnath2026spin}%
  \BibitemOpen
  \bibfield  {author} {\bibinfo {author} {\bibfnamefont {Debika}\ \bibnamefont
  {Debnath}}, \bibinfo {author} {\bibfnamefont {Arijit}\ \bibnamefont {Saha}},
  \ and\ \bibinfo {author} {\bibfnamefont {Paramita}\ \bibnamefont {Dutta}},\
  }\bibfield  {title} {\enquote {\bibinfo {title} {Spin polarization and diode
  effect in thermoelectric current through altermagnet-based superconductor
  heterostructures},}\ }\href {\doibase 10.1103/jd41-26xr} {\bibfield
  {journal} {\bibinfo  {journal} {Phys. Rev. B}\ }\textbf {\bibinfo {volume}
  {113}},\ \bibinfo {pages} {104508} (\bibinfo {year} {2026})}\BibitemShut
  {NoStop}%
\bibitem [{\citenamefont {Wu}\ and\ \citenamefont
  {Zhang}(2004)}]{wu2004dynamic}%
  \BibitemOpen
  \bibfield  {author} {\bibinfo {author} {\bibfnamefont {Congjun}\ \bibnamefont
  {Wu}}\ and\ \bibinfo {author} {\bibfnamefont {Shou-Cheng}\ \bibnamefont
  {Zhang}},\ }\bibfield  {title} {\enquote {\bibinfo {title} {{Dynamic
  Generation of Spin-Orbit Coupling}},}\ }\href {\doibase
  10.1103/PhysRevLett.93.036403} {\bibfield  {journal} {\bibinfo  {journal}
  {Phys. Rev. Lett.}\ }\textbf {\bibinfo {volume} {93}},\ \bibinfo {pages}
  {036403} (\bibinfo {year} {2004})}\BibitemShut {NoStop}%
\bibitem [{\citenamefont {Wu}\ \emph {et~al.}(2007)\citenamefont {Wu},
  \citenamefont {Sun}, \citenamefont {Fradkin},\ and\ \citenamefont
  {Zhang}}]{wu2007fermi}%
  \BibitemOpen
  \bibfield  {author} {\bibinfo {author} {\bibfnamefont {Congjun}\ \bibnamefont
  {Wu}}, \bibinfo {author} {\bibfnamefont {Kai}\ \bibnamefont {Sun}}, \bibinfo
  {author} {\bibfnamefont {Eduardo}\ \bibnamefont {Fradkin}}, \ and\ \bibinfo
  {author} {\bibfnamefont {Shou-Cheng}\ \bibnamefont {Zhang}},\ }\bibfield
  {title} {\enquote {\bibinfo {title} {Fermi liquid instabilities in the spin
  channel},}\ }\href {\doibase 10.1103/PhysRevB.75.115103} {\bibfield
  {journal} {\bibinfo  {journal} {Phys. Rev. B}\ }\textbf {\bibinfo {volume}
  {75}},\ \bibinfo {pages} {115103} (\bibinfo {year} {2007})}\BibitemShut
  {NoStop}%
\bibitem [{\citenamefont {Lee}\ and\ \citenamefont {Wu}(2009)}]{lee2009theory}%
  \BibitemOpen
  \bibfield  {author} {\bibinfo {author} {\bibfnamefont {Wei-Cheng}\
  \bibnamefont {Lee}}\ and\ \bibinfo {author} {\bibfnamefont {Congjun}\
  \bibnamefont {Wu}},\ }\bibfield  {title} {\enquote {\bibinfo {title} {Theory
  of unconventional metamagnetic electron states in orbital band systems},}\
  }\href {\doibase 10.1103/PhysRevB.80.104438} {\bibfield  {journal} {\bibinfo
  {journal} {Phys. Rev. B}\ }\textbf {\bibinfo {volume} {80}},\ \bibinfo
  {pages} {104438} (\bibinfo {year} {2009})}\BibitemShut {NoStop}%
\bibitem [{\citenamefont {Yuan}\ \emph {et~al.}(2026)\citenamefont {Yuan},
  \citenamefont {Pan},\ and\ \citenamefont {Wu}}]{yuan2026unconventional}%
  \BibitemOpen
  \bibfield  {author} {\bibinfo {author} {\bibfnamefont {Jian-Keng}\
  \bibnamefont {Yuan}}, \bibinfo {author} {\bibfnamefont {Zhiming}\
  \bibnamefont {Pan}}, \ and\ \bibinfo {author} {\bibfnamefont {Congjun}\
  \bibnamefont {Wu}},\ }\bibfield  {title} {\enquote {\bibinfo {title}
  {Unconventional magnetism in spin-orbit coupled systems},}\ }\href {\doibase
  10.1103/dy2j-mc2t} {\bibfield  {journal} {\bibinfo  {journal} {Phys. Rev. B}\
  }\textbf {\bibinfo {volume} {113}},\ \bibinfo {pages} {014426} (\bibinfo
  {year} {2026})}\BibitemShut {NoStop}%
\bibitem [{\citenamefont {Hayami}\ \emph {et~al.}(2019)\citenamefont {Hayami},
  \citenamefont {Yanagi},\ and\ \citenamefont {Kusunose}}]{hayami2019momentum}%
  \BibitemOpen
  \bibfield  {author} {\bibinfo {author} {\bibfnamefont {Satoru}\ \bibnamefont
  {Hayami}}, \bibinfo {author} {\bibfnamefont {Yuki}\ \bibnamefont {Yanagi}}, \
  and\ \bibinfo {author} {\bibfnamefont {Hiroaki}\ \bibnamefont {Kusunose}},\
  }\bibfield  {title} {\enquote {\bibinfo {title} {{Momentum-dependent spin
  splitting by collinear antiferromagnetic ordering}},}\ }\href {\doibase
  10.7566/JPSJ.88.123702} {\bibfield  {journal} {\bibinfo  {journal} {Journal
  of the Physical Society of Japan}\ }\textbf {\bibinfo {volume} {88}},\
  \bibinfo {pages} {123702} (\bibinfo {year} {2019})}\BibitemShut {NoStop}%
\bibitem [{\citenamefont {Hayami}\ \emph {et~al.}(2020)\citenamefont {Hayami},
  \citenamefont {Yanagi},\ and\ \citenamefont {Kusunose}}]{hayami2020bottom}%
  \BibitemOpen
  \bibfield  {author} {\bibinfo {author} {\bibfnamefont {Satoru}\ \bibnamefont
  {Hayami}}, \bibinfo {author} {\bibfnamefont {Yuki}\ \bibnamefont {Yanagi}}, \
  and\ \bibinfo {author} {\bibfnamefont {Hiroaki}\ \bibnamefont {Kusunose}},\
  }\bibfield  {title} {\enquote {\bibinfo {title} {{Bottom-up design of
  spin-split and reshaped electronic band structures in antiferromagnets
  without spin-orbit coupling: Procedure on the basis of augmented
  multipoles}},}\ }\href {\doibase 10.1103/PhysRevB.102.144441} {\bibfield
  {journal} {\bibinfo  {journal} {Phys. Rev. B}\ }\textbf {\bibinfo {volume}
  {102}},\ \bibinfo {pages} {144441} (\bibinfo {year} {2020})}\BibitemShut
  {NoStop}%
\bibitem [{\citenamefont {Ma}\ \emph {et~al.}(2021)\citenamefont {Ma},
  \citenamefont {Hu}, \citenamefont {Li}, \citenamefont {Liu}, \citenamefont
  {Yao}, \citenamefont {Jia},\ and\ \citenamefont
  {Liu}}]{ma2021multifunctional}%
  \BibitemOpen
  \bibfield  {author} {\bibinfo {author} {\bibfnamefont {Hai-Yang}\
  \bibnamefont {Ma}}, \bibinfo {author} {\bibfnamefont {Mengli}\ \bibnamefont
  {Hu}}, \bibinfo {author} {\bibfnamefont {Nana}\ \bibnamefont {Li}}, \bibinfo
  {author} {\bibfnamefont {Jianpeng}\ \bibnamefont {Liu}}, \bibinfo {author}
  {\bibfnamefont {Wang}\ \bibnamefont {Yao}}, \bibinfo {author} {\bibfnamefont
  {Jin-Feng}\ \bibnamefont {Jia}}, \ and\ \bibinfo {author} {\bibfnamefont
  {Junwei}\ \bibnamefont {Liu}},\ }\bibfield  {title} {\enquote {\bibinfo
  {title} {Multifunctional antiferromagnetic materials with giant
  piezomagnetism and noncollinear spin current},}\ }\href {\doibase
  10.1038/s41467-021-23127-7} {\bibfield  {journal} {\bibinfo  {journal}
  {Nature communications}\ }\textbf {\bibinfo {volume} {12}},\ \bibinfo {pages}
  {2846} (\bibinfo {year} {2021})}\BibitemShut {NoStop}%
\bibitem [{\citenamefont {Song}\ \emph {et~al.}(2025)\citenamefont {Song},
  \citenamefont {Bai}, \citenamefont {Zhou}, \citenamefont {Han}, \citenamefont
  {Reichlova}, \citenamefont {Dil}, \citenamefont {Liu}, \citenamefont {Chen},\
  and\ \citenamefont {Pan}}]{song2025altermagnets}%
  \BibitemOpen
  \bibfield  {author} {\bibinfo {author} {\bibfnamefont {Cheng}\ \bibnamefont
  {Song}}, \bibinfo {author} {\bibfnamefont {Hua}\ \bibnamefont {Bai}},
  \bibinfo {author} {\bibfnamefont {Zhiyuan}\ \bibnamefont {Zhou}}, \bibinfo
  {author} {\bibfnamefont {Lei}\ \bibnamefont {Han}}, \bibinfo {author}
  {\bibfnamefont {Helena}\ \bibnamefont {Reichlova}}, \bibinfo {author}
  {\bibfnamefont {J~Hugo}\ \bibnamefont {Dil}}, \bibinfo {author}
  {\bibfnamefont {Junwei}\ \bibnamefont {Liu}}, \bibinfo {author}
  {\bibfnamefont {Xianzhe}\ \bibnamefont {Chen}}, \ and\ \bibinfo {author}
  {\bibfnamefont {Feng}\ \bibnamefont {Pan}},\ }\bibfield  {title} {\enquote
  {\bibinfo {title} {Altermagnets as a new class of functional materials},}\
  }\href {\doibase 10.1038/s41578-025-00779-1} {\bibfield  {journal} {\bibinfo
  {journal} {Nature Reviews Materials}\ }\textbf {\bibinfo {volume} {10}},\
  \bibinfo {pages} {473} (\bibinfo {year} {2025})}\BibitemShut {NoStop}%
\bibitem [{\citenamefont {Xu}\ \emph {et~al.}(2026)\citenamefont {Xu},
  \citenamefont {Gao},\ and\ \citenamefont {Liu}}]{xu2026chemical}%
  \BibitemOpen
  \bibfield  {author} {\bibinfo {author} {\bibfnamefont {Runzhang}\
  \bibnamefont {Xu}}, \bibinfo {author} {\bibfnamefont {Yifan}\ \bibnamefont
  {Gao}}, \ and\ \bibinfo {author} {\bibfnamefont {Junwei}\ \bibnamefont
  {Liu}},\ }\bibfield  {title} {\enquote {\bibinfo {title} {Chemical design of
  monolayer altermagnets},}\ }\href {\doibase 10.1093/nsr/nwaf528} {\bibfield
  {journal} {\bibinfo  {journal} {National Science Review}\ }\textbf {\bibinfo
  {volume} {13}},\ \bibinfo {pages} {nwaf528} (\bibinfo {year}
  {2026})}\BibitemShut {NoStop}%
\bibitem [{\citenamefont {Hu}\ \emph {et~al.}(2025)\citenamefont {Hu},
  \citenamefont {Cheng}, \citenamefont {Huang},\ and\ \citenamefont
  {Liu}}]{hu2025catalog}%
  \BibitemOpen
  \bibfield  {author} {\bibinfo {author} {\bibfnamefont {Mengli}\ \bibnamefont
  {Hu}}, \bibinfo {author} {\bibfnamefont {Xingkai}\ \bibnamefont {Cheng}},
  \bibinfo {author} {\bibfnamefont {Zhenqiao}\ \bibnamefont {Huang}}, \ and\
  \bibinfo {author} {\bibfnamefont {Junwei}\ \bibnamefont {Liu}},\ }\bibfield
  {title} {\enquote {\bibinfo {title} {{Catalog of $C$-Paired Spin-Momentum
  Locking in Antiferromagnetic Systems}},}\ }\href {\doibase
  10.1103/PhysRevX.15.021083} {\bibfield  {journal} {\bibinfo  {journal} {Phys.
  Rev. X}\ }\textbf {\bibinfo {volume} {15}},\ \bibinfo {pages} {021083}
  (\bibinfo {year} {2025})}\BibitemShut {NoStop}%
\bibitem [{\citenamefont {\ifmmode~\check{S}\else \v{S}\fi{}mejkal}\ \emph
  {et~al.}(2022{\natexlab{a}})\citenamefont {\ifmmode~\check{S}\else
  \v{S}\fi{}mejkal}, \citenamefont {Hellenes}, \citenamefont
  {Gonz\'alez-Hern\'andez}, \citenamefont {Sinova},\ and\ \citenamefont
  {Jungwirth}}]{smejkal2022giant}%
  \BibitemOpen
  \bibfield  {author} {\bibinfo {author} {\bibfnamefont {Libor}\ \bibnamefont
  {\ifmmode~\check{S}\else \v{S}\fi{}mejkal}}, \bibinfo {author} {\bibfnamefont
  {Anna~Birk}\ \bibnamefont {Hellenes}}, \bibinfo {author} {\bibfnamefont
  {Rafael}\ \bibnamefont {Gonz\'alez-Hern\'andez}}, \bibinfo {author}
  {\bibfnamefont {Jairo}\ \bibnamefont {Sinova}}, \ and\ \bibinfo {author}
  {\bibfnamefont {Tomas}\ \bibnamefont {Jungwirth}},\ }\bibfield  {title}
  {\enquote {\bibinfo {title} {{Giant and Tunneling Magnetoresistance in
  Unconventional Collinear Antiferromagnets with Nonrelativistic Spin-Momentum
  Coupling}},}\ }\href {\doibase 10.1103/PhysRevX.12.011028} {\bibfield
  {journal} {\bibinfo  {journal} {Phys. Rev. X}\ }\textbf {\bibinfo {volume}
  {12}},\ \bibinfo {pages} {011028} (\bibinfo {year}
  {2022}{\natexlab{a}})}\BibitemShut {NoStop}%
\bibitem [{\citenamefont {\ifmmode~\check{S}\else \v{S}\fi{}mejkal}\ \emph
  {et~al.}(2022{\natexlab{b}})\citenamefont {\ifmmode~\check{S}\else
  \v{S}\fi{}mejkal}, \citenamefont {Sinova},\ and\ \citenamefont
  {Jungwirth}}]{smejkal2022beyond}%
  \BibitemOpen
  \bibfield  {author} {\bibinfo {author} {\bibfnamefont {Libor}\ \bibnamefont
  {\ifmmode~\check{S}\else \v{S}\fi{}mejkal}}, \bibinfo {author} {\bibfnamefont
  {Jairo}\ \bibnamefont {Sinova}}, \ and\ \bibinfo {author} {\bibfnamefont
  {Tomas}\ \bibnamefont {Jungwirth}},\ }\bibfield  {title} {\enquote {\bibinfo
  {title} {{Beyond Conventional Ferromagnetism and Antiferromagnetism: A Phase
  with Nonrelativistic Spin and Crystal Rotation Symmetry}},}\ }\href {\doibase
  10.1103/PhysRevX.12.031042} {\bibfield  {journal} {\bibinfo  {journal} {Phys.
  Rev. X}\ }\textbf {\bibinfo {volume} {12}},\ \bibinfo {pages} {031042}
  (\bibinfo {year} {2022}{\natexlab{b}})}\BibitemShut {NoStop}%
\bibitem [{\citenamefont {\ifmmode~\check{S}\else \v{S}\fi{}mejkal}\ \emph
  {et~al.}(2022{\natexlab{c}})\citenamefont {\ifmmode~\check{S}\else
  \v{S}\fi{}mejkal}, \citenamefont {Sinova},\ and\ \citenamefont
  {Jungwirth}}]{smejkal2022emerging}%
  \BibitemOpen
  \bibfield  {author} {\bibinfo {author} {\bibfnamefont {Libor}\ \bibnamefont
  {\ifmmode~\check{S}\else \v{S}\fi{}mejkal}}, \bibinfo {author} {\bibfnamefont
  {Jairo}\ \bibnamefont {Sinova}}, \ and\ \bibinfo {author} {\bibfnamefont
  {Tomas}\ \bibnamefont {Jungwirth}},\ }\bibfield  {title} {\enquote {\bibinfo
  {title} {{Emerging Research Landscape of Altermagnetism}},}\ }\href {\doibase
  10.1103/PhysRevX.12.040501} {\bibfield  {journal} {\bibinfo  {journal} {Phys.
  Rev. X}\ }\textbf {\bibinfo {volume} {12}},\ \bibinfo {pages} {040501}
  (\bibinfo {year} {2022}{\natexlab{c}})}\BibitemShut {NoStop}%
\bibitem [{\citenamefont {Smejkal}\ \emph {et~al.}(2022)\citenamefont
  {Smejkal}, \citenamefont {MacDonald}, \citenamefont {Sinova}, \citenamefont
  {Nakatsuji},\ and\ \citenamefont {Jungwirth}}]{smejkal2022anomalous}%
  \BibitemOpen
  \bibfield  {author} {\bibinfo {author} {\bibfnamefont {Libor}\ \bibnamefont
  {Smejkal}}, \bibinfo {author} {\bibfnamefont {Allan~H}\ \bibnamefont
  {MacDonald}}, \bibinfo {author} {\bibfnamefont {Jairo}\ \bibnamefont
  {Sinova}}, \bibinfo {author} {\bibfnamefont {Satoru}\ \bibnamefont
  {Nakatsuji}}, \ and\ \bibinfo {author} {\bibfnamefont {Tomas}\ \bibnamefont
  {Jungwirth}},\ }\bibfield  {title} {\enquote {\bibinfo {title} {{Anomalous
  Hall antiferromagnets}},}\ }\href {\doibase 10.1038/s41578-022-00430-3}
  {\bibfield  {journal} {\bibinfo  {journal} {Nature Reviews Materials}\
  }\textbf {\bibinfo {volume} {7}},\ \bibinfo {pages} {482} (\bibinfo {year}
  {2022})}\BibitemShut {NoStop}%
\bibitem [{\citenamefont {Jiang}\ \emph {et~al.}(2025)\citenamefont {Jiang},
  \citenamefont {Hu}, \citenamefont {Bai}, \citenamefont {Song}, \citenamefont
  {Mu}, \citenamefont {Qu}, \citenamefont {Li}, \citenamefont {Zhu},
  \citenamefont {Pi}, \citenamefont {Wei} \emph {et~al.}}]{jiang2025metallic}%
  \BibitemOpen
  \bibfield  {author} {\bibinfo {author} {\bibfnamefont {Bei}\ \bibnamefont
  {Jiang}}, \bibinfo {author} {\bibfnamefont {Mingzhe}\ \bibnamefont {Hu}},
  \bibinfo {author} {\bibfnamefont {Jianli}\ \bibnamefont {Bai}}, \bibinfo
  {author} {\bibfnamefont {Ziyin}\ \bibnamefont {Song}}, \bibinfo {author}
  {\bibfnamefont {Chao}\ \bibnamefont {Mu}}, \bibinfo {author} {\bibfnamefont
  {Gexing}\ \bibnamefont {Qu}}, \bibinfo {author} {\bibfnamefont {Wan}\
  \bibnamefont {Li}}, \bibinfo {author} {\bibfnamefont {Wenliang}\ \bibnamefont
  {Zhu}}, \bibinfo {author} {\bibfnamefont {Hanqi}\ \bibnamefont {Pi}},
  \bibinfo {author} {\bibfnamefont {Zhongxu}\ \bibnamefont {Wei}},  \emph
  {et~al.},\ }\bibfield  {title} {\enquote {\bibinfo {title} {A metallic
  room-temperature $d$-wave altermagnet},}\ }\href {\doibase
  10.1038/s41567-025-02822-y} {\bibfield  {journal} {\bibinfo  {journal}
  {Nature Physics}\ }\textbf {\bibinfo {volume} {21}},\ \bibinfo {pages} {754}
  (\bibinfo {year} {2025})}\BibitemShut {NoStop}%
\bibitem [{\citenamefont {Wang}\ \emph
  {et~al.}(2025{\natexlab{a}})\citenamefont {Wang}, \citenamefont {Yu},
  \citenamefont {Cheng}, \citenamefont {Xiao}, \citenamefont {Ma},
  \citenamefont {Quan}, \citenamefont {Song}, \citenamefont {Zhang},
  \citenamefont {Zhang}, \citenamefont {Ma} \emph {et~al.}}]{wang2025atomic}%
  \BibitemOpen
  \bibfield  {author} {\bibinfo {author} {\bibfnamefont {Zhuying}\ \bibnamefont
  {Wang}}, \bibinfo {author} {\bibfnamefont {Shuikang}\ \bibnamefont {Yu}},
  \bibinfo {author} {\bibfnamefont {Xingkai}\ \bibnamefont {Cheng}}, \bibinfo
  {author} {\bibfnamefont {Xiaoyu}\ \bibnamefont {Xiao}}, \bibinfo {author}
  {\bibfnamefont {Wanru}\ \bibnamefont {Ma}}, \bibinfo {author} {\bibfnamefont
  {Feixiong}\ \bibnamefont {Quan}}, \bibinfo {author} {\bibfnamefont {Hongxi}\
  \bibnamefont {Song}}, \bibinfo {author} {\bibfnamefont {Kunming}\
  \bibnamefont {Zhang}}, \bibinfo {author} {\bibfnamefont {Yunmei}\
  \bibnamefont {Zhang}}, \bibinfo {author} {\bibfnamefont {Yitian}\
  \bibnamefont {Ma}},  \emph {et~al.},\ }\bibfield  {title} {\enquote {\bibinfo
  {title} {{Atomic-scale spin sensing of a 2D $d$-wave altermagnet via helical
  tunneling}},}\ }\href {https://doi.org/10.48550/arXiv.2512.23290} {\bibfield
  {journal} {\bibinfo  {journal} {arXiv:2512.23290}\ } (\bibinfo {year}
  {2025}{\natexlab{a}})}\BibitemShut {NoStop}%
\bibitem [{\citenamefont {Yang}\ \emph {et~al.}(2026)\citenamefont {Yang},
  \citenamefont {Li}, \citenamefont {Wang}, \citenamefont {Zhao}, \citenamefont
  {Wan}, \citenamefont {Gui}, \citenamefont {Zeng}, \citenamefont {Cao},
  \citenamefont {Hu}, \citenamefont {Chen}, \citenamefont {Liu}, \citenamefont
  {Song}, \citenamefont {Liu}, \citenamefont {Hu}, \citenamefont {Jiao},\ and\
  \citenamefont {Yuan}}]{yang2026visualizing}%
  \BibitemOpen
  \bibfield  {author} {\bibinfo {author} {\bibfnamefont {Guofei}\ \bibnamefont
  {Yang}}, \bibinfo {author} {\bibfnamefont {Chuang}\ \bibnamefont {Li}},
  \bibinfo {author} {\bibfnamefont {Chengwei}\ \bibnamefont {Wang}}, \bibinfo
  {author} {\bibfnamefont {Xudong}\ \bibnamefont {Zhao}}, \bibinfo {author}
  {\bibfnamefont {Yifan}\ \bibnamefont {Wan}}, \bibinfo {author} {\bibfnamefont
  {Hengrui}\ \bibnamefont {Gui}}, \bibinfo {author} {\bibfnamefont {Guoqing}\
  \bibnamefont {Zeng}}, \bibinfo {author} {\bibfnamefont {Saizheng}\
  \bibnamefont {Cao}}, \bibinfo {author} {\bibfnamefont {Chuqiao}\ \bibnamefont
  {Hu}}, \bibinfo {author} {\bibfnamefont {Dong}\ \bibnamefont {Chen}},
  \bibinfo {author} {\bibfnamefont {Yu}~\bibnamefont {Liu}}, \bibinfo {author}
  {\bibfnamefont {Yu}~\bibnamefont {Song}}, \bibinfo {author} {\bibfnamefont
  {Fei}\ \bibnamefont {Liu}}, \bibinfo {author} {\bibfnamefont {Lun-Hui}\
  \bibnamefont {Hu}}, \bibinfo {author} {\bibfnamefont {Lin}\ \bibnamefont
  {Jiao}}, \ and\ \bibinfo {author} {\bibfnamefont {Huiqiu}\ \bibnamefont
  {Yuan}},\ }\bibfield  {title} {\enquote {\bibinfo {title} {{Visualizing
  spin-polarization of an altermagnet KV$_2$Se$_2$O via spin-selective
  tunneling}},}\ }\href {https://arxiv.org/abs/2603.21969} {\bibfield
  {journal} {\bibinfo  {journal} {arXiv:2603.21969}\ } (\bibinfo {year}
  {2026})}\BibitemShut {NoStop}%
\bibitem [{\citenamefont {Zhang}\ \emph {et~al.}(2025)\citenamefont {Zhang},
  \citenamefont {Cheng}, \citenamefont {Yin}, \citenamefont {Liu},
  \citenamefont {Deng}, \citenamefont {Qiao}, \citenamefont {Shi},
  \citenamefont {Zhang}, \citenamefont {Lin}, \citenamefont {Liu} \emph
  {et~al.}}]{zhang2025crystal}%
  \BibitemOpen
  \bibfield  {author} {\bibinfo {author} {\bibfnamefont {Fayuan}\ \bibnamefont
  {Zhang}}, \bibinfo {author} {\bibfnamefont {Xingkai}\ \bibnamefont {Cheng}},
  \bibinfo {author} {\bibfnamefont {Zhouyi}\ \bibnamefont {Yin}}, \bibinfo
  {author} {\bibfnamefont {Changchao}\ \bibnamefont {Liu}}, \bibinfo {author}
  {\bibfnamefont {Liwei}\ \bibnamefont {Deng}}, \bibinfo {author}
  {\bibfnamefont {Yuxi}\ \bibnamefont {Qiao}}, \bibinfo {author} {\bibfnamefont
  {Zheng}\ \bibnamefont {Shi}}, \bibinfo {author} {\bibfnamefont {Shuxuan}\
  \bibnamefont {Zhang}}, \bibinfo {author} {\bibfnamefont {Junhao}\
  \bibnamefont {Lin}}, \bibinfo {author} {\bibfnamefont {Zhengtai}\
  \bibnamefont {Liu}},  \emph {et~al.},\ }\bibfield  {title} {\enquote
  {\bibinfo {title} {Crystal-symmetry-paired spin--valley locking in a layered
  room-temperature metallic altermagnet candidate},}\ }\href {\doibase
  10.1038/s41567-025-02864-2} {\bibfield  {journal} {\bibinfo  {journal}
  {Nature Physics}\ }\textbf {\bibinfo {volume} {21}},\ \bibinfo {pages} {760}
  (\bibinfo {year} {2025})}\BibitemShut {NoStop}%
\bibitem [{\citenamefont {Hu}\ \emph {et~al.}(2026)\citenamefont {Hu},
  \citenamefont {Cheng}, \citenamefont {Duan}, \citenamefont {Hu},
  \citenamefont {Jiang}, \citenamefont {Xiao}, \citenamefont {Li},
  \citenamefont {Pan}, \citenamefont {Deng}, \citenamefont {Liu} \emph
  {et~al.}}]{hu2026observation}%
  \BibitemOpen
  \bibfield  {author} {\bibinfo {author} {\bibfnamefont {Quanxin}\ \bibnamefont
  {Hu}}, \bibinfo {author} {\bibfnamefont {Xingkai}\ \bibnamefont {Cheng}},
  \bibinfo {author} {\bibfnamefont {Qingchen}\ \bibnamefont {Duan}}, \bibinfo
  {author} {\bibfnamefont {Yudong}\ \bibnamefont {Hu}}, \bibinfo {author}
  {\bibfnamefont {Bei}\ \bibnamefont {Jiang}}, \bibinfo {author} {\bibfnamefont
  {Yusen}\ \bibnamefont {Xiao}}, \bibinfo {author} {\bibfnamefont {Yaqi}\
  \bibnamefont {Li}}, \bibinfo {author} {\bibfnamefont {Mojun}\ \bibnamefont
  {Pan}}, \bibinfo {author} {\bibfnamefont {Liwei}\ \bibnamefont {Deng}},
  \bibinfo {author} {\bibfnamefont {Changchao}\ \bibnamefont {Liu}},  \emph
  {et~al.},\ }\bibfield  {title} {\enquote {\bibinfo {title} {{Observation of
  spin-valley locked nodal lines in a quasi-2D altermagnet}},}\ }\href
  {https://doi.org/10.48550/arXiv.2601.02883} {\bibfield  {journal} {\bibinfo
  {journal} {arXiv:2601.02883}\ } (\bibinfo {year} {2026})}\BibitemShut
  {NoStop}%
\bibitem [{\citenamefont {Li}\ \emph {et~al.}(2026{\natexlab{a}})\citenamefont
  {Li}, \citenamefont {Chen}, \citenamefont {Pan}, \citenamefont {Li},
  \citenamefont {Zhang},\ and\ \citenamefont {Lu}}]{li2025exploration}%
  \BibitemOpen
  \bibfield  {author} {\bibinfo {author} {\bibfnamefont {Yu-Xin}\ \bibnamefont
  {Li}}, \bibinfo {author} {\bibfnamefont {Yiyuan}\ \bibnamefont {Chen}},
  \bibinfo {author} {\bibfnamefont {Liqing}\ \bibnamefont {Pan}}, \bibinfo
  {author} {\bibfnamefont {Shuai}\ \bibnamefont {Li}}, \bibinfo {author}
  {\bibfnamefont {Song-Bo}\ \bibnamefont {Zhang}}, \ and\ \bibinfo {author}
  {\bibfnamefont {Hai-Zhou}\ \bibnamefont {Lu}},\ }\bibfield  {title} {\enquote
  {\bibinfo {title} {{Exploration of Altermagnetism in $\mathrm{RuO_{2}}$}},}\
  }\href {\doibase 10.1007/s11433-026-2913-8} {\bibfield  {journal} {\bibinfo
  {journal} {Sci. China Phys. Mech. Astron.}\ }\textbf {\bibinfo {volume}
  {69}},\ \bibinfo {pages} {257001} (\bibinfo {year}
  {2026}{\natexlab{a}})}\BibitemShut {NoStop}%
\bibitem [{\citenamefont {Ahn}\ \emph {et~al.}(2019)\citenamefont {Ahn},
  \citenamefont {Hariki}, \citenamefont {Lee},\ and\ \citenamefont
  {Kune\ifmmode~\check{s}\else \v{s}\fi{}}}]{ahn2019antiferromagnetism}%
  \BibitemOpen
  \bibfield  {author} {\bibinfo {author} {\bibfnamefont {Kyo-Hoon}\
  \bibnamefont {Ahn}}, \bibinfo {author} {\bibfnamefont {Atsushi}\ \bibnamefont
  {Hariki}}, \bibinfo {author} {\bibfnamefont {Kwan-Woo}\ \bibnamefont {Lee}},
  \ and\ \bibinfo {author} {\bibfnamefont {Jan}\ \bibnamefont
  {Kune\ifmmode~\check{s}\else \v{s}\fi{}}},\ }\bibfield  {title} {\enquote
  {\bibinfo {title} {{Antiferromagnetism in ${\mathrm{RuO}}_{2}$ as $d$-wave
  Pomeranchuk instability}},}\ }\href {\doibase 10.1103/PhysRevB.99.184432}
  {\bibfield  {journal} {\bibinfo  {journal} {Phys. Rev. B}\ }\textbf {\bibinfo
  {volume} {99}},\ \bibinfo {pages} {184432} (\bibinfo {year}
  {2019})}\BibitemShut {NoStop}%
\bibitem [{\citenamefont {{\v{S}}mejkal}\ \emph {et~al.}(2020)\citenamefont
  {{\v{S}}mejkal}, \citenamefont {Gonz{\'a}lez-Hern{\'a}ndez}, \citenamefont
  {Jungwirth},\ and\ \citenamefont {Sinova}}]{vsmejkal2020crystal}%
  \BibitemOpen
  \bibfield  {author} {\bibinfo {author} {\bibfnamefont {Libor}\ \bibnamefont
  {{\v{S}}mejkal}}, \bibinfo {author} {\bibfnamefont {Rafael}\ \bibnamefont
  {Gonz{\'a}lez-Hern{\'a}ndez}}, \bibinfo {author} {\bibfnamefont
  {Tom{\'a}{\v{s}}}\ \bibnamefont {Jungwirth}}, \ and\ \bibinfo {author}
  {\bibfnamefont {Jairo}\ \bibnamefont {Sinova}},\ }\bibfield  {title}
  {\enquote {\bibinfo {title} {{Crystal time-reversal symmetry breaking and
  spontaneous Hall effect in collinear antiferromagnets}},}\ }\href {\doibase
  10.1126/sciadv.aaz8809} {\bibfield  {journal} {\bibinfo  {journal} {Science
  advances}\ }\textbf {\bibinfo {volume} {6}},\ \bibinfo {pages} {eaaz8809}
  (\bibinfo {year} {2020})}\BibitemShut {NoStop}%
\bibitem [{\citenamefont {Shao}\ \emph {et~al.}(2021)\citenamefont {Shao},
  \citenamefont {Zhang}, \citenamefont {Li}, \citenamefont {Eom},\ and\
  \citenamefont {Tsymbal}}]{shao2021spin}%
  \BibitemOpen
  \bibfield  {author} {\bibinfo {author} {\bibfnamefont {Ding-Fu}\ \bibnamefont
  {Shao}}, \bibinfo {author} {\bibfnamefont {Shu-Hui}\ \bibnamefont {Zhang}},
  \bibinfo {author} {\bibfnamefont {Ming}\ \bibnamefont {Li}}, \bibinfo
  {author} {\bibfnamefont {Chang-Beom}\ \bibnamefont {Eom}}, \ and\ \bibinfo
  {author} {\bibfnamefont {Evgeny~Y}\ \bibnamefont {Tsymbal}},\ }\bibfield
  {title} {\enquote {\bibinfo {title} {Spin-neutral currents for
  spintronics},}\ }\href {\doibase 10.1038/s41467-021-26915-3} {\bibfield
  {journal} {\bibinfo  {journal} {Nature Communications}\ }\textbf {\bibinfo
  {volume} {12}},\ \bibinfo {pages} {7061} (\bibinfo {year}
  {2021})}\BibitemShut {NoStop}%
\bibitem [{\citenamefont {Gonz\'alez-Hern\'andez}\ \emph
  {et~al.}(2021)\citenamefont {Gonz\'alez-Hern\'andez}, \citenamefont
  {\ifmmode~\check{S}\else \v{S}\fi{}mejkal}, \citenamefont {V\'yborn\'y},
  \citenamefont {Yahagi}, \citenamefont {Sinova}, \citenamefont {Jungwirth},\
  and\ \citenamefont {\ifmmode~\check{Z}\else
  \v{Z}\fi{}elezn\'y}}]{rafael2021efficient}%
  \BibitemOpen
  \bibfield  {author} {\bibinfo {author} {\bibfnamefont {Rafael}\ \bibnamefont
  {Gonz\'alez-Hern\'andez}}, \bibinfo {author} {\bibfnamefont {Libor}\
  \bibnamefont {\ifmmode~\check{S}\else \v{S}\fi{}mejkal}}, \bibinfo {author}
  {\bibfnamefont {Karel}\ \bibnamefont {V\'yborn\'y}}, \bibinfo {author}
  {\bibfnamefont {Yuta}\ \bibnamefont {Yahagi}}, \bibinfo {author}
  {\bibfnamefont {Jairo}\ \bibnamefont {Sinova}}, \bibinfo {author}
  {\bibfnamefont {Tom\'a\ifmmode \check{s}\else~\v{s}\fi{}}\ \bibnamefont
  {Jungwirth}}, \ and\ \bibinfo {author} {\bibfnamefont {Jakub}\ \bibnamefont
  {\ifmmode~\check{Z}\else \v{Z}\fi{}elezn\'y}},\ }\bibfield  {title} {\enquote
  {\bibinfo {title} {{Efficient Electrical Spin Splitter Based on
  Nonrelativistic Collinear Antiferromagnetism}},}\ }\href {\doibase
  10.1103/PhysRevLett.126.127701} {\bibfield  {journal} {\bibinfo  {journal}
  {Phys. Rev. Lett.}\ }\textbf {\bibinfo {volume} {126}},\ \bibinfo {pages}
  {127701} (\bibinfo {year} {2021})}\BibitemShut {NoStop}%
\bibitem [{\citenamefont {Bose}\ \emph {et~al.}(2022)\citenamefont {Bose},
  \citenamefont {Schreiber}, \citenamefont {Jain}, \citenamefont {Shao},
  \citenamefont {Nair}, \citenamefont {Sun}, \citenamefont {Zhang},
  \citenamefont {Muller}, \citenamefont {Tsymbal}, \citenamefont {Schlom} \emph
  {et~al.}}]{bose2022tilted}%
  \BibitemOpen
  \bibfield  {author} {\bibinfo {author} {\bibfnamefont {Arnab}\ \bibnamefont
  {Bose}}, \bibinfo {author} {\bibfnamefont {Nathaniel~J}\ \bibnamefont
  {Schreiber}}, \bibinfo {author} {\bibfnamefont {Rakshit}\ \bibnamefont
  {Jain}}, \bibinfo {author} {\bibfnamefont {Ding-Fu}\ \bibnamefont {Shao}},
  \bibinfo {author} {\bibfnamefont {Hari~P}\ \bibnamefont {Nair}}, \bibinfo
  {author} {\bibfnamefont {Jiaxin}\ \bibnamefont {Sun}}, \bibinfo {author}
  {\bibfnamefont {Xiyue~S}\ \bibnamefont {Zhang}}, \bibinfo {author}
  {\bibfnamefont {David~A}\ \bibnamefont {Muller}}, \bibinfo {author}
  {\bibfnamefont {Evgeny~Y}\ \bibnamefont {Tsymbal}}, \bibinfo {author}
  {\bibfnamefont {Darrell~G}\ \bibnamefont {Schlom}},  \emph {et~al.},\
  }\bibfield  {title} {\enquote {\bibinfo {title} {Tilted spin current
  generated by the collinear antiferromagnet ruthenium dioxide},}\ }\href
  {\doibase 10.1038/s41928-022-00744-8} {\bibfield  {journal} {\bibinfo
  {journal} {Nature Electronics}\ }\textbf {\bibinfo {volume} {5}},\ \bibinfo
  {pages} {267} (\bibinfo {year} {2022})}\BibitemShut {NoStop}%
\bibitem [{\citenamefont {Bai}\ \emph {et~al.}(2022)\citenamefont {Bai},
  \citenamefont {Han}, \citenamefont {Feng}, \citenamefont {Zhou},
  \citenamefont {Su}, \citenamefont {Wang}, \citenamefont {Liao}, \citenamefont
  {Zhu}, \citenamefont {Chen}, \citenamefont {Pan}, \citenamefont {Fan},\ and\
  \citenamefont {Song}}]{bai2022observation}%
  \BibitemOpen
  \bibfield  {author} {\bibinfo {author} {\bibfnamefont {H.}~\bibnamefont
  {Bai}}, \bibinfo {author} {\bibfnamefont {L.}~\bibnamefont {Han}}, \bibinfo
  {author} {\bibfnamefont {X.~Y.}\ \bibnamefont {Feng}}, \bibinfo {author}
  {\bibfnamefont {Y.~J.}\ \bibnamefont {Zhou}}, \bibinfo {author}
  {\bibfnamefont {R.~X.}\ \bibnamefont {Su}}, \bibinfo {author} {\bibfnamefont
  {Q.}~\bibnamefont {Wang}}, \bibinfo {author} {\bibfnamefont {L.~Y.}\
  \bibnamefont {Liao}}, \bibinfo {author} {\bibfnamefont {W.~X.}\ \bibnamefont
  {Zhu}}, \bibinfo {author} {\bibfnamefont {X.~Z.}\ \bibnamefont {Chen}},
  \bibinfo {author} {\bibfnamefont {F.}~\bibnamefont {Pan}}, \bibinfo {author}
  {\bibfnamefont {X.~L.}\ \bibnamefont {Fan}}, \ and\ \bibinfo {author}
  {\bibfnamefont {C.}~\bibnamefont {Song}},\ }\bibfield  {title} {\enquote
  {\bibinfo {title} {{Observation of Spin Splitting Torque in a Collinear
  Antiferromagnet ${\mathrm{RuO}}_{2}$}},}\ }\href {\doibase
  10.1103/PhysRevLett.128.197202} {\bibfield  {journal} {\bibinfo  {journal}
  {Phys. Rev. Lett.}\ }\textbf {\bibinfo {volume} {128}},\ \bibinfo {pages}
  {197202} (\bibinfo {year} {2022})}\BibitemShut {NoStop}%
\bibitem [{\citenamefont {Karube}\ \emph {et~al.}(2022)\citenamefont {Karube},
  \citenamefont {Tanaka}, \citenamefont {Sugawara}, \citenamefont {Kadoguchi},
  \citenamefont {Kohda},\ and\ \citenamefont {Nitta}}]{karube2022observation}%
  \BibitemOpen
  \bibfield  {author} {\bibinfo {author} {\bibfnamefont {Shutaro}\ \bibnamefont
  {Karube}}, \bibinfo {author} {\bibfnamefont {Takahiro}\ \bibnamefont
  {Tanaka}}, \bibinfo {author} {\bibfnamefont {Daichi}\ \bibnamefont
  {Sugawara}}, \bibinfo {author} {\bibfnamefont {Naohiro}\ \bibnamefont
  {Kadoguchi}}, \bibinfo {author} {\bibfnamefont {Makoto}\ \bibnamefont
  {Kohda}}, \ and\ \bibinfo {author} {\bibfnamefont {Junsaku}\ \bibnamefont
  {Nitta}},\ }\bibfield  {title} {\enquote {\bibinfo {title} {{Observation of
  Spin-Splitter Torque in Collinear Antiferromagnetic ${\mathrm{RuO}}_{2}$}},}\
  }\href {\doibase 10.1103/PhysRevLett.129.137201} {\bibfield  {journal}
  {\bibinfo  {journal} {Phys. Rev. Lett.}\ }\textbf {\bibinfo {volume} {129}},\
  \bibinfo {pages} {137201} (\bibinfo {year} {2022})}\BibitemShut {NoStop}%
\bibitem [{\citenamefont {Guo}\ \emph {et~al.}(2024)\citenamefont {Guo},
  \citenamefont {Zhang}, \citenamefont {Zhu}, \citenamefont {Jiang},
  \citenamefont {Jiang}, \citenamefont {Wu}, \citenamefont {Dong},
  \citenamefont {Xu}, \citenamefont {He}, \citenamefont {He} \emph
  {et~al.}}]{guo2024direct}%
  \BibitemOpen
  \bibfield  {author} {\bibinfo {author} {\bibfnamefont {Yaqin}\ \bibnamefont
  {Guo}}, \bibinfo {author} {\bibfnamefont {Jing}\ \bibnamefont {Zhang}},
  \bibinfo {author} {\bibfnamefont {Zengtai}\ \bibnamefont {Zhu}}, \bibinfo
  {author} {\bibfnamefont {Yuan-yuan}\ \bibnamefont {Jiang}}, \bibinfo {author}
  {\bibfnamefont {Longxing}\ \bibnamefont {Jiang}}, \bibinfo {author}
  {\bibfnamefont {Chuangwen}\ \bibnamefont {Wu}}, \bibinfo {author}
  {\bibfnamefont {Jing}\ \bibnamefont {Dong}}, \bibinfo {author} {\bibfnamefont
  {Xing}\ \bibnamefont {Xu}}, \bibinfo {author} {\bibfnamefont {Wenqing}\
  \bibnamefont {He}}, \bibinfo {author} {\bibfnamefont {Bin}\ \bibnamefont
  {He}},  \emph {et~al.},\ }\bibfield  {title} {\enquote {\bibinfo {title}
  {{Direct and inverse spin splitting effects in altermagnetic RuO$_2$}},}\
  }\href {\doibase 10.1002/advs.202400967} {\bibfield  {journal} {\bibinfo
  {journal} {Advanced Science}\ }\textbf {\bibinfo {volume} {11}},\ \bibinfo
  {pages} {2400967} (\bibinfo {year} {2024})}\BibitemShut {NoStop}%
\bibitem [{\citenamefont {He}\ \emph {et~al.}(2025)\citenamefont {He},
  \citenamefont {Wen}, \citenamefont {Okabayashi}, \citenamefont {Miura},
  \citenamefont {Ma}, \citenamefont {Ohkubo}, \citenamefont {Seki},
  \citenamefont {Sukegawa},\ and\ \citenamefont {Mitani}}]{he2025evidence}%
  \BibitemOpen
  \bibfield  {author} {\bibinfo {author} {\bibfnamefont {Cong}\ \bibnamefont
  {He}}, \bibinfo {author} {\bibfnamefont {Zhenchao}\ \bibnamefont {Wen}},
  \bibinfo {author} {\bibfnamefont {Jun}\ \bibnamefont {Okabayashi}}, \bibinfo
  {author} {\bibfnamefont {Yoshio}\ \bibnamefont {Miura}}, \bibinfo {author}
  {\bibfnamefont {Tianyi}\ \bibnamefont {Ma}}, \bibinfo {author} {\bibfnamefont
  {Tadakatsu}\ \bibnamefont {Ohkubo}}, \bibinfo {author} {\bibfnamefont
  {Takeshi}\ \bibnamefont {Seki}}, \bibinfo {author} {\bibfnamefont {Hiroaki}\
  \bibnamefont {Sukegawa}}, \ and\ \bibinfo {author} {\bibfnamefont {Seiji}\
  \bibnamefont {Mitani}},\ }\bibfield  {title} {\enquote {\bibinfo {title}
  {{Evidence for single variant in altermagnetic RuO$_2$ (101) thin films}},}\
  }\href {\doibase 10.1038/s41467-025-63344-y} {\bibfield  {journal} {\bibinfo
  {journal} {Nature Communications}\ }\textbf {\bibinfo {volume} {16}},\
  \bibinfo {pages} {8235} (\bibinfo {year} {2025})}\BibitemShut {NoStop}%
\bibitem [{\citenamefont {Bhowal}\ and\ \citenamefont
  {Spaldin}(2024)}]{bhowal2024ferroically}%
  \BibitemOpen
  \bibfield  {author} {\bibinfo {author} {\bibfnamefont {Sayantika}\
  \bibnamefont {Bhowal}}\ and\ \bibinfo {author} {\bibfnamefont {Nicola~A.}\
  \bibnamefont {Spaldin}},\ }\bibfield  {title} {\enquote {\bibinfo {title}
  {{Ferroically Ordered Magnetic Octupoles in $d$-Wave Altermagnets}},}\ }\href
  {\doibase 10.1103/PhysRevX.14.011019} {\bibfield  {journal} {\bibinfo
  {journal} {Phys. Rev. X}\ }\textbf {\bibinfo {volume} {14}},\ \bibinfo
  {pages} {011019} (\bibinfo {year} {2024})}\BibitemShut {NoStop}%
\bibitem [{\citenamefont {Li}\ \emph {et~al.}(2024)\citenamefont {Li},
  \citenamefont {Liu},\ and\ \citenamefont {Liu}}]{li2024creation}%
  \BibitemOpen
  \bibfield  {author} {\bibinfo {author} {\bibfnamefont {Yu-Xuan}\ \bibnamefont
  {Li}}, \bibinfo {author} {\bibfnamefont {Yichen}\ \bibnamefont {Liu}}, \ and\
  \bibinfo {author} {\bibfnamefont {Cheng-Cheng}\ \bibnamefont {Liu}},\
  }\bibfield  {title} {\enquote {\bibinfo {title} {Creation and manipulation of
  higher-order topological states by altermagnets},}\ }\href {\doibase
  10.1103/PhysRevB.109.L201109} {\bibfield  {journal} {\bibinfo  {journal}
  {Phys. Rev. B}\ }\textbf {\bibinfo {volume} {109}},\ \bibinfo {pages}
  {L201109} (\bibinfo {year} {2024})}\BibitemShut {NoStop}%
\bibitem [{\citenamefont {Mazin}(2023)}]{mazin2023altermagnetism}%
  \BibitemOpen
  \bibfield  {author} {\bibinfo {author} {\bibfnamefont {I.~I.}\ \bibnamefont
  {Mazin}},\ }\bibfield  {title} {\enquote {\bibinfo {title} {{Altermagnetism
  in MnTe: Origin, predicted manifestations, and routes to detwinning}},}\
  }\href {\doibase 10.1103/PhysRevB.107.L100418} {\bibfield  {journal}
  {\bibinfo  {journal} {Phys. Rev. B}\ }\textbf {\bibinfo {volume} {107}},\
  \bibinfo {pages} {L100418} (\bibinfo {year} {2023})}\BibitemShut {NoStop}%
\bibitem [{\citenamefont {Krempask{\`y}}\ \emph {et~al.}(2024)\citenamefont
  {Krempask{\`y}}, \citenamefont {{\v{S}}mejkal}, \citenamefont {D’souza},
  \citenamefont {Hajlaoui}, \citenamefont {Springholz}, \citenamefont
  {Uhl{\'\i}{\v{r}}ov{\'a}}, \citenamefont {Alarab}, \citenamefont
  {Constantinou}, \citenamefont {Strocov}, \citenamefont {Usanov} \emph
  {et~al.}}]{krempasky2024altermagnetic}%
  \BibitemOpen
  \bibfield  {author} {\bibinfo {author} {\bibfnamefont {J}~\bibnamefont
  {Krempask{\`y}}}, \bibinfo {author} {\bibfnamefont {L}~\bibnamefont
  {{\v{S}}mejkal}}, \bibinfo {author} {\bibfnamefont {SW}~\bibnamefont
  {D’souza}}, \bibinfo {author} {\bibfnamefont {M}~\bibnamefont {Hajlaoui}},
  \bibinfo {author} {\bibfnamefont {G}~\bibnamefont {Springholz}}, \bibinfo
  {author} {\bibfnamefont {K}~\bibnamefont {Uhl{\'\i}{\v{r}}ov{\'a}}}, \bibinfo
  {author} {\bibfnamefont {F}~\bibnamefont {Alarab}}, \bibinfo {author}
  {\bibfnamefont {PC}~\bibnamefont {Constantinou}}, \bibinfo {author}
  {\bibfnamefont {V}~\bibnamefont {Strocov}}, \bibinfo {author} {\bibfnamefont
  {D}~\bibnamefont {Usanov}},  \emph {et~al.},\ }\bibfield  {title} {\enquote
  {\bibinfo {title} {{Altermagnetic lifting of Kramers spin degeneracy}},}\
  }\href {\doibase 10.1038/s41586-023-06907-7} {\bibfield  {journal} {\bibinfo
  {journal} {Nature}\ }\textbf {\bibinfo {volume} {626}},\ \bibinfo {pages}
  {517} (\bibinfo {year} {2024})}\BibitemShut {NoStop}%
\bibitem [{\citenamefont {Lee}\ \emph {et~al.}(2024)\citenamefont {Lee},
  \citenamefont {Lee}, \citenamefont {Jung}, \citenamefont {Jung},
  \citenamefont {Kim}, \citenamefont {Lee}, \citenamefont {Seok}, \citenamefont
  {Kim}, \citenamefont {Park}, \citenamefont {\ifmmode~\check{S}\else
  \v{S}\fi{}mejkal}, \citenamefont {Kang},\ and\ \citenamefont
  {Kim}}]{lee2024broken}%
  \BibitemOpen
  \bibfield  {author} {\bibinfo {author} {\bibfnamefont {Suyoung}\ \bibnamefont
  {Lee}}, \bibinfo {author} {\bibfnamefont {Sangjae}\ \bibnamefont {Lee}},
  \bibinfo {author} {\bibfnamefont {Saegyeol}\ \bibnamefont {Jung}}, \bibinfo
  {author} {\bibfnamefont {Jiwon}\ \bibnamefont {Jung}}, \bibinfo {author}
  {\bibfnamefont {Donghan}\ \bibnamefont {Kim}}, \bibinfo {author}
  {\bibfnamefont {Yeonjae}\ \bibnamefont {Lee}}, \bibinfo {author}
  {\bibfnamefont {Byeongjun}\ \bibnamefont {Seok}}, \bibinfo {author}
  {\bibfnamefont {Jaeyoung}\ \bibnamefont {Kim}}, \bibinfo {author}
  {\bibfnamefont {Byeong~Gyu}\ \bibnamefont {Park}}, \bibinfo {author}
  {\bibfnamefont {Libor}\ \bibnamefont {\ifmmode~\check{S}\else
  \v{S}\fi{}mejkal}}, \bibinfo {author} {\bibfnamefont {Chang-Jong}\
  \bibnamefont {Kang}}, \ and\ \bibinfo {author} {\bibfnamefont {Changyoung}\
  \bibnamefont {Kim}},\ }\bibfield  {title} {\enquote {\bibinfo {title}
  {{Broken Kramers Degeneracy in Altermagnetic MnTe}},}\ }\href {\doibase
  10.1103/PhysRevLett.132.036702} {\bibfield  {journal} {\bibinfo  {journal}
  {Phys. Rev. Lett.}\ }\textbf {\bibinfo {volume} {132}},\ \bibinfo {pages}
  {036702} (\bibinfo {year} {2024})}\BibitemShut {NoStop}%
\bibitem [{\citenamefont {Osumi}\ \emph {et~al.}(2024)\citenamefont {Osumi},
  \citenamefont {Souma}, \citenamefont {Aoyama}, \citenamefont {Yamauchi},
  \citenamefont {Honma}, \citenamefont {Nakayama}, \citenamefont {Takahashi},
  \citenamefont {Ohgushi},\ and\ \citenamefont {Sato}}]{osumi2024observation}%
  \BibitemOpen
  \bibfield  {author} {\bibinfo {author} {\bibfnamefont {T.}~\bibnamefont
  {Osumi}}, \bibinfo {author} {\bibfnamefont {S.}~\bibnamefont {Souma}},
  \bibinfo {author} {\bibfnamefont {T.}~\bibnamefont {Aoyama}}, \bibinfo
  {author} {\bibfnamefont {K.}~\bibnamefont {Yamauchi}}, \bibinfo {author}
  {\bibfnamefont {A.}~\bibnamefont {Honma}}, \bibinfo {author} {\bibfnamefont
  {K.}~\bibnamefont {Nakayama}}, \bibinfo {author} {\bibfnamefont
  {T.}~\bibnamefont {Takahashi}}, \bibinfo {author} {\bibfnamefont
  {K.}~\bibnamefont {Ohgushi}}, \ and\ \bibinfo {author} {\bibfnamefont
  {T.}~\bibnamefont {Sato}},\ }\bibfield  {title} {\enquote {\bibinfo {title}
  {{Observation of a giant band splitting in altermagnetic MnTe}},}\ }\href
  {\doibase 10.1103/PhysRevB.109.115102} {\bibfield  {journal} {\bibinfo
  {journal} {Phys. Rev. B}\ }\textbf {\bibinfo {volume} {109}},\ \bibinfo
  {pages} {115102} (\bibinfo {year} {2024})}\BibitemShut {NoStop}%
\bibitem [{\citenamefont {Orlova}\ \emph {et~al.}(2025)\citenamefont {Orlova},
  \citenamefont {Esin}, \citenamefont {Timonina}, \citenamefont {Kolesnikov},\
  and\ \citenamefont {Deviatov}}]{orlova2025magnetocaloric}%
  \BibitemOpen
  \bibfield  {author} {\bibinfo {author} {\bibfnamefont {N.~N.}\ \bibnamefont
  {Orlova}}, \bibinfo {author} {\bibfnamefont {V.~D.}\ \bibnamefont {Esin}},
  \bibinfo {author} {\bibfnamefont {A.~V.}\ \bibnamefont {Timonina}}, \bibinfo
  {author} {\bibfnamefont {N.~N.}\ \bibnamefont {Kolesnikov}}, \ and\ \bibinfo
  {author} {\bibfnamefont {E.~V.}\ \bibnamefont {Deviatov}},\ }\bibfield
  {title} {\enquote {\bibinfo {title} {{Magnetocaloric effect for the
  altermagnetic candidate MnTe}},}\ }\href {https://arxiv.org/abs/2510.02777}
  {\bibfield  {journal} {\bibinfo  {journal} {arXiv:2510.02777}\ } (\bibinfo
  {year} {2025})}\BibitemShut {NoStop}%
\bibitem [{\citenamefont {Belashchenko}(2025)}]{belashchenko2025giant}%
  \BibitemOpen
  \bibfield  {author} {\bibinfo {author} {\bibfnamefont {K.~D.}\ \bibnamefont
  {Belashchenko}},\ }\bibfield  {title} {\enquote {\bibinfo {title} {{Giant
  Strain-Induced Spin Splitting Effect in MnTe, a $g$-Wave Altermagnetic
  Semiconductor}},}\ }\href {\doibase 10.1103/PhysRevLett.134.086701}
  {\bibfield  {journal} {\bibinfo  {journal} {Phys. Rev. Lett.}\ }\textbf
  {\bibinfo {volume} {134}},\ \bibinfo {pages} {086701} (\bibinfo {year}
  {2025})}\BibitemShut {NoStop}%
\bibitem [{\citenamefont {Reimers}\ \emph {et~al.}(2024)\citenamefont
  {Reimers}, \citenamefont {Odenbreit}, \citenamefont {{\v{S}}mejkal},
  \citenamefont {Strocov}, \citenamefont {Constantinou}, \citenamefont
  {Hellenes}, \citenamefont {Jaeschke~Ubiergo}, \citenamefont {Campos},
  \citenamefont {Bharadwaj}, \citenamefont {Chakraborty} \emph
  {et~al.}}]{reimers2024direct}%
  \BibitemOpen
  \bibfield  {author} {\bibinfo {author} {\bibfnamefont {Sonka}\ \bibnamefont
  {Reimers}}, \bibinfo {author} {\bibfnamefont {Lukas}\ \bibnamefont
  {Odenbreit}}, \bibinfo {author} {\bibfnamefont {Libor}\ \bibnamefont
  {{\v{S}}mejkal}}, \bibinfo {author} {\bibfnamefont {Vladimir~N}\ \bibnamefont
  {Strocov}}, \bibinfo {author} {\bibfnamefont {Procopios}\ \bibnamefont
  {Constantinou}}, \bibinfo {author} {\bibfnamefont {Anna~B}\ \bibnamefont
  {Hellenes}}, \bibinfo {author} {\bibfnamefont {Rodrigo}\ \bibnamefont
  {Jaeschke~Ubiergo}}, \bibinfo {author} {\bibfnamefont {Warlley~H}\
  \bibnamefont {Campos}}, \bibinfo {author} {\bibfnamefont {Venkata~K}\
  \bibnamefont {Bharadwaj}}, \bibinfo {author} {\bibfnamefont {Atasi}\
  \bibnamefont {Chakraborty}},  \emph {et~al.},\ }\bibfield  {title} {\enquote
  {\bibinfo {title} {{Direct observation of altermagnetic band splitting in
  CrSb thin films}},}\ }\href {\doibase 10.1038/s41467-024-46476-5} {\bibfield
  {journal} {\bibinfo  {journal} {Nature Communications}\ }\textbf {\bibinfo
  {volume} {15}},\ \bibinfo {pages} {2116} (\bibinfo {year}
  {2024})}\BibitemShut {NoStop}%
\bibitem [{\citenamefont {Ding}\ \emph {et~al.}(2024)\citenamefont {Ding},
  \citenamefont {Jiang}, \citenamefont {Chen}, \citenamefont {Tao},
  \citenamefont {Liu}, \citenamefont {Li}, \citenamefont {Liu}, \citenamefont
  {Sun}, \citenamefont {Cheng}, \citenamefont {Liu}, \citenamefont {Yang},
  \citenamefont {Zhang}, \citenamefont {Deng}, \citenamefont {Jing},
  \citenamefont {Huang}, \citenamefont {Shi}, \citenamefont {Ye}, \citenamefont
  {Qiao}, \citenamefont {Wang}, \citenamefont {Guo}, \citenamefont {Feng},\
  and\ \citenamefont {Shen}}]{ding2024large}%
  \BibitemOpen
  \bibfield  {author} {\bibinfo {author} {\bibfnamefont {Jianyang}\
  \bibnamefont {Ding}}, \bibinfo {author} {\bibfnamefont {Zhicheng}\
  \bibnamefont {Jiang}}, \bibinfo {author} {\bibfnamefont {Xiuhua}\
  \bibnamefont {Chen}}, \bibinfo {author} {\bibfnamefont {Zicheng}\
  \bibnamefont {Tao}}, \bibinfo {author} {\bibfnamefont {Zhengtai}\
  \bibnamefont {Liu}}, \bibinfo {author} {\bibfnamefont {Tongrui}\ \bibnamefont
  {Li}}, \bibinfo {author} {\bibfnamefont {Jishan}\ \bibnamefont {Liu}},
  \bibinfo {author} {\bibfnamefont {Jianping}\ \bibnamefont {Sun}}, \bibinfo
  {author} {\bibfnamefont {Jinguang}\ \bibnamefont {Cheng}}, \bibinfo {author}
  {\bibfnamefont {Jiayu}\ \bibnamefont {Liu}}, \bibinfo {author} {\bibfnamefont
  {Yichen}\ \bibnamefont {Yang}}, \bibinfo {author} {\bibfnamefont {Runfeng}\
  \bibnamefont {Zhang}}, \bibinfo {author} {\bibfnamefont {Liwei}\ \bibnamefont
  {Deng}}, \bibinfo {author} {\bibfnamefont {Wenchuan}\ \bibnamefont {Jing}},
  \bibinfo {author} {\bibfnamefont {Yu}~\bibnamefont {Huang}}, \bibinfo
  {author} {\bibfnamefont {Yuming}\ \bibnamefont {Shi}}, \bibinfo {author}
  {\bibfnamefont {Mao}\ \bibnamefont {Ye}}, \bibinfo {author} {\bibfnamefont
  {Shan}\ \bibnamefont {Qiao}}, \bibinfo {author} {\bibfnamefont {Yilin}\
  \bibnamefont {Wang}}, \bibinfo {author} {\bibfnamefont {Yanfeng}\
  \bibnamefont {Guo}}, \bibinfo {author} {\bibfnamefont {Donglai}\ \bibnamefont
  {Feng}}, \ and\ \bibinfo {author} {\bibfnamefont {Dawei}\ \bibnamefont
  {Shen}},\ }\bibfield  {title} {\enquote {\bibinfo {title} {{Large Band
  Splitting in $g$-Wave Altermagnet CrSb}},}\ }\href {\doibase
  10.1103/PhysRevLett.133.206401} {\bibfield  {journal} {\bibinfo  {journal}
  {Phys. Rev. Lett.}\ }\textbf {\bibinfo {volume} {133}},\ \bibinfo {pages}
  {206401} (\bibinfo {year} {2024})}\BibitemShut {NoStop}%
\bibitem [{\citenamefont {Peng}\ \emph {et~al.}(2025)\citenamefont {Peng},
  \citenamefont {Wang}, \citenamefont {Zhang}, \citenamefont {Zhou},
  \citenamefont {Sun}, \citenamefont {Su}, \citenamefont {Wu}, \citenamefont
  {Zhou}, \citenamefont {Liu}, \citenamefont {Wang}, \citenamefont {Yang},
  \citenamefont {Chen}, \citenamefont {Fang}, \citenamefont {Du}, \citenamefont
  {Jiao}, \citenamefont {Wu},\ and\ \citenamefont {Fang}}]{peng2025scaling}%
  \BibitemOpen
  \bibfield  {author} {\bibinfo {author} {\bibfnamefont {Xin}\ \bibnamefont
  {Peng}}, \bibinfo {author} {\bibfnamefont {Yuzhi}\ \bibnamefont {Wang}},
  \bibinfo {author} {\bibfnamefont {Shengnan}\ \bibnamefont {Zhang}}, \bibinfo
  {author} {\bibfnamefont {Yi}~\bibnamefont {Zhou}}, \bibinfo {author}
  {\bibfnamefont {Yuran}\ \bibnamefont {Sun}}, \bibinfo {author} {\bibfnamefont
  {Yahui}\ \bibnamefont {Su}}, \bibinfo {author} {\bibfnamefont {Chunxiang}\
  \bibnamefont {Wu}}, \bibinfo {author} {\bibfnamefont {Tingyu}\ \bibnamefont
  {Zhou}}, \bibinfo {author} {\bibfnamefont {Le}~\bibnamefont {Liu}}, \bibinfo
  {author} {\bibfnamefont {Hangdong}\ \bibnamefont {Wang}}, \bibinfo {author}
  {\bibfnamefont {Jinhu}\ \bibnamefont {Yang}}, \bibinfo {author}
  {\bibfnamefont {Bin}\ \bibnamefont {Chen}}, \bibinfo {author} {\bibfnamefont
  {Zhong}\ \bibnamefont {Fang}}, \bibinfo {author} {\bibfnamefont {Jianhua}\
  \bibnamefont {Du}}, \bibinfo {author} {\bibfnamefont {Zhiwei}\ \bibnamefont
  {Jiao}}, \bibinfo {author} {\bibfnamefont {Quansheng}\ \bibnamefont {Wu}}, \
  and\ \bibinfo {author} {\bibfnamefont {Minghu}\ \bibnamefont {Fang}},\
  }\bibfield  {title} {\enquote {\bibinfo {title} {{Scaling behavior of
  magnetoresistance and Hall resistivity in the altermagnet CrSb}},}\ }\href
  {\doibase 10.1103/PhysRevB.111.144402} {\bibfield  {journal} {\bibinfo
  {journal} {Phys. Rev. B}\ }\textbf {\bibinfo {volume} {111}},\ \bibinfo
  {pages} {144402} (\bibinfo {year} {2025})}\BibitemShut {NoStop}%
\bibitem [{\citenamefont {Zhou}\ \emph {et~al.}(2025)\citenamefont {Zhou},
  \citenamefont {Cheng}, \citenamefont {Hu}, \citenamefont {Chu}, \citenamefont
  {Bai}, \citenamefont {Han}, \citenamefont {Liu}, \citenamefont {Pan},\ and\
  \citenamefont {Song}}]{zhou2025manipulation}%
  \BibitemOpen
  \bibfield  {author} {\bibinfo {author} {\bibfnamefont {Zhiyuan}\ \bibnamefont
  {Zhou}}, \bibinfo {author} {\bibfnamefont {Xingkai}\ \bibnamefont {Cheng}},
  \bibinfo {author} {\bibfnamefont {Mengli}\ \bibnamefont {Hu}}, \bibinfo
  {author} {\bibfnamefont {Ruiyue}\ \bibnamefont {Chu}}, \bibinfo {author}
  {\bibfnamefont {Hua}\ \bibnamefont {Bai}}, \bibinfo {author} {\bibfnamefont
  {Lei}\ \bibnamefont {Han}}, \bibinfo {author} {\bibfnamefont {Junwei}\
  \bibnamefont {Liu}}, \bibinfo {author} {\bibfnamefont {Feng}\ \bibnamefont
  {Pan}}, \ and\ \bibinfo {author} {\bibfnamefont {Cheng}\ \bibnamefont
  {Song}},\ }\bibfield  {title} {\enquote {\bibinfo {title} {{Manipulation of
  the altermagnetic order in CrSb via crystal symmetry}},}\ }\href {\doibase
  10.1038/s41586-024-08436-3} {\bibfield  {journal} {\bibinfo  {journal}
  {Nature}\ }\textbf {\bibinfo {volume} {638}},\ \bibinfo {pages} {645}
  (\bibinfo {year} {2025})}\BibitemShut {NoStop}%
\bibitem [{\citenamefont {Mazin}\ \emph {et~al.}(2021)\citenamefont {Mazin},
  \citenamefont {Koepernik}, \citenamefont {Johannes}, \citenamefont
  {Gonz{\'a}lez-Hern{\'a}ndez},\ and\ \citenamefont
  {{\v{S}}mejkal}}]{mazin2021prediction}%
  \BibitemOpen
  \bibfield  {author} {\bibinfo {author} {\bibfnamefont {Igor~I}\ \bibnamefont
  {Mazin}}, \bibinfo {author} {\bibfnamefont {Klaus}\ \bibnamefont
  {Koepernik}}, \bibinfo {author} {\bibfnamefont {Michelle~D}\ \bibnamefont
  {Johannes}}, \bibinfo {author} {\bibfnamefont {Rafael}\ \bibnamefont
  {Gonz{\'a}lez-Hern{\'a}ndez}}, \ and\ \bibinfo {author} {\bibfnamefont
  {Libor}\ \bibnamefont {{\v{S}}mejkal}},\ }\bibfield  {title} {\enquote
  {\bibinfo {title} {{Prediction of unconventional magnetism in doped
  FeSb$_2$}},}\ }\href {\doibase 10.1073/pnas.2108924118} {\bibfield  {journal}
  {\bibinfo  {journal} {Proceedings of the National Academy of Sciences}\
  }\textbf {\bibinfo {volume} {118}},\ \bibinfo {pages} {e2108924118} (\bibinfo
  {year} {2021})}\BibitemShut {NoStop}%
\bibitem [{\citenamefont {Attias}\ \emph {et~al.}(2024)\citenamefont {Attias},
  \citenamefont {Levchenko},\ and\ \citenamefont
  {Khodas}}]{attias2024intrinsic}%
  \BibitemOpen
  \bibfield  {author} {\bibinfo {author} {\bibfnamefont {Lotan}\ \bibnamefont
  {Attias}}, \bibinfo {author} {\bibfnamefont {Alex}\ \bibnamefont
  {Levchenko}}, \ and\ \bibinfo {author} {\bibfnamefont {Maxim}\ \bibnamefont
  {Khodas}},\ }\bibfield  {title} {\enquote {\bibinfo {title} {{Intrinsic
  anomalous Hall effect in altermagnets}},}\ }\href {\doibase
  10.1103/PhysRevB.110.094425} {\bibfield  {journal} {\bibinfo  {journal}
  {Phys. Rev. B}\ }\textbf {\bibinfo {volume} {110}},\ \bibinfo {pages}
  {094425} (\bibinfo {year} {2024})}\BibitemShut {NoStop}%
\bibitem [{\citenamefont {Phillips}\ \emph {et~al.}(2025)\citenamefont
  {Phillips}, \citenamefont {Pokharel}, \citenamefont {Shtefiienko},
  \citenamefont {Bhandari}, \citenamefont {Graf}, \citenamefont {Rai},\ and\
  \citenamefont {Shrestha}}]{phillips2025electronic}%
  \BibitemOpen
  \bibfield  {author} {\bibinfo {author} {\bibfnamefont {Cole}\ \bibnamefont
  {Phillips}}, \bibinfo {author} {\bibfnamefont {Ganesh}\ \bibnamefont
  {Pokharel}}, \bibinfo {author} {\bibfnamefont {Kyryl}\ \bibnamefont
  {Shtefiienko}}, \bibinfo {author} {\bibfnamefont {Shalika~R.}\ \bibnamefont
  {Bhandari}}, \bibinfo {author} {\bibfnamefont {David~E.}\ \bibnamefont
  {Graf}}, \bibinfo {author} {\bibfnamefont {D.~P.}\ \bibnamefont {Rai}}, \
  and\ \bibinfo {author} {\bibfnamefont {Keshav}\ \bibnamefont {Shrestha}},\
  }\bibfield  {title} {\enquote {\bibinfo {title} {{Electronic structure of the
  altermagnet candidate ${\mathrm{FeSb}}_{2}$: High-field torque magnetometry
  and density functional theory studies}},}\ }\href {\doibase
  10.1103/PhysRevB.111.075141} {\bibfield  {journal} {\bibinfo  {journal}
  {Phys. Rev. B}\ }\textbf {\bibinfo {volume} {111}},\ \bibinfo {pages}
  {075141} (\bibinfo {year} {2025})}\BibitemShut {NoStop}%
\bibitem [{\citenamefont {Leivisk\"a}\ \emph {et~al.}(2024)\citenamefont
  {Leivisk\"a}, \citenamefont {Rial}, \citenamefont {Bad'ura}, \citenamefont
  {Seeger}, \citenamefont {Kounta}, \citenamefont {Beckert}, \citenamefont
  {Kriegner}, \citenamefont {Joumard}, \citenamefont {Schmoranzerov\'a},
  \citenamefont {Sinova}, \citenamefont {Gomonay}, \citenamefont {Thomas},
  \citenamefont {Goennenwein}, \citenamefont {Reichlov\'a}, \citenamefont
  {\ifmmode~\check{S}\else \v{S}\fi{}mejkal}, \citenamefont {Michez},
  \citenamefont {Jungwirth},\ and\ \citenamefont
  {Baltz}}]{leiviska2024anisotropy}%
  \BibitemOpen
  \bibfield  {author} {\bibinfo {author} {\bibfnamefont {Miina}\ \bibnamefont
  {Leivisk\"a}}, \bibinfo {author} {\bibfnamefont {Javier}\ \bibnamefont
  {Rial}}, \bibinfo {author} {\bibfnamefont {Anton\'{\i}n}\ \bibnamefont
  {Bad'ura}}, \bibinfo {author} {\bibfnamefont {Rafael~Lopes}\ \bibnamefont
  {Seeger}}, \bibinfo {author} {\bibfnamefont {Isma\"{\i}la}\ \bibnamefont
  {Kounta}}, \bibinfo {author} {\bibfnamefont {Sebastian}\ \bibnamefont
  {Beckert}}, \bibinfo {author} {\bibfnamefont {Dominik}\ \bibnamefont
  {Kriegner}}, \bibinfo {author} {\bibfnamefont {Isabelle}\ \bibnamefont
  {Joumard}}, \bibinfo {author} {\bibfnamefont {Eva}\ \bibnamefont
  {Schmoranzerov\'a}}, \bibinfo {author} {\bibfnamefont {Jairo}\ \bibnamefont
  {Sinova}}, \bibinfo {author} {\bibfnamefont {Olena}\ \bibnamefont {Gomonay}},
  \bibinfo {author} {\bibfnamefont {Andy}\ \bibnamefont {Thomas}}, \bibinfo
  {author} {\bibfnamefont {Sebastian T.~B.}\ \bibnamefont {Goennenwein}},
  \bibinfo {author} {\bibfnamefont {Helena}\ \bibnamefont {Reichlov\'a}},
  \bibinfo {author} {\bibfnamefont {Libor}\ \bibnamefont
  {\ifmmode~\check{S}\else \v{S}\fi{}mejkal}}, \bibinfo {author} {\bibfnamefont
  {Lisa}\ \bibnamefont {Michez}}, \bibinfo {author} {\bibfnamefont
  {Tom\'a\ifmmode~\check{s}\else\v{s}\fi{}}\ \bibnamefont {Jungwirth}}, \ and\
  \bibinfo {author} {\bibfnamefont {Vincent}\ \bibnamefont {Baltz}},\
  }\bibfield  {title} {\enquote {\bibinfo {title} {{Anisotropy of the anomalous
  Hall effect in thin films of the altermagnet candidate
  ${\mathrm{Mn}}_{5}{\mathrm{Si}}_{3}$}},}\ }\href {\doibase
  10.1103/PhysRevB.109.224430} {\bibfield  {journal} {\bibinfo  {journal}
  {Phys. Rev. B}\ }\textbf {\bibinfo {volume} {109}},\ \bibinfo {pages}
  {224430} (\bibinfo {year} {2024})}\BibitemShut {NoStop}%
\bibitem [{\citenamefont {Reichlova}\ \emph {et~al.}(2024)\citenamefont
  {Reichlova}, \citenamefont {Lopes~Seeger}, \citenamefont
  {Gonz{\'a}lez-Hern{\'a}ndez}, \citenamefont {Kounta}, \citenamefont
  {Schlitz}, \citenamefont {Kriegner}, \citenamefont {Ritzinger}, \citenamefont
  {Lammel}, \citenamefont {Leivisk{\"a}}, \citenamefont {Birk~Hellenes} \emph
  {et~al.}}]{reichlova2024observation}%
  \BibitemOpen
  \bibfield  {author} {\bibinfo {author} {\bibfnamefont {Helena}\ \bibnamefont
  {Reichlova}}, \bibinfo {author} {\bibfnamefont {Rafael}\ \bibnamefont
  {Lopes~Seeger}}, \bibinfo {author} {\bibfnamefont {Rafael}\ \bibnamefont
  {Gonz{\'a}lez-Hern{\'a}ndez}}, \bibinfo {author} {\bibfnamefont {Ismaila}\
  \bibnamefont {Kounta}}, \bibinfo {author} {\bibfnamefont {Richard}\
  \bibnamefont {Schlitz}}, \bibinfo {author} {\bibfnamefont {Dominik}\
  \bibnamefont {Kriegner}}, \bibinfo {author} {\bibfnamefont {Philipp}\
  \bibnamefont {Ritzinger}}, \bibinfo {author} {\bibfnamefont {Michaela}\
  \bibnamefont {Lammel}}, \bibinfo {author} {\bibfnamefont {Miina}\
  \bibnamefont {Leivisk{\"a}}}, \bibinfo {author} {\bibfnamefont {Anna}\
  \bibnamefont {Birk~Hellenes}},  \emph {et~al.},\ }\bibfield  {title}
  {\enquote {\bibinfo {title} {{Observation of a spontaneous anomalous Hall
  response in the Mn$_5$Si$_3$ $d$-wave altermagnet candidate}},}\ }\href
  {\doibase 10.1038/s41467-024-48493-w} {\bibfield  {journal} {\bibinfo
  {journal} {Nature Communications}\ }\textbf {\bibinfo {volume} {15}},\
  \bibinfo {pages} {4961} (\bibinfo {year} {2024})}\BibitemShut {NoStop}%
\bibitem [{\citenamefont {Rial}\ \emph {et~al.}(2024)\citenamefont {Rial},
  \citenamefont {Leivisk\"a}, \citenamefont {Skobjin}, \citenamefont {Bad'ura},
  \citenamefont {Gaudin}, \citenamefont {Disdier}, \citenamefont {Schlitz},
  \citenamefont {Kounta}, \citenamefont {Beckert}, \citenamefont {Kriegner},
  \citenamefont {Thomas}, \citenamefont {Schmoranzerov\'a}, \citenamefont
  {\ifmmode~\check{S}\else \v{S}\fi{}mejkal}, \citenamefont {Sinova},
  \citenamefont {Jungwirth}, \citenamefont {Michez}, \citenamefont
  {Reichlov\'a}, \citenamefont {Goennenwein}, \citenamefont {Gomonay},\ and\
  \citenamefont {Baltz}}]{rial2024altermagnetic}%
  \BibitemOpen
  \bibfield  {author} {\bibinfo {author} {\bibfnamefont {Javier}\ \bibnamefont
  {Rial}}, \bibinfo {author} {\bibfnamefont {Miina}\ \bibnamefont
  {Leivisk\"a}}, \bibinfo {author} {\bibfnamefont {Gregor}\ \bibnamefont
  {Skobjin}}, \bibinfo {author} {\bibfnamefont {Anton\'{\i}n}\ \bibnamefont
  {Bad'ura}}, \bibinfo {author} {\bibfnamefont {Gilles}\ \bibnamefont
  {Gaudin}}, \bibinfo {author} {\bibfnamefont {Florian}\ \bibnamefont
  {Disdier}}, \bibinfo {author} {\bibfnamefont {Richard}\ \bibnamefont
  {Schlitz}}, \bibinfo {author} {\bibfnamefont {Isma\"{\i}la}\ \bibnamefont
  {Kounta}}, \bibinfo {author} {\bibfnamefont {Sebastian}\ \bibnamefont
  {Beckert}}, \bibinfo {author} {\bibfnamefont {Dominik}\ \bibnamefont
  {Kriegner}}, \bibinfo {author} {\bibfnamefont {Andy}\ \bibnamefont {Thomas}},
  \bibinfo {author} {\bibfnamefont {Eva}\ \bibnamefont {Schmoranzerov\'a}},
  \bibinfo {author} {\bibfnamefont {Libor}\ \bibnamefont
  {\ifmmode~\check{S}\else \v{S}\fi{}mejkal}}, \bibinfo {author} {\bibfnamefont
  {Jairo}\ \bibnamefont {Sinova}}, \bibinfo {author} {\bibfnamefont
  {Tom\'a\ifmmode \check{s}\else~\v{s}\fi{}}\ \bibnamefont {Jungwirth}},
  \bibinfo {author} {\bibfnamefont {Lisa}\ \bibnamefont {Michez}}, \bibinfo
  {author} {\bibfnamefont {Helena}\ \bibnamefont {Reichlov\'a}}, \bibinfo
  {author} {\bibfnamefont {Sebastian T.~B.}\ \bibnamefont {Goennenwein}},
  \bibinfo {author} {\bibfnamefont {Olena}\ \bibnamefont {Gomonay}}, \ and\
  \bibinfo {author} {\bibfnamefont {Vincent}\ \bibnamefont {Baltz}},\
  }\bibfield  {title} {\enquote {\bibinfo {title} {{Altermagnetic variants in
  thin films of $\mathrm{M}{\mathrm{n}}_{5}\mathrm{S}{\mathrm{i}}_{3}$}},}\
  }\href {\doibase 10.1103/PhysRevB.110.L220411} {\bibfield  {journal}
  {\bibinfo  {journal} {Phys. Rev. B}\ }\textbf {\bibinfo {volume} {110}},\
  \bibinfo {pages} {L220411} (\bibinfo {year} {2024})}\BibitemShut {NoStop}%
\bibitem [{\citenamefont {Han}\ \emph {et~al.}(2024)\citenamefont {Han},
  \citenamefont {Fu}, \citenamefont {Peng}, \citenamefont {Cheng},
  \citenamefont {Dai}, \citenamefont {Liu}, \citenamefont {Li}, \citenamefont
  {Zhang}, \citenamefont {Zhu}, \citenamefont {Bai} \emph
  {et~al.}}]{han2024electrical}%
  \BibitemOpen
  \bibfield  {author} {\bibinfo {author} {\bibfnamefont {Lei}\ \bibnamefont
  {Han}}, \bibinfo {author} {\bibfnamefont {Xizhi}\ \bibnamefont {Fu}},
  \bibinfo {author} {\bibfnamefont {Rui}\ \bibnamefont {Peng}}, \bibinfo
  {author} {\bibfnamefont {Xingkai}\ \bibnamefont {Cheng}}, \bibinfo {author}
  {\bibfnamefont {Jiankun}\ \bibnamefont {Dai}}, \bibinfo {author}
  {\bibfnamefont {Liangyang}\ \bibnamefont {Liu}}, \bibinfo {author}
  {\bibfnamefont {Yidian}\ \bibnamefont {Li}}, \bibinfo {author} {\bibfnamefont
  {Yichi}\ \bibnamefont {Zhang}}, \bibinfo {author} {\bibfnamefont {Wenxuan}\
  \bibnamefont {Zhu}}, \bibinfo {author} {\bibfnamefont {Hua}\ \bibnamefont
  {Bai}},  \emph {et~al.},\ }\bibfield  {title} {\enquote {\bibinfo {title}
  {{Electrical 180 switching of N{\'e}el vector in spin-splitting
  antiferromagnet}},}\ }\href {\doibase 10.1126/sciadv.adn0479} {\bibfield
  {journal} {\bibinfo  {journal} {Science Advances}\ }\textbf {\bibinfo
  {volume} {10}},\ \bibinfo {pages} {eadn0479} (\bibinfo {year}
  {2024})}\BibitemShut {NoStop}%
\bibitem [{\citenamefont {Urru}\ \emph {et~al.}(2025)\citenamefont {Urru},
  \citenamefont {Seleznev}, \citenamefont {Teng}, \citenamefont {Park},
  \citenamefont {Reyes-Lillo},\ and\ \citenamefont {Rabe}}]{urru2025g}%
  \BibitemOpen
  \bibfield  {author} {\bibinfo {author} {\bibfnamefont {Andrea}\ \bibnamefont
  {Urru}}, \bibinfo {author} {\bibfnamefont {Daniel}\ \bibnamefont {Seleznev}},
  \bibinfo {author} {\bibfnamefont {Yujia}\ \bibnamefont {Teng}}, \bibinfo
  {author} {\bibfnamefont {Se~Young}\ \bibnamefont {Park}}, \bibinfo {author}
  {\bibfnamefont {Sebastian~E.}\ \bibnamefont {Reyes-Lillo}}, \ and\ \bibinfo
  {author} {\bibfnamefont {Karin~M.}\ \bibnamefont {Rabe}},\ }\bibfield
  {title} {\enquote {\bibinfo {title} {{$G$-type antiferromagnetic
  ${\mathrm{BiFeO}}_{3}$ is a multiferroic $g$-wave altermagnet}},}\ }\href
  {\doibase 10.1103/v3fg-6smc} {\bibfield  {journal} {\bibinfo  {journal}
  {Phys. Rev. B}\ }\textbf {\bibinfo {volume} {112}},\ \bibinfo {pages}
  {104411} (\bibinfo {year} {2025})}\BibitemShut {NoStop}%
\bibitem [{\citenamefont {Fratian}\ \emph {et~al.}(2026)\citenamefont
  {Fratian}, \citenamefont {Ramesh}, \citenamefont {Li}, \citenamefont
  {Golias}, \citenamefont {Nahas}, \citenamefont {Schultheis}, \citenamefont
  {Skolaut}, \citenamefont {Checa}, \citenamefont {Ghosal}, \citenamefont
  {Priessnitz}, \citenamefont {Mbognou}, \citenamefont {Ojha}, \citenamefont
  {Zhou}, \citenamefont {Qualls}, \citenamefont {Litzius}, \citenamefont
  {Klewe}, \citenamefont {Meisenheimer}, \citenamefont {Bellaiche},
  \citenamefont {Šmejkal}, \citenamefont {Schlom}, \citenamefont {Han},
  \citenamefont {Prokhorenko}, \citenamefont {Ramesh}, \citenamefont
  {Stevenson}, \citenamefont {Wittmann},\ and\ \citenamefont
  {Caretta}}]{george2026topological}%
  \BibitemOpen
  \bibfield  {author} {\bibinfo {author} {\bibfnamefont {George}\ \bibnamefont
  {Fratian}}, \bibinfo {author} {\bibfnamefont {Maya}\ \bibnamefont {Ramesh}},
  \bibinfo {author} {\bibfnamefont {Xinyan}\ \bibnamefont {Li}}, \bibinfo
  {author} {\bibfnamefont {Evangelos}\ \bibnamefont {Golias}}, \bibinfo
  {author} {\bibfnamefont {Yousra}\ \bibnamefont {Nahas}}, \bibinfo {author}
  {\bibfnamefont {Sebastian Maria~Ulrich}\ \bibnamefont {Schultheis}}, \bibinfo
  {author} {\bibfnamefont {Julian}\ \bibnamefont {Skolaut}}, \bibinfo {author}
  {\bibfnamefont {Marti}\ \bibnamefont {Checa}}, \bibinfo {author}
  {\bibfnamefont {Arundhati}\ \bibnamefont {Ghosal}}, \bibinfo {author}
  {\bibfnamefont {Jan}\ \bibnamefont {Priessnitz}}, \bibinfo {author}
  {\bibfnamefont {F.~C.~Fobasso}\ \bibnamefont {Mbognou}}, \bibinfo {author}
  {\bibfnamefont {Shashank~Kumar}\ \bibnamefont {Ojha}}, \bibinfo {author}
  {\bibfnamefont {Shiyu}\ \bibnamefont {Zhou}}, \bibinfo {author}
  {\bibfnamefont {Alexander}\ \bibnamefont {Qualls}}, \bibinfo {author}
  {\bibfnamefont {Kai}\ \bibnamefont {Litzius}}, \bibinfo {author}
  {\bibfnamefont {Christoph}\ \bibnamefont {Klewe}}, \bibinfo {author}
  {\bibfnamefont {Peter}\ \bibnamefont {Meisenheimer}}, \bibinfo {author}
  {\bibfnamefont {Laurent}\ \bibnamefont {Bellaiche}}, \bibinfo {author}
  {\bibfnamefont {Libor}\ \bibnamefont {Šmejkal}}, \bibinfo {author}
  {\bibfnamefont {Darrell~G.}\ \bibnamefont {Schlom}}, \bibinfo {author}
  {\bibfnamefont {Yimo}\ \bibnamefont {Han}}, \bibinfo {author} {\bibfnamefont
  {Sergei}\ \bibnamefont {Prokhorenko}}, \bibinfo {author} {\bibfnamefont
  {Ramamoorthy}\ \bibnamefont {Ramesh}}, \bibinfo {author} {\bibfnamefont
  {Paul}\ \bibnamefont {Stevenson}}, \bibinfo {author} {\bibfnamefont {Angela}\
  \bibnamefont {Wittmann}}, \ and\ \bibinfo {author} {\bibfnamefont {Lucas}\
  \bibnamefont {Caretta}},\ }\bibfield  {title} {\enquote {\bibinfo {title}
  {{Topological textures and emergent altermagnetic signatures in ultrathin
  BiFeO$_3$}},}\ }\href {https://doi.org/10.48550/arXiv.2601.09970} {\bibfield
  {journal} {\bibinfo  {journal} {arXiv:2601.09970}\ } (\bibinfo {year}
  {2026})}\BibitemShut {NoStop}%
\bibitem [{\citenamefont {Husain}\ \emph {et~al.}(2026)\citenamefont {Husain},
  \citenamefont {Ramesh}, \citenamefont {Song}, \citenamefont {Prokhorenko},
  \citenamefont {Ojha}, \citenamefont {Panda}, \citenamefont {Li},
  \citenamefont {Nahas}, \citenamefont {Kumar}, \citenamefont {Gupta},
  \citenamefont {Chang}, \citenamefont {Jung}, \citenamefont {de~Sousa},
  \citenamefont {Analytis}, \citenamefont {Martin}, \citenamefont {Yao},
  \citenamefont {Han}, \citenamefont {Cheong}, \citenamefont {Bellaiche},
  \citenamefont {Bibes}, \citenamefont {Schlom},\ and\ \citenamefont
  {Ramesh}}]{sajid2026signatures}%
  \BibitemOpen
  \bibfield  {author} {\bibinfo {author} {\bibfnamefont {Sajid}\ \bibnamefont
  {Husain}}, \bibinfo {author} {\bibfnamefont {Maya}\ \bibnamefont {Ramesh}},
  \bibinfo {author} {\bibfnamefont {Qian}\ \bibnamefont {Song}}, \bibinfo
  {author} {\bibfnamefont {Sergei}\ \bibnamefont {Prokhorenko}}, \bibinfo
  {author} {\bibfnamefont {Shashank~Kumar}\ \bibnamefont {Ojha}}, \bibinfo
  {author} {\bibfnamefont {Surya~Narayan}\ \bibnamefont {Panda}}, \bibinfo
  {author} {\bibfnamefont {Xinyan}\ \bibnamefont {Li}}, \bibinfo {author}
  {\bibfnamefont {Yousra}\ \bibnamefont {Nahas}}, \bibinfo {author}
  {\bibfnamefont {Yogesh}\ \bibnamefont {Kumar}}, \bibinfo {author}
  {\bibfnamefont {Pushpendra}\ \bibnamefont {Gupta}}, \bibinfo {author}
  {\bibfnamefont {Tenzin}\ \bibnamefont {Chang}}, \bibinfo {author}
  {\bibfnamefont {Alan Ji-in}\ \bibnamefont {Jung}}, \bibinfo {author}
  {\bibfnamefont {Rog\'erio}\ \bibnamefont {de~Sousa}}, \bibinfo {author}
  {\bibfnamefont {James~G.}\ \bibnamefont {Analytis}}, \bibinfo {author}
  {\bibfnamefont {Lane~W.}\ \bibnamefont {Martin}}, \bibinfo {author}
  {\bibfnamefont {Zhi}\ \bibnamefont {Yao}}, \bibinfo {author} {\bibfnamefont
  {Yimo}\ \bibnamefont {Han}}, \bibinfo {author} {\bibfnamefont {Sang-Wook}\
  \bibnamefont {Cheong}}, \bibinfo {author} {\bibfnamefont {Laurent}\
  \bibnamefont {Bellaiche}}, \bibinfo {author} {\bibfnamefont {Manuel}\
  \bibnamefont {Bibes}}, \bibinfo {author} {\bibfnamefont {Darrell~G.}\
  \bibnamefont {Schlom}}, \ and\ \bibinfo {author} {\bibfnamefont
  {Ramamoorthy}\ \bibnamefont {Ramesh}},\ }\bibfield  {title} {\enquote
  {\bibinfo {title} {{Signatures of Altermagnetism in
  ${\mathrm{BiFeO}}_{3}$}},}\ }\href {\doibase 10.1103/1wyj-nrgk} {\bibfield
  {journal} {\bibinfo  {journal} {Phys. Rev. Lett.}\ }\textbf {\bibinfo
  {volume} {137}},\ \bibinfo {pages} {036702} (\bibinfo {year}
  {2026})}\BibitemShut {NoStop}%
\bibitem [{\citenamefont {Wang}\ \emph {et~al.}(2026)\citenamefont {Wang},
  \citenamefont {Li}, \citenamefont {Li}, \citenamefont {Chen}, \citenamefont
  {Dong}, \citenamefont {Chen}, \citenamefont {Chen}, \citenamefont {Zheng},
  \citenamefont {Guo}, \citenamefont {Tong}, \citenamefont {Bai}, \citenamefont
  {Zhang}, \citenamefont {Gao}, \citenamefont {Shen}, \citenamefont {Zhu},
  \citenamefont {Han}, \citenamefont {Wei}, \citenamefont {Jiang},
  \citenamefont {Zhang}, \citenamefont {Wang}, \citenamefont {Xu},
  \citenamefont {Shi}, \citenamefont {Wang}, \citenamefont {Zhang},
  \citenamefont {Liu}, \citenamefont {Song}, \citenamefont {Liu}, \citenamefont
  {Xie},\ and\ \citenamefont {Liu}}]{gui2026electric}%
  \BibitemOpen
  \bibfield  {author} {\bibinfo {author} {\bibfnamefont {Gui}\ \bibnamefont
  {Wang}}, \bibinfo {author} {\bibfnamefont {Yuhang}\ \bibnamefont {Li}},
  \bibinfo {author} {\bibfnamefont {Bin}\ \bibnamefont {Li}}, \bibinfo {author}
  {\bibfnamefont {Xianzhe}\ \bibnamefont {Chen}}, \bibinfo {author}
  {\bibfnamefont {Jianting}\ \bibnamefont {Dong}}, \bibinfo {author}
  {\bibfnamefont {Weizhao}\ \bibnamefont {Chen}}, \bibinfo {author}
  {\bibfnamefont {Xiaobing}\ \bibnamefont {Chen}}, \bibinfo {author}
  {\bibfnamefont {Naifu}\ \bibnamefont {Zheng}}, \bibinfo {author}
  {\bibfnamefont {Maosen}\ \bibnamefont {Guo}}, \bibinfo {author}
  {\bibfnamefont {Aomei}\ \bibnamefont {Tong}}, \bibinfo {author}
  {\bibfnamefont {Hua}\ \bibnamefont {Bai}}, \bibinfo {author} {\bibfnamefont
  {Hongrui}\ \bibnamefont {Zhang}}, \bibinfo {author} {\bibfnamefont {Yifan}\
  \bibnamefont {Gao}}, \bibinfo {author} {\bibfnamefont {Kaiwen}\ \bibnamefont
  {Shen}}, \bibinfo {author} {\bibfnamefont {Jiangyuan}\ \bibnamefont {Zhu}},
  \bibinfo {author} {\bibfnamefont {Jiahao}\ \bibnamefont {Han}}, \bibinfo
  {author} {\bibfnamefont {Yingfen}\ \bibnamefont {Wei}}, \bibinfo {author}
  {\bibfnamefont {Hao}\ \bibnamefont {Jiang}}, \bibinfo {author} {\bibfnamefont
  {Xumeng}\ \bibnamefont {Zhang}}, \bibinfo {author} {\bibfnamefont {Ming}\
  \bibnamefont {Wang}}, \bibinfo {author} {\bibfnamefont {Kebiao}\ \bibnamefont
  {Xu}}, \bibinfo {author} {\bibfnamefont {Wu}~\bibnamefont {Shi}}, \bibinfo
  {author} {\bibfnamefont {Pengfei}\ \bibnamefont {Wang}}, \bibinfo {author}
  {\bibfnamefont {Jia}\ \bibnamefont {Zhang}}, \bibinfo {author} {\bibfnamefont
  {Qihang}\ \bibnamefont {Liu}}, \bibinfo {author} {\bibfnamefont {Cheng}\
  \bibnamefont {Song}}, \bibinfo {author} {\bibfnamefont {Qi}~\bibnamefont
  {Liu}}, \bibinfo {author} {\bibfnamefont {Xincheng}\ \bibnamefont {Xie}}, \
  and\ \bibinfo {author} {\bibfnamefont {Ming}\ \bibnamefont {Liu}},\
  }\bibfield  {title} {\enquote {\bibinfo {title} {Electric field switching of
  altermagnetic spin-splitting in multiferroic skyrmions},}\ }\href
  {https://doi.org/10.48550/arXiv.2601.06738} {\bibfield  {journal} {\bibinfo
  {journal} {arXiv:2601.06738}\ } (\bibinfo {year} {2026})}\BibitemShut
  {NoStop}%
\bibitem [{\citenamefont {Ouassou}\ \emph {et~al.}(2023)\citenamefont
  {Ouassou}, \citenamefont {Brataas},\ and\ \citenamefont
  {Linder}}]{ouassou2023dc}%
  \BibitemOpen
  \bibfield  {author} {\bibinfo {author} {\bibfnamefont {Jabir~Ali}\
  \bibnamefont {Ouassou}}, \bibinfo {author} {\bibfnamefont {Arne}\
  \bibnamefont {Brataas}}, \ and\ \bibinfo {author} {\bibfnamefont {Jacob}\
  \bibnamefont {Linder}},\ }\bibfield  {title} {\enquote {\bibinfo {title} {{dc
  Josephson Effect in Altermagnets}},}\ }\href {\doibase
  10.1103/PhysRevLett.131.076003} {\bibfield  {journal} {\bibinfo  {journal}
  {Phys. Rev. Lett.}\ }\textbf {\bibinfo {volume} {131}},\ \bibinfo {pages}
  {076003} (\bibinfo {year} {2023})}\BibitemShut {NoStop}%
\bibitem [{\citenamefont {Zhang}\ \emph {et~al.}(2024)\citenamefont {Zhang},
  \citenamefont {Hu},\ and\ \citenamefont {Neupert}}]{zhang2024finite}%
  \BibitemOpen
  \bibfield  {author} {\bibinfo {author} {\bibfnamefont {Song-Bo}\ \bibnamefont
  {Zhang}}, \bibinfo {author} {\bibfnamefont {Lun-Hui}\ \bibnamefont {Hu}}, \
  and\ \bibinfo {author} {\bibfnamefont {Titus}\ \bibnamefont {Neupert}},\
  }\bibfield  {title} {\enquote {\bibinfo {title} {{Finite-momentum Cooper
  pairing in proximitized altermagnets}},}\ }\href {\doibase
  10.1038/s41467-024-45951-3} {\bibfield  {journal} {\bibinfo  {journal}
  {Nature Communications}\ }\textbf {\bibinfo {volume} {15}},\ \bibinfo {pages}
  {136301} (\bibinfo {year} {2024})}\BibitemShut {NoStop}%
\bibitem [{\citenamefont {Cheng}\ and\ \citenamefont
  {Sun}(2024)}]{cheng2024orientation}%
  \BibitemOpen
  \bibfield  {author} {\bibinfo {author} {\bibfnamefont {Qiang}\ \bibnamefont
  {Cheng}}\ and\ \bibinfo {author} {\bibfnamefont {Qing-Feng}\ \bibnamefont
  {Sun}},\ }\bibfield  {title} {\enquote {\bibinfo {title}
  {Orientation-dependent josephson effect in spin-singlet
  superconductor/altermagnet/spin-triplet superconductor junctions},}\ }\href
  {\doibase 10.1103/PhysRevB.109.024517} {\bibfield  {journal} {\bibinfo
  {journal} {Phys. Rev. B}\ }\textbf {\bibinfo {volume} {109}},\ \bibinfo
  {pages} {024517} (\bibinfo {year} {2024})}\BibitemShut {NoStop}%
\bibitem [{\citenamefont {Beenakker}\ and\ \citenamefont
  {Vakhtel}(2023)}]{beenakker2023phase}%
  \BibitemOpen
  \bibfield  {author} {\bibinfo {author} {\bibfnamefont {C.~W.~J.}\
  \bibnamefont {Beenakker}}\ and\ \bibinfo {author} {\bibfnamefont
  {T.}~\bibnamefont {Vakhtel}},\ }\bibfield  {title} {\enquote {\bibinfo
  {title} {{Phase-shifted Andreev levels in an altermagnet Josephson
  junction}},}\ }\href {\doibase 10.1103/PhysRevB.108.075425} {\bibfield
  {journal} {\bibinfo  {journal} {Phys. Rev. B}\ }\textbf {\bibinfo {volume}
  {108}},\ \bibinfo {pages} {075425} (\bibinfo {year} {2023})}\BibitemShut
  {NoStop}%
\bibitem [{\citenamefont {Lu}\ \emph {et~al.}(2024)\citenamefont {Lu},
  \citenamefont {Maeda}, \citenamefont {Ito}, \citenamefont {Yada},\ and\
  \citenamefont {Tanaka}}]{lu2024varphi}%
  \BibitemOpen
  \bibfield  {author} {\bibinfo {author} {\bibfnamefont {Bo}~\bibnamefont
  {Lu}}, \bibinfo {author} {\bibfnamefont {Kazuki}\ \bibnamefont {Maeda}},
  \bibinfo {author} {\bibfnamefont {Hiroyuki}\ \bibnamefont {Ito}}, \bibinfo
  {author} {\bibfnamefont {Keiji}\ \bibnamefont {Yada}}, \ and\ \bibinfo
  {author} {\bibfnamefont {Yukio}\ \bibnamefont {Tanaka}},\ }\bibfield  {title}
  {\enquote {\bibinfo {title} {{$\ensuremath{\varphi}$ Josephson Junction
  Induced by Altermagnetism}},}\ }\href {\doibase
  10.1103/PhysRevLett.133.226002} {\bibfield  {journal} {\bibinfo  {journal}
  {Phys. Rev. Lett.}\ }\textbf {\bibinfo {volume} {133}},\ \bibinfo {pages}
  {226002} (\bibinfo {year} {2024})}\BibitemShut {NoStop}%
\bibitem [{\citenamefont {Sun}\ \emph {et~al.}(2025{\natexlab{a}})\citenamefont
  {Sun}, \citenamefont {Zhang}, \citenamefont {Li},\ and\ \citenamefont
  {Trauzettel}}]{sun2025tunable}%
  \BibitemOpen
  \bibfield  {author} {\bibinfo {author} {\bibfnamefont {Hai-Peng}\
  \bibnamefont {Sun}}, \bibinfo {author} {\bibfnamefont {Song-Bo}\ \bibnamefont
  {Zhang}}, \bibinfo {author} {\bibfnamefont {Chang-An}\ \bibnamefont {Li}}, \
  and\ \bibinfo {author} {\bibfnamefont {Bj\"orn}\ \bibnamefont {Trauzettel}},\
  }\bibfield  {title} {\enquote {\bibinfo {title} {Tunable second harmonic in
  altermagnetic josephson junctions},}\ }\href {\doibase
  10.1103/PhysRevB.111.165406} {\bibfield  {journal} {\bibinfo  {journal}
  {Phys. Rev. B}\ }\textbf {\bibinfo {volume} {111}},\ \bibinfo {pages}
  {165406} (\bibinfo {year} {2025}{\natexlab{a}})}\BibitemShut {NoStop}%
\bibitem [{\citenamefont {Fukaya}\ \emph
  {et~al.}(2025{\natexlab{a}})\citenamefont {Fukaya}, \citenamefont {Maeda},
  \citenamefont {Yada}, \citenamefont {Cayao}, \citenamefont {Tanaka},\ and\
  \citenamefont {Lu}}]{fukaya2025josephson}%
  \BibitemOpen
  \bibfield  {author} {\bibinfo {author} {\bibfnamefont {Yuri}\ \bibnamefont
  {Fukaya}}, \bibinfo {author} {\bibfnamefont {Kazuki}\ \bibnamefont {Maeda}},
  \bibinfo {author} {\bibfnamefont {Keiji}\ \bibnamefont {Yada}}, \bibinfo
  {author} {\bibfnamefont {Jorge}\ \bibnamefont {Cayao}}, \bibinfo {author}
  {\bibfnamefont {Yukio}\ \bibnamefont {Tanaka}}, \ and\ \bibinfo {author}
  {\bibfnamefont {Bo}~\bibnamefont {Lu}},\ }\bibfield  {title} {\enquote
  {\bibinfo {title} {Josephson effect and odd-frequency pairing in
  superconducting junctions with unconventional magnets},}\ }\href {\doibase
  10.1103/PhysRevB.111.064502} {\bibfield  {journal} {\bibinfo  {journal}
  {Phys. Rev. B}\ }\textbf {\bibinfo {volume} {111}},\ \bibinfo {pages}
  {064502} (\bibinfo {year} {2025}{\natexlab{a}})}\BibitemShut {NoStop}%
\bibitem [{\citenamefont {Pal}\ \emph {et~al.}(2025)\citenamefont {Pal},
  \citenamefont {Mondal}, \citenamefont {Nag},\ and\ \citenamefont
  {Saha}}]{pal2025josephson}%
  \BibitemOpen
  \bibfield  {author} {\bibinfo {author} {\bibfnamefont {Amartya}\ \bibnamefont
  {Pal}}, \bibinfo {author} {\bibfnamefont {Debashish}\ \bibnamefont {Mondal}},
  \bibinfo {author} {\bibfnamefont {Tanay}\ \bibnamefont {Nag}}, \ and\
  \bibinfo {author} {\bibfnamefont {Arijit}\ \bibnamefont {Saha}},\ }\bibfield
  {title} {\enquote {\bibinfo {title} {{Josephson current signature of Floquet
  Majorana and topological accidental zero modes in altermagnet
  heterostructures}},}\ }\href {\doibase 10.1103/prnx-47mk} {\bibfield
  {journal} {\bibinfo  {journal} {Phys. Rev. B}\ }\textbf {\bibinfo {volume}
  {112}},\ \bibinfo {pages} {L201408} (\bibinfo {year} {2025})}\BibitemShut
  {NoStop}%
\bibitem [{\citenamefont {Sun}\ \emph {et~al.}(2023)\citenamefont {Sun},
  \citenamefont {Brataas},\ and\ \citenamefont {Linder}}]{sun2023andreev}%
  \BibitemOpen
  \bibfield  {author} {\bibinfo {author} {\bibfnamefont {Chi}\ \bibnamefont
  {Sun}}, \bibinfo {author} {\bibfnamefont {Arne}\ \bibnamefont {Brataas}}, \
  and\ \bibinfo {author} {\bibfnamefont {Jacob}\ \bibnamefont {Linder}},\
  }\bibfield  {title} {\enquote {\bibinfo {title} {{Andreev reflection in
  altermagnets}},}\ }\href {\doibase 10.1103/PhysRevB.108.054511} {\bibfield
  {journal} {\bibinfo  {journal} {Phys. Rev. B}\ }\textbf {\bibinfo {volume}
  {108}},\ \bibinfo {pages} {054511} (\bibinfo {year} {2023})}\BibitemShut
  {NoStop}%
\bibitem [{\citenamefont {Papaj}(2023)}]{papaj2023andreev}%
  \BibitemOpen
  \bibfield  {author} {\bibinfo {author} {\bibfnamefont {Micha\l{}}\
  \bibnamefont {Papaj}},\ }\bibfield  {title} {\enquote {\bibinfo {title}
  {Andreev reflection at the altermagnet-superconductor interface},}\ }\href
  {\doibase 10.1103/PhysRevB.108.L060508} {\bibfield  {journal} {\bibinfo
  {journal} {Phys. Rev. B}\ }\textbf {\bibinfo {volume} {108}},\ \bibinfo
  {pages} {L060508} (\bibinfo {year} {2023})}\BibitemShut {NoStop}%
\bibitem [{\citenamefont {Ezawa}(2024{\natexlab{a}})}]{ezawa2024intrinsic}%
  \BibitemOpen
  \bibfield  {author} {\bibinfo {author} {\bibfnamefont {Motohiko}\
  \bibnamefont {Ezawa}},\ }\bibfield  {title} {\enquote {\bibinfo {title}
  {{Intrinsic nonlinear conductivity induced by quantum geometry in
  altermagnets and measurement of the in-plane N\'eel vector}},}\ }\href
  {\doibase 10.1103/PhysRevB.110.L241405} {\bibfield  {journal} {\bibinfo
  {journal} {Phys. Rev. B}\ }\textbf {\bibinfo {volume} {110}},\ \bibinfo
  {pages} {L241405} (\bibinfo {year} {2024}{\natexlab{a}})}\BibitemShut
  {NoStop}%
\bibitem [{\citenamefont {Fang}\ \emph {et~al.}(2024)\citenamefont {Fang},
  \citenamefont {Cano},\ and\ \citenamefont {Ghorashi}}]{fang2024quantum}%
  \BibitemOpen
  \bibfield  {author} {\bibinfo {author} {\bibfnamefont {Yuan}\ \bibnamefont
  {Fang}}, \bibinfo {author} {\bibfnamefont {Jennifer}\ \bibnamefont {Cano}}, \
  and\ \bibinfo {author} {\bibfnamefont {Sayed Ali~Akbar}\ \bibnamefont
  {Ghorashi}},\ }\bibfield  {title} {\enquote {\bibinfo {title} {{Quantum
  Geometry Induced Nonlinear Transport in Altermagnets}},}\ }\href {\doibase
  10.1103/PhysRevLett.133.106701} {\bibfield  {journal} {\bibinfo  {journal}
  {Phys. Rev. Lett.}\ }\textbf {\bibinfo {volume} {133}},\ \bibinfo {pages}
  {106701} (\bibinfo {year} {2024})}\BibitemShut {NoStop}%
\bibitem [{\citenamefont {Liu}\ \emph {et~al.}(2025)\citenamefont {Liu},
  \citenamefont {Chen}, \citenamefont {Xiao}, \citenamefont {Duan},
  \citenamefont {Wang},\ and\ \citenamefont {Deng}}]{liu2025enhancement}%
  \BibitemOpen
  \bibfield  {author} {\bibinfo {author} {\bibfnamefont {Tian-Xin}\
  \bibnamefont {Liu}}, \bibinfo {author} {\bibfnamefont {Fu-Yang}\ \bibnamefont
  {Chen}}, \bibinfo {author} {\bibfnamefont {Xin}\ \bibnamefont {Xiao}},
  \bibinfo {author} {\bibfnamefont {Hou-Jian}\ \bibnamefont {Duan}}, \bibinfo
  {author} {\bibfnamefont {Rui-Qiang}\ \bibnamefont {Wang}}, \ and\ \bibinfo
  {author} {\bibfnamefont {Ming-Xun}\ \bibnamefont {Deng}},\ }\bibfield
  {title} {\enquote {\bibinfo {title} {{Enhancement of the linear and nonlinear
  planar Hall effect by altermagnets on the surface of topological
  insulators}},}\ }\href {\doibase 10.1103/PhysRevB.111.155124} {\bibfield
  {journal} {\bibinfo  {journal} {Phys. Rev. B}\ }\textbf {\bibinfo {volume}
  {111}},\ \bibinfo {pages} {155124} (\bibinfo {year} {2025})}\BibitemShut
  {NoStop}%
\bibitem [{\citenamefont {Sun}\ \emph {et~al.}(2025{\natexlab{b}})\citenamefont
  {Sun}, \citenamefont {Mao}, \citenamefont {Zhuang},\ and\ \citenamefont
  {Sun}}]{sun2025tunneling}%
  \BibitemOpen
  \bibfield  {author} {\bibinfo {author} {\bibfnamefont {Yu-Fei}\ \bibnamefont
  {Sun}}, \bibinfo {author} {\bibfnamefont {Yue}\ \bibnamefont {Mao}}, \bibinfo
  {author} {\bibfnamefont {Yu-Chen}\ \bibnamefont {Zhuang}}, \ and\ \bibinfo
  {author} {\bibfnamefont {Qing-Feng}\ \bibnamefont {Sun}},\ }\bibfield
  {title} {\enquote {\bibinfo {title} {Tunneling magnetoresistance effect in
  altermagnets},}\ }\href {\doibase 10.1103/t8b5-l859} {\bibfield  {journal}
  {\bibinfo  {journal} {Phys. Rev. B}\ }\textbf {\bibinfo {volume} {112}},\
  \bibinfo {pages} {094411} (\bibinfo {year} {2025}{\natexlab{b}})}\BibitemShut
  {NoStop}%
\bibitem [{\citenamefont {Wan}\ and\ \citenamefont
  {Sun}(2025)}]{wan2025altermagnetism}%
  \BibitemOpen
  \bibfield  {author} {\bibinfo {author} {\bibfnamefont {Yu-Hao}\ \bibnamefont
  {Wan}}\ and\ \bibinfo {author} {\bibfnamefont {Qing-Feng}\ \bibnamefont
  {Sun}},\ }\bibfield  {title} {\enquote {\bibinfo {title}
  {Altermagnetism-induced parity anomaly in weak topological insulators},}\
  }\href {\doibase 10.1103/PhysRevB.111.045407} {\bibfield  {journal} {\bibinfo
   {journal} {Phys. Rev. B}\ }\textbf {\bibinfo {volume} {111}},\ \bibinfo
  {pages} {045407} (\bibinfo {year} {2025})}\BibitemShut {NoStop}%
\bibitem [{\citenamefont {Ezawa}(2025)}]{ezawa2025bulk}%
  \BibitemOpen
  \bibfield  {author} {\bibinfo {author} {\bibfnamefont {Motohiko}\
  \bibnamefont {Ezawa}},\ }\bibfield  {title} {\enquote {\bibinfo {title} {Bulk
  photovoltaic effects in altermagnets},}\ }\href {\doibase
  10.1103/PhysRevB.111.L201405} {\bibfield  {journal} {\bibinfo  {journal}
  {Phys. Rev. B}\ }\textbf {\bibinfo {volume} {111}},\ \bibinfo {pages}
  {L201405} (\bibinfo {year} {2025})}\BibitemShut {NoStop}%
\bibitem [{\citenamefont {Zhu}\ \emph {et~al.}(2026{\natexlab{a}})\citenamefont
  {Zhu}, \citenamefont {Huang}, \citenamefont {Chen}, \citenamefont {Cui},
  \citenamefont {Duan}, \citenamefont {Zhang}, \citenamefont {\ifmmode
  \check{Z}\else \v{Z}\fi{}uti\ifmmode~\acute{c}\else \'{c}\fi{}},\ and\
  \citenamefont {Zhou}}]{zhu2026altermagnetic}%
  \BibitemOpen
  \bibfield  {author} {\bibinfo {author} {\bibfnamefont {Ziye}\ \bibnamefont
  {Zhu}}, \bibinfo {author} {\bibfnamefont {Richang}\ \bibnamefont {Huang}},
  \bibinfo {author} {\bibfnamefont {Xianzhang}\ \bibnamefont {Chen}}, \bibinfo
  {author} {\bibfnamefont {Zhou}\ \bibnamefont {Cui}}, \bibinfo {author}
  {\bibfnamefont {Xunkai}\ \bibnamefont {Duan}}, \bibinfo {author}
  {\bibfnamefont {Jiayong}\ \bibnamefont {Zhang}}, \bibinfo {author}
  {\bibfnamefont {Igor}\ \bibnamefont {\ifmmode \check{Z}\else
  \v{Z}\fi{}uti\ifmmode~\acute{c}\else \'{c}\fi{}}}, \ and\ \bibinfo {author}
  {\bibfnamefont {Tong}\ \bibnamefont {Zhou}},\ }\bibfield  {title} {\enquote
  {\bibinfo {title} {{Altermagnetic Proximity Effect}},}\ }\href {\doibase
  10.1103/kqy8-myz1} {\bibfield  {journal} {\bibinfo  {journal} {Phys. Rev.
  Lett.}\ }\textbf {\bibinfo {volume} {136}},\ \bibinfo {pages} {186702}
  (\bibinfo {year} {2026}{\natexlab{a}})}\BibitemShut {NoStop}%
\bibitem [{\citenamefont {Qin}\ and\ \citenamefont
  {Chen}(2026)}]{qin2026layer}%
  \BibitemOpen
  \bibfield  {author} {\bibinfo {author} {\bibfnamefont {Fang}\ \bibnamefont
  {Qin}}\ and\ \bibinfo {author} {\bibfnamefont {Rui}\ \bibnamefont {Chen}},\
  }\bibfield  {title} {\enquote {\bibinfo {title} {{Layer Hall effect induced
  by altermagnetism}},}\ }\href {\doibase 10.1103/vwtc-klg7} {\bibfield
  {journal} {\bibinfo  {journal} {Phys. Rev. B}\ }\textbf {\bibinfo {volume}
  {113}},\ \bibinfo {pages} {165418} (\bibinfo {year} {2026})}\BibitemShut
  {NoStop}%
\bibitem [{\citenamefont {Chen}\ \emph {et~al.}(2026)\citenamefont {Chen},
  \citenamefont {Zhou},\ and\ \citenamefont {Xu}}]{chen2026quasicrystalline}%
  \BibitemOpen
  \bibfield  {author} {\bibinfo {author} {\bibfnamefont {Rui}\ \bibnamefont
  {Chen}}, \bibinfo {author} {\bibfnamefont {Bin}\ \bibnamefont {Zhou}}, \ and\
  \bibinfo {author} {\bibfnamefont {Dong-Hui}\ \bibnamefont {Xu}},\ }\bibfield
  {title} {\enquote {\bibinfo {title} {{Altermagnetism in Quasicrystals}},}\
  }\href {\doibase 10.1103/src1-b2xt} {\bibfield  {journal} {\bibinfo
  {journal} {Phys. Rev. Lett.}\ }\textbf {\bibinfo {volume} {137}},\ \bibinfo
  {pages} {046705} (\bibinfo {year} {2026})}\BibitemShut {NoStop}%
\bibitem [{\citenamefont {Shao}\ \emph {et~al.}(2025)\citenamefont {Shao},
  \citenamefont {Lu}, \citenamefont {Pan}, \citenamefont {Liu},\ and\
  \citenamefont {Yang}}]{shao2025classification}%
  \BibitemOpen
  \bibfield  {author} {\bibinfo {author} {\bibfnamefont {Zhi-Yan}\ \bibnamefont
  {Shao}}, \bibinfo {author} {\bibfnamefont {Chen}\ \bibnamefont {Lu}},
  \bibinfo {author} {\bibfnamefont {Zhiming}\ \bibnamefont {Pan}}, \bibinfo
  {author} {\bibfnamefont {Yu-Bo}\ \bibnamefont {Liu}}, \ and\ \bibinfo
  {author} {\bibfnamefont {Fan}\ \bibnamefont {Yang}},\ }\bibfield  {title}
  {\enquote {\bibinfo {title} {{Classification of Magnetism and Altermagnetism
  in Quasicrystals}},}\ }\href {https://arxiv.org/abs/2508.15702} {\bibfield
  {journal} {\bibinfo  {journal} {arXiv:2508.15702}\ } (\bibinfo {year}
  {2025})}\BibitemShut {NoStop}%
\bibitem [{\citenamefont {Li}\ \emph {et~al.}(2026{\natexlab{b}})\citenamefont
  {Li}, \citenamefont {Pan}, \citenamefont {Leng}, \citenamefont {Chen},\ and\
  \citenamefont {Huang}}]{li2026unconventional}%
  \BibitemOpen
  \bibfield  {author} {\bibinfo {author} {\bibfnamefont {Yiming}\ \bibnamefont
  {Li}}, \bibinfo {author} {\bibfnamefont {Mingxiang}\ \bibnamefont {Pan}},
  \bibinfo {author} {\bibfnamefont {Jun}\ \bibnamefont {Leng}}, \bibinfo
  {author} {\bibfnamefont {Yuxiao}\ \bibnamefont {Chen}}, \ and\ \bibinfo
  {author} {\bibfnamefont {Huaqing}\ \bibnamefont {Huang}},\ }\bibfield
  {title} {\enquote {\bibinfo {title} {{Unconventional Altermagnetism in
  Quasicrystals: A Hyperspatial Projective Construction}},}\ }\href {\doibase
  10.1103/hx9n-v1yg} {\bibfield  {journal} {\bibinfo  {journal} {Phys. Rev.
  Lett.}\ }\textbf {\bibinfo {volume} {137}},\ \bibinfo {pages} {046704}
  (\bibinfo {year} {2026}{\natexlab{b}})}\BibitemShut {NoStop}%
\bibitem [{\citenamefont {Lin}\ \emph {et~al.}(2025)\citenamefont {Lin},
  \citenamefont {Zhang}, \citenamefont {Lu},\ and\ \citenamefont
  {Xie}}]{lin2025coulomb}%
  \BibitemOpen
  \bibfield  {author} {\bibinfo {author} {\bibfnamefont {Hao-Jie}\ \bibnamefont
  {Lin}}, \bibinfo {author} {\bibfnamefont {Song-Bo}\ \bibnamefont {Zhang}},
  \bibinfo {author} {\bibfnamefont {Hai-Zhou}\ \bibnamefont {Lu}}, \ and\
  \bibinfo {author} {\bibfnamefont {X.~C.}\ \bibnamefont {Xie}},\ }\bibfield
  {title} {\enquote {\bibinfo {title} {{Coulomb Drag in Altermagnets}},}\
  }\href {\doibase 10.1103/PhysRevLett.134.136301} {\bibfield  {journal}
  {\bibinfo  {journal} {Phys. Rev. Lett.}\ }\textbf {\bibinfo {volume} {134}},\
  \bibinfo {pages} {136301} (\bibinfo {year} {2025})}\BibitemShut {NoStop}%
\bibitem [{\citenamefont {Wang}\ \emph
  {et~al.}(2025{\natexlab{b}})\citenamefont {Wang}, \citenamefont {Zhang},
  \citenamefont {Zhang}, \citenamefont {Sun}, \citenamefont {Dagotto},
  \citenamefont {Xu},\ and\ \citenamefont {Hu}}]{wang2025spin}%
  \BibitemOpen
  \bibfield  {author} {\bibinfo {author} {\bibfnamefont {Zi-Ming}\ \bibnamefont
  {Wang}}, \bibinfo {author} {\bibfnamefont {Yang}\ \bibnamefont {Zhang}},
  \bibinfo {author} {\bibfnamefont {Song-Bo}\ \bibnamefont {Zhang}}, \bibinfo
  {author} {\bibfnamefont {Jin-Hua}\ \bibnamefont {Sun}}, \bibinfo {author}
  {\bibfnamefont {Elbio}\ \bibnamefont {Dagotto}}, \bibinfo {author}
  {\bibfnamefont {Dong-Hui}\ \bibnamefont {Xu}}, \ and\ \bibinfo {author}
  {\bibfnamefont {Lun-Hui}\ \bibnamefont {Hu}},\ }\bibfield  {title} {\enquote
  {\bibinfo {title} {{Spin-Orbital Altermagnetism}},}\ }\href {\doibase
  10.1103/cjzw-j4v7} {\bibfield  {journal} {\bibinfo  {journal} {Phys. Rev.
  Lett.}\ }\textbf {\bibinfo {volume} {135}},\ \bibinfo {pages} {176705}
  (\bibinfo {year} {2025}{\natexlab{b}})}\BibitemShut {NoStop}%
\bibitem [{\citenamefont {Chen}\ \emph {et~al.}(2025)\citenamefont {Chen},
  \citenamefont {Wang}, \citenamefont {Wu}, \citenamefont {Sun}, \citenamefont
  {Zhou}, \citenamefont {Wang},\ and\ \citenamefont {Xu}}]{chen2025probing}%
  \BibitemOpen
  \bibfield  {author} {\bibinfo {author} {\bibfnamefont {Rui}\ \bibnamefont
  {Chen}}, \bibinfo {author} {\bibfnamefont {Zi-Ming}\ \bibnamefont {Wang}},
  \bibinfo {author} {\bibfnamefont {Ke}~\bibnamefont {Wu}}, \bibinfo {author}
  {\bibfnamefont {Hai-Peng}\ \bibnamefont {Sun}}, \bibinfo {author}
  {\bibfnamefont {Bin}\ \bibnamefont {Zhou}}, \bibinfo {author} {\bibfnamefont
  {Rui}\ \bibnamefont {Wang}}, \ and\ \bibinfo {author} {\bibfnamefont
  {Dong-Hui}\ \bibnamefont {Xu}},\ }\bibfield  {title} {\enquote {\bibinfo
  {title} {{Probing $k$-Space Alternating Spin Polarization via the Anomalous
  Hall Effect}},}\ }\href {\doibase 10.1103/yrs7-m6zy} {\bibfield  {journal}
  {\bibinfo  {journal} {Phys. Rev. Lett.}\ }\textbf {\bibinfo {volume} {135}},\
  \bibinfo {pages} {096602} (\bibinfo {year} {2025})}\BibitemShut {NoStop}%
\bibitem [{\citenamefont {Fu}\ \emph {et~al.}(2026{\natexlab{a}})\citenamefont
  {Fu}, \citenamefont {Mondal}, \citenamefont {Liu}, \citenamefont {Tanaka},\
  and\ \citenamefont {Cayao}}]{fu2026floquet}%
  \BibitemOpen
  \bibfield  {author} {\bibinfo {author} {\bibfnamefont {Pei-Hao}\ \bibnamefont
  {Fu}}, \bibinfo {author} {\bibfnamefont {Sayan}\ \bibnamefont {Mondal}},
  \bibinfo {author} {\bibfnamefont {Jun-Feng}\ \bibnamefont {Liu}}, \bibinfo
  {author} {\bibfnamefont {Yukio}\ \bibnamefont {Tanaka}}, \ and\ \bibinfo
  {author} {\bibfnamefont {Jorge}\ \bibnamefont {Cayao}},\ }\bibfield  {title}
  {\enquote {\bibinfo {title} {Floquet engineering spin triplet states in
  unconventional magnets},}\ }\href {\doibase 10.1103/lkf9-jgv6} {\bibfield
  {journal} {\bibinfo  {journal} {Phys. Rev. Lett.}\ }\textbf {\bibinfo
  {volume} {136}},\ \bibinfo {pages} {066703} (\bibinfo {year}
  {2026}{\natexlab{a}})}\BibitemShut {NoStop}%
\bibitem [{\citenamefont {Fu}\ \emph {et~al.}(2026{\natexlab{b}})\citenamefont
  {Fu}, \citenamefont {Mondal}, \citenamefont {Liu},\ and\ \citenamefont
  {Cayao}}]{fu2025light}%
  \BibitemOpen
  \bibfield  {author} {\bibinfo {author} {\bibfnamefont {Pei-Hao}\ \bibnamefont
  {Fu}}, \bibinfo {author} {\bibfnamefont {Sayan}\ \bibnamefont {Mondal}},
  \bibinfo {author} {\bibfnamefont {Jun-Feng}\ \bibnamefont {Liu}}, \ and\
  \bibinfo {author} {\bibfnamefont {Jorge}\ \bibnamefont {Cayao}},\ }\bibfield
  {title} {\enquote {\bibinfo {title} {{Light-induced Floquet spin-triplet
  Cooper pairs in unconventional magnets}},}\ }\href {\doibase
  10.21468/SciPostPhys.20.2.059} {\bibfield  {journal} {\bibinfo  {journal}
  {SciPost Phys.}\ }\textbf {\bibinfo {volume} {20}},\ \bibinfo {pages} {059}
  (\bibinfo {year} {2026}{\natexlab{b}})}\BibitemShut {NoStop}%
\bibitem [{\citenamefont {Fukaya}\ \emph
  {et~al.}(2025{\natexlab{b}})\citenamefont {Fukaya}, \citenamefont {Lu},
  \citenamefont {Yada}, \citenamefont {Tanaka},\ and\ \citenamefont
  {Cayao}}]{fukaya2025superconducting}%
  \BibitemOpen
  \bibfield  {author} {\bibinfo {author} {\bibfnamefont {Yuri}\ \bibnamefont
  {Fukaya}}, \bibinfo {author} {\bibfnamefont {Bo}~\bibnamefont {Lu}}, \bibinfo
  {author} {\bibfnamefont {Keiji}\ \bibnamefont {Yada}}, \bibinfo {author}
  {\bibfnamefont {Yukio}\ \bibnamefont {Tanaka}}, \ and\ \bibinfo {author}
  {\bibfnamefont {Jorge}\ \bibnamefont {Cayao}},\ }\bibfield  {title} {\enquote
  {\bibinfo {title} {Superconducting phenomena in systems with unconventional
  magnets},}\ }\href {\doibase 10.1088/1361-648X/adf1cf} {\bibfield  {journal}
  {\bibinfo  {journal} {Journal of Physics: Condensed Matter}\ }\textbf
  {\bibinfo {volume} {37}},\ \bibinfo {pages} {313003} (\bibinfo {year}
  {2025}{\natexlab{b}})}\BibitemShut {NoStop}%
\bibitem [{\citenamefont {Maeda}\ \emph {et~al.}(2025)\citenamefont {Maeda},
  \citenamefont {Fukaya}, \citenamefont {Yada}, \citenamefont {Lu},
  \citenamefont {Tanaka},\ and\ \citenamefont
  {Cayao}}]{maeda2025classification}%
  \BibitemOpen
  \bibfield  {author} {\bibinfo {author} {\bibfnamefont {Kazuki}\ \bibnamefont
  {Maeda}}, \bibinfo {author} {\bibfnamefont {Yuri}\ \bibnamefont {Fukaya}},
  \bibinfo {author} {\bibfnamefont {Keiji}\ \bibnamefont {Yada}}, \bibinfo
  {author} {\bibfnamefont {Bo}~\bibnamefont {Lu}}, \bibinfo {author}
  {\bibfnamefont {Yukio}\ \bibnamefont {Tanaka}}, \ and\ \bibinfo {author}
  {\bibfnamefont {Jorge}\ \bibnamefont {Cayao}},\ }\bibfield  {title} {\enquote
  {\bibinfo {title} {Classification of pair symmetries in superconductors with
  unconventional magnetism},}\ }\href {\doibase 10.1103/PhysRevB.111.144508}
  {\bibfield  {journal} {\bibinfo  {journal} {Phys. Rev. B}\ }\textbf {\bibinfo
  {volume} {111}},\ \bibinfo {pages} {144508} (\bibinfo {year}
  {2025})}\BibitemShut {NoStop}%
\bibitem [{\citenamefont {Fu}\ \emph {et~al.}(2025)\citenamefont {Fu},
  \citenamefont {Lv}, \citenamefont {Xu}, \citenamefont {Cayao}, \citenamefont
  {Liu},\ and\ \citenamefont {Yu}}]{fu2025all}%
  \BibitemOpen
  \bibfield  {author} {\bibinfo {author} {\bibfnamefont {Pei-Hao}\ \bibnamefont
  {Fu}}, \bibinfo {author} {\bibfnamefont {Qianqian}\ \bibnamefont {Lv}},
  \bibinfo {author} {\bibfnamefont {Yong}\ \bibnamefont {Xu}}, \bibinfo
  {author} {\bibfnamefont {Jorge}\ \bibnamefont {Cayao}}, \bibinfo {author}
  {\bibfnamefont {Jun-Feng}\ \bibnamefont {Liu}}, \ and\ \bibinfo {author}
  {\bibfnamefont {Xiang-Long}\ \bibnamefont {Yu}},\ }\bibfield  {title}
  {\enquote {\bibinfo {title} {All-electrically controlled spintronics in
  altermagnetic heterostructures},}\ }\href {\doibase
  10.1038/s41535-025-00827-7} {\bibfield  {journal} {\bibinfo  {journal} {npj
  Quantum Materials}\ }\textbf {\bibinfo {volume} {10}},\ \bibinfo {pages}
  {111} (\bibinfo {year} {2025})}\BibitemShut {NoStop}%
\bibitem [{\citenamefont {Ezawa}(2024{\natexlab{b}})}]{ezawa2024detecting}%
  \BibitemOpen
  \bibfield  {author} {\bibinfo {author} {\bibfnamefont {Motohiko}\
  \bibnamefont {Ezawa}},\ }\bibfield  {title} {\enquote {\bibinfo {title}
  {{Detecting the N\'eel vector of altermagnets in heterostructures with a
  topological insulator and a crystalline valley-edge insulator}},}\ }\href
  {\doibase 10.1103/PhysRevB.109.245306} {\bibfield  {journal} {\bibinfo
  {journal} {Phys. Rev. B}\ }\textbf {\bibinfo {volume} {109}},\ \bibinfo
  {pages} {245306} (\bibinfo {year} {2024}{\natexlab{b}})}\BibitemShut
  {NoStop}%
\bibitem [{\citenamefont {Ma}\ and\ \citenamefont
  {Jia}(2024)}]{ma2024altermagnetic}%
  \BibitemOpen
  \bibfield  {author} {\bibinfo {author} {\bibfnamefont {Hai-Yang}\
  \bibnamefont {Ma}}\ and\ \bibinfo {author} {\bibfnamefont {Jin-Feng}\
  \bibnamefont {Jia}},\ }\bibfield  {title} {\enquote {\bibinfo {title}
  {Altermagnetic topological insulator and the selection rules},}\ }\href
  {\doibase 10.1103/PhysRevB.110.064426} {\bibfield  {journal} {\bibinfo
  {journal} {Phys. Rev. B}\ }\textbf {\bibinfo {volume} {110}},\ \bibinfo
  {pages} {064426} (\bibinfo {year} {2024})}\BibitemShut {NoStop}%
\bibitem [{\citenamefont {Antonenko}\ \emph {et~al.}(2025)\citenamefont
  {Antonenko}, \citenamefont {Fernandes},\ and\ \citenamefont
  {Venderbos}}]{antonenko2025mirror}%
  \BibitemOpen
  \bibfield  {author} {\bibinfo {author} {\bibfnamefont {Daniil~S.}\
  \bibnamefont {Antonenko}}, \bibinfo {author} {\bibfnamefont {Rafael~M.}\
  \bibnamefont {Fernandes}}, \ and\ \bibinfo {author} {\bibfnamefont {J\"orn
  W.~F.}\ \bibnamefont {Venderbos}},\ }\bibfield  {title} {\enquote {\bibinfo
  {title} {{Mirror Chern Bands and Weyl Nodal Loops in Altermagnets}},}\ }\href
  {\doibase 10.1103/PhysRevLett.134.096703} {\bibfield  {journal} {\bibinfo
  {journal} {Phys. Rev. Lett.}\ }\textbf {\bibinfo {volume} {134}},\ \bibinfo
  {pages} {096703} (\bibinfo {year} {2025})}\BibitemShut {NoStop}%
\bibitem [{\citenamefont {Qu}\ \emph {et~al.}(2025)\citenamefont {Qu},
  \citenamefont {Hou}, \citenamefont {Liu}, \citenamefont {Guo},\ and\
  \citenamefont {Lu}}]{qu2025altermagnetic}%
  \BibitemOpen
  \bibfield  {author} {\bibinfo {author} {\bibfnamefont {Shuai}\ \bibnamefont
  {Qu}}, \bibinfo {author} {\bibfnamefont {Xiao-Yao}\ \bibnamefont {Hou}},
  \bibinfo {author} {\bibfnamefont {Zheng-Xin}\ \bibnamefont {Liu}}, \bibinfo
  {author} {\bibfnamefont {Peng-Jie}\ \bibnamefont {Guo}}, \ and\ \bibinfo
  {author} {\bibfnamefont {Zhong-Yi}\ \bibnamefont {Lu}},\ }\bibfield  {title}
  {\enquote {\bibinfo {title} {{Altermagnetic Weyl node-network metals
  protected by spin symmetry}},}\ }\href {\doibase 10.1103/PhysRevB.111.195138}
  {\bibfield  {journal} {\bibinfo  {journal} {Phys. Rev. B}\ }\textbf {\bibinfo
  {volume} {111}},\ \bibinfo {pages} {195138} (\bibinfo {year}
  {2025})}\BibitemShut {NoStop}%
\bibitem [{\citenamefont {Parshukov}\ \emph {et~al.}(2025)\citenamefont
  {Parshukov}, \citenamefont {Wiedmann},\ and\ \citenamefont
  {Schnyder}}]{parshukov2025topological}%
  \BibitemOpen
  \bibfield  {author} {\bibinfo {author} {\bibfnamefont {Kirill}\ \bibnamefont
  {Parshukov}}, \bibinfo {author} {\bibfnamefont {Raymond}\ \bibnamefont
  {Wiedmann}}, \ and\ \bibinfo {author} {\bibfnamefont {Andreas~P.}\
  \bibnamefont {Schnyder}},\ }\bibfield  {title} {\enquote {\bibinfo {title}
  {{Topological crossings in two-dimensional altermagnets: Symmetry
  classification and topological responses}},}\ }\href {\doibase
  10.1103/PhysRevB.111.224406} {\bibfield  {journal} {\bibinfo  {journal}
  {Phys. Rev. B}\ }\textbf {\bibinfo {volume} {111}},\ \bibinfo {pages}
  {224406} (\bibinfo {year} {2025})}\BibitemShut {NoStop}%
\bibitem [{\citenamefont {Fernandes}\ \emph {et~al.}(2024)\citenamefont
  {Fernandes}, \citenamefont {de~Carvalho}, \citenamefont {Birol},\ and\
  \citenamefont {Pereira}}]{fernandes2024topological}%
  \BibitemOpen
  \bibfield  {author} {\bibinfo {author} {\bibfnamefont {Rafael~M.}\
  \bibnamefont {Fernandes}}, \bibinfo {author} {\bibfnamefont {Vanuildo~S.}\
  \bibnamefont {de~Carvalho}}, \bibinfo {author} {\bibfnamefont {Turan}\
  \bibnamefont {Birol}}, \ and\ \bibinfo {author} {\bibfnamefont {Rodrigo~G.}\
  \bibnamefont {Pereira}},\ }\bibfield  {title} {\enquote {\bibinfo {title}
  {{Topological transition from nodal to nodeless Zeeman splitting in
  altermagnets}},}\ }\href {\doibase 10.1103/PhysRevB.109.024404} {\bibfield
  {journal} {\bibinfo  {journal} {Phys. Rev. B}\ }\textbf {\bibinfo {volume}
  {109}},\ \bibinfo {pages} {024404} (\bibinfo {year} {2024})}\BibitemShut
  {NoStop}%
\bibitem [{\citenamefont {Zhuang}\ \emph {et~al.}(2025)\citenamefont {Zhuang},
  \citenamefont {Zhu}, \citenamefont {Liu}, \citenamefont {Wu},\ and\
  \citenamefont {Yan}}]{zhuang2025odd}%
  \BibitemOpen
  \bibfield  {author} {\bibinfo {author} {\bibfnamefont {Zheng-Yang}\
  \bibnamefont {Zhuang}}, \bibinfo {author} {\bibfnamefont {Di}~\bibnamefont
  {Zhu}}, \bibinfo {author} {\bibfnamefont {Dongling}\ \bibnamefont {Liu}},
  \bibinfo {author} {\bibfnamefont {Zhigang}\ \bibnamefont {Wu}}, \ and\
  \bibinfo {author} {\bibfnamefont {Zhongbo}\ \bibnamefont {Yan}},\ }\bibfield
  {title} {\enquote {\bibinfo {title} {{Odd-Parity Altermagnetism Originated
  from Orbital Orders}},}\ }\href {https://arxiv.org/abs/2508.18361} {\bibfield
   {journal} {\bibinfo  {journal} {arXiv:2508.18361}\ } (\bibinfo {year}
  {2025})}\BibitemShut {NoStop}%
\bibitem [{\citenamefont {Huang}\ \emph {et~al.}(2026)\citenamefont {Huang},
  \citenamefont {Qin}, \citenamefont {Zhan}, \citenamefont {Xu}, \citenamefont
  {Ma},\ and\ \citenamefont {Wang}}]{huang2025light}%
  \BibitemOpen
  \bibfield  {author} {\bibinfo {author} {\bibfnamefont {Shengpu}\ \bibnamefont
  {Huang}}, \bibinfo {author} {\bibfnamefont {Zheng}\ \bibnamefont {Qin}},
  \bibinfo {author} {\bibfnamefont {Fangyang}\ \bibnamefont {Zhan}}, \bibinfo
  {author} {\bibfnamefont {Dong-Hui}\ \bibnamefont {Xu}}, \bibinfo {author}
  {\bibfnamefont {Da-Shuai}\ \bibnamefont {Ma}}, \ and\ \bibinfo {author}
  {\bibfnamefont {Rui}\ \bibnamefont {Wang}},\ }\bibfield  {title} {\enquote
  {\bibinfo {title} {{Light-Induced Odd-Parity Magnetism in Conventional
  Antiferromagnetism}},}\ }\href {\doibase 10.1103/9346-9jpf} {\bibfield
  {journal} {\bibinfo  {journal} {Phys. Rev. Lett.}\ }\textbf {\bibinfo
  {volume} {136}},\ \bibinfo {pages} {126703} (\bibinfo {year}
  {2026})}\BibitemShut {NoStop}%
\bibitem [{\citenamefont {Zhu}\ \emph {et~al.}(2026{\natexlab{b}})\citenamefont
  {Zhu}, \citenamefont {Zhou}, \citenamefont {Wang}, \citenamefont {Wei},\ and\
  \citenamefont {Ruan}}]{zhu2025floquet}%
  \BibitemOpen
  \bibfield  {author} {\bibinfo {author} {\bibfnamefont {Tongshuai}\
  \bibnamefont {Zhu}}, \bibinfo {author} {\bibfnamefont {Di}~\bibnamefont
  {Zhou}}, \bibinfo {author} {\bibfnamefont {Huaiqiang}\ \bibnamefont {Wang}},
  \bibinfo {author} {\bibfnamefont {Su-Huai}\ \bibnamefont {Wei}}, \ and\
  \bibinfo {author} {\bibfnamefont {Jiawei}\ \bibnamefont {Ruan}},\ }\bibfield
  {title} {\enquote {\bibinfo {title} {{Floquet Odd-Parity Collinear
  Magnets}},}\ }\href {\doibase 10.1103/7ywb-ml2q} {\bibfield  {journal}
  {\bibinfo  {journal} {Phys. Rev. Lett.}\ }\textbf {\bibinfo {volume} {136}},\
  \bibinfo {pages} {126704} (\bibinfo {year} {2026}{\natexlab{b}})}\BibitemShut
  {NoStop}%
\bibitem [{\citenamefont {Liu}\ \emph {et~al.}(2026)\citenamefont {Liu},
  \citenamefont {Zhuang}, \citenamefont {Zhu}, \citenamefont {Wu},\ and\
  \citenamefont {Yan}}]{liu2026light}%
  \BibitemOpen
  \bibfield  {author} {\bibinfo {author} {\bibfnamefont {Dongling}\
  \bibnamefont {Liu}}, \bibinfo {author} {\bibfnamefont {Zheng-Yang}\
  \bibnamefont {Zhuang}}, \bibinfo {author} {\bibfnamefont {Di}~\bibnamefont
  {Zhu}}, \bibinfo {author} {\bibfnamefont {Zhigang}\ \bibnamefont {Wu}}, \
  and\ \bibinfo {author} {\bibfnamefont {Zhongbo}\ \bibnamefont {Yan}},\
  }\bibfield  {title} {\enquote {\bibinfo {title} {Light-induced odd-parity
  altermagnets on dimerized lattices},}\ }\href {\doibase 10.1103/wnqs-3djt}
  {\bibfield  {journal} {\bibinfo  {journal} {Phys. Rev. B}\ }\textbf {\bibinfo
  {volume} {113}},\ \bibinfo {pages} {L060409} (\bibinfo {year}
  {2026})}\BibitemShut {NoStop}%
\bibitem [{\citenamefont {Zhu}\ \emph {et~al.}(2026{\natexlab{c}})\citenamefont
  {Zhu}, \citenamefont {Liu}, \citenamefont {Zhuang}, \citenamefont {Wu},\ and\
  \citenamefont {Yan}}]{zhu2026light}%
  \BibitemOpen
  \bibfield  {author} {\bibinfo {author} {\bibfnamefont {Di}~\bibnamefont
  {Zhu}}, \bibinfo {author} {\bibfnamefont {Dongling}\ \bibnamefont {Liu}},
  \bibinfo {author} {\bibfnamefont {Zheng-Yang}\ \bibnamefont {Zhuang}},
  \bibinfo {author} {\bibfnamefont {Zhigang}\ \bibnamefont {Wu}}, \ and\
  \bibinfo {author} {\bibfnamefont {Zhongbo}\ \bibnamefont {Yan}},\ }\bibfield
  {title} {\enquote {\bibinfo {title} {{Light-Induced Even-Parity
  Unidirectional Spin Splitting in Coplanar Antiferromagnets}},}\ }\href
  {https://doi.org/10.48550/arXiv.2601.03358} {\bibfield  {journal} {\bibinfo
  {journal} {arXiv:2601.03358}\ } (\bibinfo {year}
  {2026}{\natexlab{c}})}\BibitemShut {NoStop}%
\bibitem [{\citenamefont {Zhu}\ \emph {et~al.}(2025)\citenamefont {Zhu},
  \citenamefont {Liu}, \citenamefont {Chen},\ and\ \citenamefont
  {Zhou}}]{zhu2025floquet_Chen}%
  \BibitemOpen
  \bibfield  {author} {\bibinfo {author} {\bibfnamefont {Jiong-Yi}\
  \bibnamefont {Zhu}}, \bibinfo {author} {\bibfnamefont {Zheng-Rong}\
  \bibnamefont {Liu}}, \bibinfo {author} {\bibfnamefont {Rui}\ \bibnamefont
  {Chen}}, \ and\ \bibinfo {author} {\bibfnamefont {Bin}\ \bibnamefont
  {Zhou}},\ }\bibfield  {title} {\enquote {\bibinfo {title} {Floquet-induced
  two-dimensional weak topological insulator phase},}\ }\href {\doibase
  10.1103/kqly-w5d3} {\bibfield  {journal} {\bibinfo  {journal} {Phys. Rev. B}\
  }\textbf {\bibinfo {volume} {112}},\ \bibinfo {pages} {115436} (\bibinfo
  {year} {2025})}\BibitemShut {NoStop}%
\bibitem [{\citenamefont {Ganguli}\ \emph {et~al.}(2026)\citenamefont
  {Ganguli}, \citenamefont {Jana},\ and\ \citenamefont
  {Narayan}}]{ganguli2026tunable}%
  \BibitemOpen
  \bibfield  {author} {\bibinfo {author} {\bibfnamefont {Maitri}\ \bibnamefont
  {Ganguli}}, \bibinfo {author} {\bibfnamefont {Aneek}\ \bibnamefont {Jana}}, \
  and\ \bibinfo {author} {\bibfnamefont {Awadhesh}\ \bibnamefont {Narayan}},\
  }\bibfield  {title} {\enquote {\bibinfo {title} {{Tunable topology, Hall
  response, and spin textures in bicircularly polarized light illuminated
  altermagnets}},}\ }\href {\doibase 10.1103/6jv2-nmnn} {\bibfield  {journal}
  {\bibinfo  {journal} {Phys. Rev. B}\ }\textbf {\bibinfo {volume} {114}},\
  \bibinfo {pages} {065411} (\bibinfo {year} {2026})}\BibitemShut {NoStop}%
\bibitem [{\citenamefont {Ghorashi}\ and\ \citenamefont
  {Li}(2025)}]{ghorashi2025dynamical}%
  \BibitemOpen
  \bibfield  {author} {\bibinfo {author} {\bibfnamefont {Sayed Ali~Akbar}\
  \bibnamefont {Ghorashi}}\ and\ \bibinfo {author} {\bibfnamefont {Qiang}\
  \bibnamefont {Li}},\ }\bibfield  {title} {\enquote {\bibinfo {title}
  {{Dynamical Generation of Higher-Order Spin-Orbit Coupling, Topology, and
  Persistent Spin Texture in Light-Irradiated Altermagnets}},}\ }\href
  {\doibase 10.1103/tm58-lbdl} {\bibfield  {journal} {\bibinfo  {journal}
  {Phys. Rev. Lett.}\ }\textbf {\bibinfo {volume} {135}},\ \bibinfo {pages}
  {236702} (\bibinfo {year} {2025})}\BibitemShut {NoStop}%
\bibitem [{\citenamefont {Cheraghchi}(2026)}]{cheraghchi2026floquet}%
  \BibitemOpen
  \bibfield  {author} {\bibinfo {author} {\bibfnamefont {Hosein}\ \bibnamefont
  {Cheraghchi}},\ }\bibfield  {title} {\enquote {\bibinfo {title}
  {{Floquet-engineered Chern insulator in two-dimensional
  ${d}_{{x}^{2}\ensuremath{-}{y}^{2}}$-wave altermagnets}},}\ }\href {\doibase
  10.1103/mslj-zz6t} {\bibfield  {journal} {\bibinfo  {journal} {Phys. Rev. B}\
  }\textbf {\bibinfo {volume} {113}},\ \bibinfo {pages} {155439} (\bibinfo
  {year} {2026})}\BibitemShut {NoStop}%
\bibitem [{\citenamefont {Yarmohammadi}\ \emph {et~al.}(2026)\citenamefont
  {Yarmohammadi}, \citenamefont {Gunnink}, \citenamefont {Sinova},\ and\
  \citenamefont {Freericks}}]{yarmohammadi2026floquet}%
  \BibitemOpen
  \bibfield  {author} {\bibinfo {author} {\bibfnamefont {Mohsen}\ \bibnamefont
  {Yarmohammadi}}, \bibinfo {author} {\bibfnamefont {Pieter~M}\ \bibnamefont
  {Gunnink}}, \bibinfo {author} {\bibfnamefont {Jairo}\ \bibnamefont {Sinova}},
  \ and\ \bibinfo {author} {\bibfnamefont {James~K}\ \bibnamefont
  {Freericks}},\ }\bibfield  {title} {\enquote {\bibinfo {title}
  {{Floquet-induced anisotropic magnetoresistance and anomalous Hall effect in
  2D $d$-wave altermagnets with Rashba spin-orbit coupling}},}\ }\href
  {https://doi.org/10.48550/arXiv.2606.21281} {\bibfield  {journal} {\bibinfo
  {journal} {arXiv:2606.21281}\ } (\bibinfo {year} {2026})}\BibitemShut
  {NoStop}%
\bibitem [{\citenamefont {Chang}\ and\ \citenamefont
  {Niu}(1995)}]{chang1995berry}%
  \BibitemOpen
  \bibfield  {author} {\bibinfo {author} {\bibfnamefont {Ming-Che}\
  \bibnamefont {Chang}}\ and\ \bibinfo {author} {\bibfnamefont {Qian}\
  \bibnamefont {Niu}},\ }\bibfield  {title} {\enquote {\bibinfo {title} {{Berry
  Phase, Hyperorbits, and the Hofstadter Spectrum}},}\ }\href {\doibase
  10.1103/PhysRevLett.75.1348} {\bibfield  {journal} {\bibinfo  {journal}
  {Phys. Rev. Lett.}\ }\textbf {\bibinfo {volume} {75}},\ \bibinfo {pages}
  {1348} (\bibinfo {year} {1995})}\BibitemShut {NoStop}%
\bibitem [{\citenamefont {Sundaram}\ and\ \citenamefont
  {Niu}(1999)}]{sundaram1999wave}%
  \BibitemOpen
  \bibfield  {author} {\bibinfo {author} {\bibfnamefont {Ganesh}\ \bibnamefont
  {Sundaram}}\ and\ \bibinfo {author} {\bibfnamefont {Qian}\ \bibnamefont
  {Niu}},\ }\bibfield  {title} {\enquote {\bibinfo {title} {{Wave-packet
  dynamics in slowly perturbed crystals: Gradient corrections and Berry-phase
  effects}},}\ }\href {\doibase 10.1103/PhysRevB.59.14915} {\bibfield
  {journal} {\bibinfo  {journal} {Phys. Rev. B}\ }\textbf {\bibinfo {volume}
  {59}},\ \bibinfo {pages} {14915} (\bibinfo {year} {1999})}\BibitemShut
  {NoStop}%
\bibitem [{\citenamefont {Xiao}\ \emph {et~al.}(2010)\citenamefont {Xiao},
  \citenamefont {Chang},\ and\ \citenamefont {Niu}}]{xiao2010berry}%
  \BibitemOpen
  \bibfield  {author} {\bibinfo {author} {\bibfnamefont {Di}~\bibnamefont
  {Xiao}}, \bibinfo {author} {\bibfnamefont {Ming-Che}\ \bibnamefont {Chang}},
  \ and\ \bibinfo {author} {\bibfnamefont {Qian}\ \bibnamefont {Niu}},\
  }\bibfield  {title} {\enquote {\bibinfo {title} {Berry phase effects on
  electronic properties},}\ }\href {\doibase 10.1103/RevModPhys.82.1959}
  {\bibfield  {journal} {\bibinfo  {journal} {Rev. Mod. Phys.}\ }\textbf
  {\bibinfo {volume} {82}},\ \bibinfo {pages} {1959} (\bibinfo {year}
  {2010})}\BibitemShut {NoStop}%
\bibitem [{\citenamefont {Shen}(2017)}]{shen2017topological}%
  \BibitemOpen
  \bibfield  {author} {\bibinfo {author} {\bibfnamefont {Shun-Qing}\
  \bibnamefont {Shen}},\ }\href@noop {} {\emph {\bibinfo {title} {Topological
  insulators}}}\ (\bibinfo  {publisher} {Springer, Singapore},\ \bibinfo {year}
  {2017})\BibitemShut {NoStop}%
\bibitem [{\citenamefont {Qin}\ \emph {et~al.}(2024{\natexlab{a}})\citenamefont
  {Qin}, \citenamefont {Chen},\ and\ \citenamefont {Lee}}]{qin2024light}%
  \BibitemOpen
  \bibfield  {author} {\bibinfo {author} {\bibfnamefont {Fang}\ \bibnamefont
  {Qin}}, \bibinfo {author} {\bibfnamefont {Rui}\ \bibnamefont {Chen}}, \ and\
  \bibinfo {author} {\bibfnamefont {Ching~Hua}\ \bibnamefont {Lee}},\
  }\bibfield  {title} {\enquote {\bibinfo {title} {{Light-enhanced nonlinear
  Hall effect}},}\ }\href {\doibase 10.1038/s42005-024-01820-5} {\bibfield
  {journal} {\bibinfo  {journal} {Communications Physics}\ }\textbf {\bibinfo
  {volume} {7}},\ \bibinfo {pages} {368} (\bibinfo {year}
  {2024}{\natexlab{a}})}\BibitemShut {NoStop}%
\bibitem [{\citenamefont {Li}\ \emph {et~al.}(2016)\citenamefont {Li},
  \citenamefont {He}, \citenamefont {Lu}, \citenamefont {Zhang}, \citenamefont
  {Liu}, \citenamefont {Ma}, \citenamefont {Fan}, \citenamefont {Shen},\ and\
  \citenamefont {Wang}}]{li2016negative}%
  \BibitemOpen
  \bibfield  {author} {\bibinfo {author} {\bibfnamefont {Hui}\ \bibnamefont
  {Li}}, \bibinfo {author} {\bibfnamefont {Hongtao}\ \bibnamefont {He}},
  \bibinfo {author} {\bibfnamefont {Hai-Zhou}\ \bibnamefont {Lu}}, \bibinfo
  {author} {\bibfnamefont {Huachen}\ \bibnamefont {Zhang}}, \bibinfo {author}
  {\bibfnamefont {Hongchao}\ \bibnamefont {Liu}}, \bibinfo {author}
  {\bibfnamefont {Rong}\ \bibnamefont {Ma}}, \bibinfo {author} {\bibfnamefont
  {Zhiyong}\ \bibnamefont {Fan}}, \bibinfo {author} {\bibfnamefont {Shun-Qing}\
  \bibnamefont {Shen}}, \ and\ \bibinfo {author} {\bibfnamefont {Jiannong}\
  \bibnamefont {Wang}},\ }\bibfield  {title} {\enquote {\bibinfo {title}
  {{Negative magnetoresistance in Dirac semimetal Cd$_3$As$_2$}},}\ }\href
  {\doibase 10.1038/ncomms10301} {\bibfield  {journal} {\bibinfo  {journal}
  {Nature communications}\ }\textbf {\bibinfo {volume} {7}},\ \bibinfo {pages}
  {10301} (\bibinfo {year} {2016})}\BibitemShut {NoStop}%
\bibitem [{\citenamefont {Lu}\ and\ \citenamefont
  {Shen}(2017)}]{lu2017quantum}%
  \BibitemOpen
  \bibfield  {author} {\bibinfo {author} {\bibfnamefont {Hai-Zhou}\
  \bibnamefont {Lu}}\ and\ \bibinfo {author} {\bibfnamefont {Shun-Qing}\
  \bibnamefont {Shen}},\ }\bibfield  {title} {\enquote {\bibinfo {title}
  {Quantum transport in topological semimetals under magnetic fields},}\ }\href
  {\doibase 10.1007/s11467-016-0609-y} {\bibfield  {journal} {\bibinfo
  {journal} {Frontiers of Physics}\ }\textbf {\bibinfo {volume} {12}},\
  \bibinfo {pages} {127201} (\bibinfo {year} {2017})}\BibitemShut {NoStop}%
\bibitem [{\citenamefont {Qin}\ and\ \citenamefont
  {Chen}(2025)}]{qin2025emergent}%
  \BibitemOpen
  \bibfield  {author} {\bibinfo {author} {\bibfnamefont {Fang}\ \bibnamefont
  {Qin}}\ and\ \bibinfo {author} {\bibfnamefont {Rui}\ \bibnamefont {Chen}},\
  }\bibfield  {title} {\enquote {\bibinfo {title} {{Emergent Weyl-like points
  in periodically modulated systems}},}\ }\href {\doibase 10.1103/g25n-phmj}
  {\bibfield  {journal} {\bibinfo  {journal} {Phys. Rev. B}\ }\textbf {\bibinfo
  {volume} {112}},\ \bibinfo {pages} {115432} (\bibinfo {year}
  {2025})}\BibitemShut {NoStop}%
\bibitem [{\citenamefont {Qin}\ \emph {et~al.}(2025)\citenamefont {Qin},
  \citenamefont {Shen},\ and\ \citenamefont {Lee}}]{qin2025nonlinear}%
  \BibitemOpen
  \bibfield  {author} {\bibinfo {author} {\bibfnamefont {Fang}\ \bibnamefont
  {Qin}}, \bibinfo {author} {\bibfnamefont {Ruizhe}\ \bibnamefont {Shen}}, \
  and\ \bibinfo {author} {\bibfnamefont {Ching~Hua}\ \bibnamefont {Lee}},\
  }\bibfield  {title} {\enquote {\bibinfo {title} {{Nonlinear Hall effects with
  an exceptional ring}},}\ }\href {\doibase 10.1103/8g3q-qrpg} {\bibfield
  {journal} {\bibinfo  {journal} {Phys. Rev. B}\ }\textbf {\bibinfo {volume}
  {111}},\ \bibinfo {pages} {245413} (\bibinfo {year} {2025})}\BibitemShut
  {NoStop}%
\bibitem [{\citenamefont {Qin}\ \emph {et~al.}(2014)\citenamefont {Qin},
  \citenamefont {Wen},\ and\ \citenamefont {Chen}}]{qin2014thermal}%
  \BibitemOpen
  \bibfield  {author} {\bibinfo {author} {\bibfnamefont {Fang}\ \bibnamefont
  {Qin}}, \bibinfo {author} {\bibfnamefont {Wen}\ \bibnamefont {Wen}}, \ and\
  \bibinfo {author} {\bibfnamefont {Ji-Sheng}\ \bibnamefont {Chen}},\
  }\bibfield  {title} {\enquote {\bibinfo {title} {Thermal and electrical
  conductivities of a three-dimensional ideal anyon gas with fractional
  exclusion statistics},}\ }\href {\doibase 10.1088/0253-6102/62/1/14}
  {\bibfield  {journal} {\bibinfo  {journal} {Communications in Theoretical
  Physics}\ }\textbf {\bibinfo {volume} {62}},\ \bibinfo {pages} {81} (\bibinfo
  {year} {2014})}\BibitemShut {NoStop}%
\bibitem [{Sup()}]{SuppMat}%
  \BibitemOpen
  \href@noop {} {\ }\bibinfo {note} {See Supplemental Material at
  {\color{blue}http://link.aps.org/supplemental/10.1103/dqs3-935d} for the
  derivations of the anomalous thermoelectric and thermal current densities
  from the Boltzmann equation, including
  Refs.~\cite{chang1995berry,sundaram1999wave,xiao2010berry,shen2017topological,qiang2023topological,qin2024light,qin2025nonlinear,qin2014thermal,li2016negative,lu2017quantum,qin2025emergent};
  the derivations of the transport coefficients from intrinsic contributions,
  including
  Refs.~\cite{xiao2006berry,bergman2010theory,yokoyama2011transverse,pathria1996statistical};
  the derivations of the low-temperature limits of the transport coefficients,
  including
  Refs.~\cite{qin2015three,qin2016universal,qin2018high,qin2009comparative,qin2010finite,qin2011joule,qin2012adiabatic,qin2012pauli};
  the derivations of the Floquet Hamiltonian; the derivations of the
  tight-binding model; and the derivations of the Berry curvature, including
  Refs.~\cite{thouless1982quantized,qin2023light,qin2024kinked}.}\BibitemShut
  {Stop}%
\bibitem [{\citenamefont {Qin}\ \emph {et~al.}(2015)\citenamefont {Qin},
  \citenamefont {Wu}, \citenamefont {Zhang}, \citenamefont {Yi},\ and\
  \citenamefont {Guo}}]{qin2015three}%
  \BibitemOpen
  \bibfield  {author} {\bibinfo {author} {\bibfnamefont {Fang}\ \bibnamefont
  {Qin}}, \bibinfo {author} {\bibfnamefont {Fan}\ \bibnamefont {Wu}}, \bibinfo
  {author} {\bibfnamefont {Wei}\ \bibnamefont {Zhang}}, \bibinfo {author}
  {\bibfnamefont {Wei}\ \bibnamefont {Yi}}, \ and\ \bibinfo {author}
  {\bibfnamefont {Guang-Can}\ \bibnamefont {Guo}},\ }\bibfield  {title}
  {\enquote {\bibinfo {title} {{Three-component Fulde-Ferrell superfluids in a
  two-dimensional Fermi gas with spin-orbit coupling}},}\ }\href {\doibase
  10.1103/PhysRevA.92.023604} {\bibfield  {journal} {\bibinfo  {journal} {Phys.
  Rev. A}\ }\textbf {\bibinfo {volume} {92}},\ \bibinfo {pages} {023604}
  (\bibinfo {year} {2015})}\BibitemShut {NoStop}%
\bibitem [{\citenamefont {Qin}\ \emph {et~al.}(2016)\citenamefont {Qin},
  \citenamefont {Cui},\ and\ \citenamefont {Yi}}]{qin2016universal}%
  \BibitemOpen
  \bibfield  {author} {\bibinfo {author} {\bibfnamefont {Fang}\ \bibnamefont
  {Qin}}, \bibinfo {author} {\bibfnamefont {Xiaoling}\ \bibnamefont {Cui}}, \
  and\ \bibinfo {author} {\bibfnamefont {Wei}\ \bibnamefont {Yi}},\ }\bibfield
  {title} {\enquote {\bibinfo {title} {{Universal relations and normal phase of
  an ultracold Fermi gas with coexisting $s$- and $p$-wave interactions}},}\
  }\href {\doibase 10.1103/PhysRevA.94.063616} {\bibfield  {journal} {\bibinfo
  {journal} {Phys. Rev. A}\ }\textbf {\bibinfo {volume} {94}},\ \bibinfo
  {pages} {063616} (\bibinfo {year} {2016})}\BibitemShut {NoStop}%
\bibitem [{\citenamefont {Qin}\ \emph {et~al.}(2018)\citenamefont {Qin},
  \citenamefont {Jie}, \citenamefont {Yi},\ and\ \citenamefont
  {Guo}}]{qin2018high}%
  \BibitemOpen
  \bibfield  {author} {\bibinfo {author} {\bibfnamefont {Fang}\ \bibnamefont
  {Qin}}, \bibinfo {author} {\bibfnamefont {Jianwen}\ \bibnamefont {Jie}},
  \bibinfo {author} {\bibfnamefont {Wei}\ \bibnamefont {Yi}}, \ and\ \bibinfo
  {author} {\bibfnamefont {Guang-Can}\ \bibnamefont {Guo}},\ }\bibfield
  {title} {\enquote {\bibinfo {title} {{High-momentum tail and universal
  relations of a Fermi gas near a Raman-dressed Feshbach resonance}},}\ }\href
  {\doibase 10.1103/PhysRevA.97.033610} {\bibfield  {journal} {\bibinfo
  {journal} {Phys. Rev. A}\ }\textbf {\bibinfo {volume} {97}},\ \bibinfo
  {pages} {033610} (\bibinfo {year} {2018})}\BibitemShut {NoStop}%
\bibitem [{\citenamefont {Pathria}(1996)}]{pathria1996statistical}%
  \BibitemOpen
  \bibfield  {author} {\bibinfo {author} {\bibfnamefont {R.~K.}\ \bibnamefont
  {Pathria}},\ }\href@noop {} {\emph {\bibinfo {title} {{Statistical
  Mechanics}}}}\ (\bibinfo  {publisher} {Butterworth-Heinemann, Oxford},\
  \bibinfo {year} {1996})\BibitemShut {NoStop}%
\bibitem [{\citenamefont {Qin}\ and\ \citenamefont
  {Chen}(2009)}]{qin2009comparative}%
  \BibitemOpen
  \bibfield  {author} {\bibinfo {author} {\bibfnamefont {Fang}\ \bibnamefont
  {Qin}}\ and\ \bibinfo {author} {\bibfnamefont {Ji-sheng}\ \bibnamefont
  {Chen}},\ }\bibfield  {title} {\enquote {\bibinfo {title} {Comparative study
  of the finite-temperature thermodynamics of a unitary fermi gas},}\ }\href
  {\doibase 10.1103/PhysRevA.79.043625} {\bibfield  {journal} {\bibinfo
  {journal} {Phys. Rev. A}\ }\textbf {\bibinfo {volume} {79}},\ \bibinfo
  {pages} {043625} (\bibinfo {year} {2009})}\BibitemShut {NoStop}%
\bibitem [{\citenamefont {Qin}\ and\ \citenamefont
  {Chen}(2010)}]{qin2010finite}%
  \BibitemOpen
  \bibfield  {author} {\bibinfo {author} {\bibfnamefont {Fang}\ \bibnamefont
  {Qin}}\ and\ \bibinfo {author} {\bibfnamefont {Ji-sheng}\ \bibnamefont
  {Chen}},\ }\bibfield  {title} {\enquote {\bibinfo {title} {The
  finite-temperature thermodynamics of a trapped unitary fermi gas within
  fractional exclusion statistics},}\ }\href {\doibase
  10.1088/0953-4075/43/5/055302} {\bibfield  {journal} {\bibinfo  {journal}
  {Journal of Physics B: Atomic, Molecular and Optical Physics}\ }\textbf
  {\bibinfo {volume} {43}},\ \bibinfo {pages} {055302} (\bibinfo {year}
  {2010})}\BibitemShut {NoStop}%
\bibitem [{\citenamefont {Qin}\ and\ \citenamefont
  {Chen}(2011)}]{qin2011joule}%
  \BibitemOpen
  \bibfield  {author} {\bibinfo {author} {\bibfnamefont {Fang}\ \bibnamefont
  {Qin}}\ and\ \bibinfo {author} {\bibfnamefont {Ji-sheng}\ \bibnamefont
  {Chen}},\ }\bibfield  {title} {\enquote {\bibinfo {title} {{Joule-Thomson
  coefficient of ideal anyons within fractional exclusion statistics}},}\
  }\href {\doibase 10.1103/PhysRevE.83.021111} {\bibfield  {journal} {\bibinfo
  {journal} {Phys. Rev. E}\ }\textbf {\bibinfo {volume} {83}},\ \bibinfo
  {pages} {021111} (\bibinfo {year} {2011})}\BibitemShut {NoStop}%
\bibitem [{\citenamefont {Qin}\ and\ \citenamefont
  {Chen}(2012{\natexlab{a}})}]{qin2012adiabatic}%
  \BibitemOpen
  \bibfield  {author} {\bibinfo {author} {\bibfnamefont {Fang}\ \bibnamefont
  {Qin}}\ and\ \bibinfo {author} {\bibfnamefont {Ji-sheng}\ \bibnamefont
  {Chen}},\ }\bibfield  {title} {\enquote {\bibinfo {title} {Adiabatic sound
  velocity and compressibility of a trapped $d$-dimensional ideal anyon gas},}\
  }\href {\doibase 10.1016/j.physleta.2012.02.034} {\bibfield  {journal}
  {\bibinfo  {journal} {Physics Letters A}\ }\textbf {\bibinfo {volume}
  {376}},\ \bibinfo {pages} {1191} (\bibinfo {year}
  {2012}{\natexlab{a}})}\BibitemShut {NoStop}%
\bibitem [{\citenamefont {Qin}\ and\ \citenamefont
  {Chen}(2012{\natexlab{b}})}]{qin2012pauli}%
  \BibitemOpen
  \bibfield  {author} {\bibinfo {author} {\bibfnamefont {Fang}\ \bibnamefont
  {Qin}}\ and\ \bibinfo {author} {\bibfnamefont {Ji-Sheng}\ \bibnamefont
  {Chen}},\ }\bibfield  {title} {\enquote {\bibinfo {title} {Pauli paramagnetic
  susceptibility of an ideal anyon gas within haldane fractional exclusion
  statistics},}\ }\href {\doibase 10.1088/0253-6102/58/4/22} {\bibfield
  {journal} {\bibinfo  {journal} {Communications in Theoretical Physics}\
  }\textbf {\bibinfo {volume} {58}},\ \bibinfo {pages} {573} (\bibinfo {year}
  {2012}{\natexlab{b}})}\BibitemShut {NoStop}%
\bibitem [{\citenamefont {Magnus}(1954)}]{magnus1954exponential}%
  \BibitemOpen
  \bibfield  {author} {\bibinfo {author} {\bibfnamefont {Wilhelm}\ \bibnamefont
  {Magnus}},\ }\bibfield  {title} {\enquote {\bibinfo {title} {On the
  exponential solution of differential equations for a linear operator},}\
  }\href {\doibase 10.1002/cpa.3160070404} {\bibfield  {journal} {\bibinfo
  {journal} {Communications on pure and applied mathematics}\ }\textbf
  {\bibinfo {volume} {7}},\ \bibinfo {pages} {649} (\bibinfo {year}
  {1954})}\BibitemShut {NoStop}%
\bibitem [{\citenamefont {Blanes}\ \emph {et~al.}(2009)\citenamefont {Blanes},
  \citenamefont {Casas}, \citenamefont {Oteo},\ and\ \citenamefont
  {Ros}}]{blanes2009magnus}%
  \BibitemOpen
  \bibfield  {author} {\bibinfo {author} {\bibfnamefont {Sergio}\ \bibnamefont
  {Blanes}}, \bibinfo {author} {\bibfnamefont {Fernando}\ \bibnamefont
  {Casas}}, \bibinfo {author} {\bibfnamefont {Jose-Angel}\ \bibnamefont
  {Oteo}}, \ and\ \bibinfo {author} {\bibfnamefont {Jos{\'e}}\ \bibnamefont
  {Ros}},\ }\bibfield  {title} {\enquote {\bibinfo {title} {The magnus
  expansion and some of its applications},}\ }\href {\doibase
  10.1016/j.physrep.2008.11.001} {\bibfield  {journal} {\bibinfo  {journal}
  {Physics reports}\ }\textbf {\bibinfo {volume} {470}},\ \bibinfo {pages}
  {151} (\bibinfo {year} {2009})}\BibitemShut {NoStop}%
\bibitem [{\citenamefont {Kuwahara}\ \emph {et~al.}(2016)\citenamefont
  {Kuwahara}, \citenamefont {Mori},\ and\ \citenamefont
  {Saito}}]{kuwahara2016floquet}%
  \BibitemOpen
  \bibfield  {author} {\bibinfo {author} {\bibfnamefont {Tomotaka}\
  \bibnamefont {Kuwahara}}, \bibinfo {author} {\bibfnamefont {Takashi}\
  \bibnamefont {Mori}}, \ and\ \bibinfo {author} {\bibfnamefont {Keiji}\
  \bibnamefont {Saito}},\ }\bibfield  {title} {\enquote {\bibinfo {title}
  {{Floquet--Magnus theory and generic transient dynamics in periodically
  driven many-body quantum systems}},}\ }\href {\doibase
  10.1016/j.aop.2016.01.012} {\bibfield  {journal} {\bibinfo  {journal} {Annals
  of Physics}\ }\textbf {\bibinfo {volume} {367}},\ \bibinfo {pages} {96}
  (\bibinfo {year} {2016})}\BibitemShut {NoStop}%
\bibitem [{\citenamefont {Qin}\ \emph {et~al.}(2022{\natexlab{a}})\citenamefont
  {Qin}, \citenamefont {Chen},\ and\ \citenamefont {Lu}}]{qin2022phase}%
  \BibitemOpen
  \bibfield  {author} {\bibinfo {author} {\bibfnamefont {Fang}\ \bibnamefont
  {Qin}}, \bibinfo {author} {\bibfnamefont {Rui}\ \bibnamefont {Chen}}, \ and\
  \bibinfo {author} {\bibfnamefont {Hai-Zhou}\ \bibnamefont {Lu}},\ }\bibfield
  {title} {\enquote {\bibinfo {title} {Phase transitions in intrinsic magnetic
  topological insulator with high-frequency pumping},}\ }\href {\doibase
  10.1088/1361-648X/ac530f} {\bibfield  {journal} {\bibinfo  {journal} {Journal
  of Physics: Condensed Matter}\ }\textbf {\bibinfo {volume} {34}},\ \bibinfo
  {pages} {225001} (\bibinfo {year} {2022}{\natexlab{a}})}\BibitemShut
  {NoStop}%
\bibitem [{\citenamefont {Qin}\ \emph {et~al.}(2023)\citenamefont {Qin},
  \citenamefont {Lee},\ and\ \citenamefont {Chen}}]{qin2023light}%
  \BibitemOpen
  \bibfield  {author} {\bibinfo {author} {\bibfnamefont {Fang}\ \bibnamefont
  {Qin}}, \bibinfo {author} {\bibfnamefont {Ching~Hua}\ \bibnamefont {Lee}}, \
  and\ \bibinfo {author} {\bibfnamefont {Rui}\ \bibnamefont {Chen}},\
  }\bibfield  {title} {\enquote {\bibinfo {title} {{Light-induced
  half-quantized Hall effect and axion insulator}},}\ }\href {\doibase
  10.1103/PhysRevB.108.075435} {\bibfield  {journal} {\bibinfo  {journal}
  {Phys. Rev. B}\ }\textbf {\bibinfo {volume} {108}},\ \bibinfo {pages}
  {075435} (\bibinfo {year} {2023})}\BibitemShut {NoStop}%
\bibitem [{\citenamefont {Qin}\ \emph {et~al.}(2022{\natexlab{b}})\citenamefont
  {Qin}, \citenamefont {Lee},\ and\ \citenamefont {Chen}}]{qin2022light}%
  \BibitemOpen
  \bibfield  {author} {\bibinfo {author} {\bibfnamefont {Fang}\ \bibnamefont
  {Qin}}, \bibinfo {author} {\bibfnamefont {Ching~Hua}\ \bibnamefont {Lee}}, \
  and\ \bibinfo {author} {\bibfnamefont {Rui}\ \bibnamefont {Chen}},\
  }\bibfield  {title} {\enquote {\bibinfo {title} {{Light-induced phase
  crossovers in a quantum spin Hall system}},}\ }\href {\doibase
  10.1103/PhysRevB.106.235405} {\bibfield  {journal} {\bibinfo  {journal}
  {Phys. Rev. B}\ }\textbf {\bibinfo {volume} {106}},\ \bibinfo {pages}
  {235405} (\bibinfo {year} {2022}{\natexlab{b}})}\BibitemShut {NoStop}%
\bibitem [{\citenamefont {Oka}\ and\ \citenamefont
  {Aoki}(2009)}]{oka2009photovoltaic}%
  \BibitemOpen
  \bibfield  {author} {\bibinfo {author} {\bibfnamefont {Takashi}\ \bibnamefont
  {Oka}}\ and\ \bibinfo {author} {\bibfnamefont {Hideo}\ \bibnamefont {Aoki}},\
  }\bibfield  {title} {\enquote {\bibinfo {title} {{Photovoltaic Hall effect in
  graphene}},}\ }\href {\doibase 10.1103/PhysRevB.79.081406} {\bibfield
  {journal} {\bibinfo  {journal} {Phys. Rev. B}\ }\textbf {\bibinfo {volume}
  {79}},\ \bibinfo {pages} {081406} (\bibinfo {year} {2009})}\BibitemShut
  {NoStop}%
\bibitem [{\citenamefont {Grutter}\ and\ \citenamefont
  {He}(2021)}]{grutter2021magnetic}%
  \BibitemOpen
  \bibfield  {author} {\bibinfo {author} {\bibfnamefont {A.~J.}\ \bibnamefont
  {Grutter}}\ and\ \bibinfo {author} {\bibfnamefont {Q.~L.}\ \bibnamefont
  {He}},\ }\bibfield  {title} {\enquote {\bibinfo {title} {Magnetic proximity
  effects in topological insulator heterostructures: Implementation and
  characterization},}\ }\href {\doibase 10.1103/PhysRevMaterials.5.090301}
  {\bibfield  {journal} {\bibinfo  {journal} {Phys. Rev. Mater.}\ }\textbf
  {\bibinfo {volume} {5}},\ \bibinfo {pages} {090301} (\bibinfo {year}
  {2021})}\BibitemShut {NoStop}%
\bibitem [{\citenamefont {Zhang}\ \emph {et~al.}(2009)\citenamefont {Zhang},
  \citenamefont {Liu}, \citenamefont {Qi}, \citenamefont {Dai}, \citenamefont
  {Fang},\ and\ \citenamefont {Zhang}}]{zhang2009topological}%
  \BibitemOpen
  \bibfield  {author} {\bibinfo {author} {\bibfnamefont {Haijun}\ \bibnamefont
  {Zhang}}, \bibinfo {author} {\bibfnamefont {Chao-Xing}\ \bibnamefont {Liu}},
  \bibinfo {author} {\bibfnamefont {Xiao-Liang}\ \bibnamefont {Qi}}, \bibinfo
  {author} {\bibfnamefont {Xi}~\bibnamefont {Dai}}, \bibinfo {author}
  {\bibfnamefont {Zhong}\ \bibnamefont {Fang}}, \ and\ \bibinfo {author}
  {\bibfnamefont {Shou-Cheng}\ \bibnamefont {Zhang}},\ }\bibfield  {title}
  {\enquote {\bibinfo {title} {{Topological insulators in Bi$_2$Se$_3$,
  Bi$_2$Te$_3$ and Sb$_2$Te$_3$ with a single Dirac cone on the surface}},}\
  }\href {\doibase 10.1038/nphys1270} {\bibfield  {journal} {\bibinfo
  {journal} {Nature physics}\ }\textbf {\bibinfo {volume} {5}},\ \bibinfo
  {pages} {438} (\bibinfo {year} {2009})}\BibitemShut {NoStop}%
\bibitem [{\citenamefont {Chang}\ \emph {et~al.}(2013)\citenamefont {Chang},
  \citenamefont {Zhang}, \citenamefont {Feng}, \citenamefont {Shen},
  \citenamefont {Zhang}, \citenamefont {Guo}, \citenamefont {Li}, \citenamefont
  {Ou}, \citenamefont {Wei}, \citenamefont {Wang} \emph
  {et~al.}}]{chang2013experimental}%
  \BibitemOpen
  \bibfield  {author} {\bibinfo {author} {\bibfnamefont {Cui-Zu}\ \bibnamefont
  {Chang}}, \bibinfo {author} {\bibfnamefont {Jinsong}\ \bibnamefont {Zhang}},
  \bibinfo {author} {\bibfnamefont {Xiao}\ \bibnamefont {Feng}}, \bibinfo
  {author} {\bibfnamefont {Jie}\ \bibnamefont {Shen}}, \bibinfo {author}
  {\bibfnamefont {Zuocheng}\ \bibnamefont {Zhang}}, \bibinfo {author}
  {\bibfnamefont {Minghua}\ \bibnamefont {Guo}}, \bibinfo {author}
  {\bibfnamefont {Kang}\ \bibnamefont {Li}}, \bibinfo {author} {\bibfnamefont
  {Yunbo}\ \bibnamefont {Ou}}, \bibinfo {author} {\bibfnamefont {Pang}\
  \bibnamefont {Wei}}, \bibinfo {author} {\bibfnamefont {Li-Li}\ \bibnamefont
  {Wang}},  \emph {et~al.},\ }\bibfield  {title} {\enquote {\bibinfo {title}
  {{Experimental observation of the quantum anomalous Hall effect in a magnetic
  topological insulator}},}\ }\href {\doibase 10.1126/science.1234414}
  {\bibfield  {journal} {\bibinfo  {journal} {Science}\ }\textbf {\bibinfo
  {volume} {340}},\ \bibinfo {pages} {167} (\bibinfo {year}
  {2013})}\BibitemShut {NoStop}%
\bibitem [{\citenamefont {Mogi}\ \emph {et~al.}(2022)\citenamefont {Mogi},
  \citenamefont {Okamura}, \citenamefont {Kawamura}, \citenamefont {Yoshimi},
  \citenamefont {Yasuda}, \citenamefont {Tsukazaki}, \citenamefont {Takahashi},
  \citenamefont {Morimoto}, \citenamefont {Nagaosa}, \citenamefont {Kawasaki}
  \emph {et~al.}}]{mogi2022experimental}%
  \BibitemOpen
  \bibfield  {author} {\bibinfo {author} {\bibfnamefont {M}~\bibnamefont
  {Mogi}}, \bibinfo {author} {\bibfnamefont {Y}~\bibnamefont {Okamura}},
  \bibinfo {author} {\bibfnamefont {M}~\bibnamefont {Kawamura}}, \bibinfo
  {author} {\bibfnamefont {R}~\bibnamefont {Yoshimi}}, \bibinfo {author}
  {\bibfnamefont {K}~\bibnamefont {Yasuda}}, \bibinfo {author} {\bibfnamefont
  {A}~\bibnamefont {Tsukazaki}}, \bibinfo {author} {\bibfnamefont
  {KS}~\bibnamefont {Takahashi}}, \bibinfo {author} {\bibfnamefont
  {T}~\bibnamefont {Morimoto}}, \bibinfo {author} {\bibfnamefont
  {N}~\bibnamefont {Nagaosa}}, \bibinfo {author} {\bibfnamefont
  {M}~\bibnamefont {Kawasaki}},  \emph {et~al.},\ }\bibfield  {title} {\enquote
  {\bibinfo {title} {Experimental signature of the parity anomaly in a
  semi-magnetic topological insulator},}\ }\href {\doibase
  10.1038/s41567-021-01490-y} {\bibfield  {journal} {\bibinfo  {journal}
  {Nature Physics}\ }\textbf {\bibinfo {volume} {18}},\ \bibinfo {pages} {390}
  (\bibinfo {year} {2022})}\BibitemShut {NoStop}%
\bibitem [{\citenamefont {Marfoua}\ and\ \citenamefont
  {Hong}(2025)}]{marfoua2025strain}%
  \BibitemOpen
  \bibfield  {author} {\bibinfo {author} {\bibfnamefont {Brahim}\ \bibnamefont
  {Marfoua}}\ and\ \bibinfo {author} {\bibfnamefont {Jisang}\ \bibnamefont
  {Hong}},\ }\bibfield  {title} {\enquote {\bibinfo {title} {{Strain-dependent
  Rashba effect, and spin Hall conductivity in the altermagnetic Janus
  V$_2$SeTeO monolayer}},}\ }\href {\doibase 10.1016/j.cap.2024.10.014}
  {\bibfield  {journal} {\bibinfo  {journal} {Current Applied Physics}\
  }\textbf {\bibinfo {volume} {69}},\ \bibinfo {pages} {47} (\bibinfo {year}
  {2025})}\BibitemShut {NoStop}%
\bibitem [{\citenamefont {Wang}\ \emph {et~al.}(2018)\citenamefont {Wang},
  \citenamefont {Liu}, \citenamefont {Yang},\ and\ \citenamefont
  {Liu}}]{wang2018light}%
  \BibitemOpen
  \bibfield  {author} {\bibinfo {author} {\bibfnamefont {Z.~F.}\ \bibnamefont
  {Wang}}, \bibinfo {author} {\bibfnamefont {Zhao}\ \bibnamefont {Liu}},
  \bibinfo {author} {\bibfnamefont {Jinlong}\ \bibnamefont {Yang}}, \ and\
  \bibinfo {author} {\bibfnamefont {Feng}\ \bibnamefont {Liu}},\ }\bibfield
  {title} {\enquote {\bibinfo {title} {{Light-Induced Type-II Band Inversion
  and Quantum Anomalous Hall State in Monolayer FeSe}},}\ }\href {\doibase
  10.1103/PhysRevLett.120.156406} {\bibfield  {journal} {\bibinfo  {journal}
  {Phys. Rev. Lett.}\ }\textbf {\bibinfo {volume} {120}},\ \bibinfo {pages}
  {156406} (\bibinfo {year} {2018})}\BibitemShut {NoStop}%
\bibitem [{\citenamefont {Bao}\ \emph {et~al.}(2022)\citenamefont {Bao},
  \citenamefont {Tang}, \citenamefont {Sun},\ and\ \citenamefont
  {Zhou}}]{bao2022light}%
  \BibitemOpen
  \bibfield  {author} {\bibinfo {author} {\bibfnamefont {Changhua}\
  \bibnamefont {Bao}}, \bibinfo {author} {\bibfnamefont {Peizhe}\ \bibnamefont
  {Tang}}, \bibinfo {author} {\bibfnamefont {Dong}\ \bibnamefont {Sun}}, \ and\
  \bibinfo {author} {\bibfnamefont {Shuyun}\ \bibnamefont {Zhou}},\ }\bibfield
  {title} {\enquote {\bibinfo {title} {{Light-induced emergent phenomena in 2D
  materials and topological materials}},}\ }\href {\doibase
  10.1038/s42254-021-00388-1} {\bibfield  {journal} {\bibinfo  {journal}
  {Nature Reviews Physics}\ }\textbf {\bibinfo {volume} {4}},\ \bibinfo {pages}
  {33} (\bibinfo {year} {2022})}\BibitemShut {NoStop}%
\bibitem [{\citenamefont {Sentef}\ \emph {et~al.}(2015)\citenamefont {Sentef},
  \citenamefont {Claassen}, \citenamefont {Kemper}, \citenamefont {Moritz},
  \citenamefont {Oka}, \citenamefont {Freericks},\ and\ \citenamefont
  {Devereaux}}]{sentef2015theory}%
  \BibitemOpen
  \bibfield  {author} {\bibinfo {author} {\bibfnamefont {M.~A.}\ \bibnamefont
  {Sentef}}, \bibinfo {author} {\bibfnamefont {M.}~\bibnamefont {Claassen}},
  \bibinfo {author} {\bibfnamefont {A.~F.}\ \bibnamefont {Kemper}}, \bibinfo
  {author} {\bibfnamefont {B.}~\bibnamefont {Moritz}}, \bibinfo {author}
  {\bibfnamefont {T.}~\bibnamefont {Oka}}, \bibinfo {author} {\bibfnamefont
  {J.~K.}\ \bibnamefont {Freericks}}, \ and\ \bibinfo {author} {\bibfnamefont
  {T.~P.}\ \bibnamefont {Devereaux}},\ }\bibfield  {title} {\enquote {\bibinfo
  {title} {{Theory of Floquet band formation and local pseudospin textures in
  pump-probe photoemission of graphene}},}\ }\href {\doibase
  10.1038/ncomms8047} {\bibfield  {journal} {\bibinfo  {journal} {Nature
  communications}\ }\textbf {\bibinfo {volume} {6}},\ \bibinfo {pages} {7047}
  (\bibinfo {year} {2015})}\BibitemShut {NoStop}%
\bibitem [{\citenamefont {Thouless}\ \emph {et~al.}(1982)\citenamefont
  {Thouless}, \citenamefont {Kohmoto}, \citenamefont {Nightingale},\ and\
  \citenamefont {den Nijs}}]{thouless1982quantized}%
  \BibitemOpen
  \bibfield  {author} {\bibinfo {author} {\bibfnamefont {D.~J.}\ \bibnamefont
  {Thouless}}, \bibinfo {author} {\bibfnamefont {M.}~\bibnamefont {Kohmoto}},
  \bibinfo {author} {\bibfnamefont {M.~P.}\ \bibnamefont {Nightingale}}, \ and\
  \bibinfo {author} {\bibfnamefont {M.}~\bibnamefont {den Nijs}},\ }\bibfield
  {title} {\enquote {\bibinfo {title} {{Quantized Hall Conductance in a
  Two-Dimensional Periodic Potential}},}\ }\href {\doibase
  10.1103/PhysRevLett.49.405} {\bibfield  {journal} {\bibinfo  {journal} {Phys.
  Rev. Lett.}\ }\textbf {\bibinfo {volume} {49}},\ \bibinfo {pages} {405}
  (\bibinfo {year} {1982})}\BibitemShut {NoStop}%
\bibitem [{\citenamefont {Qin}\ \emph {et~al.}(2024{\natexlab{b}})\citenamefont
  {Qin}, \citenamefont {Shen}, \citenamefont {Li},\ and\ \citenamefont
  {Lee}}]{qin2024kinked}%
  \BibitemOpen
  \bibfield  {author} {\bibinfo {author} {\bibfnamefont {Fang}\ \bibnamefont
  {Qin}}, \bibinfo {author} {\bibfnamefont {Ruizhe}\ \bibnamefont {Shen}},
  \bibinfo {author} {\bibfnamefont {Linhu}\ \bibnamefont {Li}}, \ and\ \bibinfo
  {author} {\bibfnamefont {Ching~Hua}\ \bibnamefont {Lee}},\ }\bibfield
  {title} {\enquote {\bibinfo {title} {{Kinked linear response from
  non-Hermitian cold-atom pumping}},}\ }\href {\doibase
  10.1103/PhysRevA.109.053311} {\bibfield  {journal} {\bibinfo  {journal}
  {Phys. Rev. A}\ }\textbf {\bibinfo {volume} {109}},\ \bibinfo {pages}
  {053311} (\bibinfo {year} {2024}{\natexlab{b}})}\BibitemShut {NoStop}%
\bibitem [{\citenamefont {Abanin}\ \emph {et~al.}(2015)\citenamefont {Abanin},
  \citenamefont {De~Roeck},\ and\ \citenamefont
  {Huveneers}}]{abanin2015exponentially}%
  \BibitemOpen
  \bibfield  {author} {\bibinfo {author} {\bibfnamefont {Dmitry~A.}\
  \bibnamefont {Abanin}}, \bibinfo {author} {\bibfnamefont {Wojciech}\
  \bibnamefont {De~Roeck}}, \ and\ \bibinfo {author} {\bibfnamefont
  {Fran\ifmmode \mbox{\c{c}}\else~\c{c}\fi{}ois}\ \bibnamefont {Huveneers}},\
  }\bibfield  {title} {\enquote {\bibinfo {title} {{Exponentially Slow Heating
  in Periodically Driven Many-Body Systems}},}\ }\href {\doibase
  10.1103/PhysRevLett.115.256803} {\bibfield  {journal} {\bibinfo  {journal}
  {Phys. Rev. Lett.}\ }\textbf {\bibinfo {volume} {115}},\ \bibinfo {pages}
  {256803} (\bibinfo {year} {2015})}\BibitemShut {NoStop}%
\bibitem [{\citenamefont {Mori}\ \emph {et~al.}(2016)\citenamefont {Mori},
  \citenamefont {Kuwahara},\ and\ \citenamefont {Saito}}]{mori2016rigorous}%
  \BibitemOpen
  \bibfield  {author} {\bibinfo {author} {\bibfnamefont {Takashi}\ \bibnamefont
  {Mori}}, \bibinfo {author} {\bibfnamefont {Tomotaka}\ \bibnamefont
  {Kuwahara}}, \ and\ \bibinfo {author} {\bibfnamefont {Keiji}\ \bibnamefont
  {Saito}},\ }\bibfield  {title} {\enquote {\bibinfo {title} {{Rigorous Bound
  on Energy Absorption and Generic Relaxation in Periodically Driven Quantum
  Systems}},}\ }\href {\doibase 10.1103/PhysRevLett.116.120401} {\bibfield
  {journal} {\bibinfo  {journal} {Phys. Rev. Lett.}\ }\textbf {\bibinfo
  {volume} {116}},\ \bibinfo {pages} {120401} (\bibinfo {year}
  {2016})}\BibitemShut {NoStop}%
\bibitem [{\citenamefont {Mori}\ \emph {et~al.}(2018)\citenamefont {Mori},
  \citenamefont {Ikeda}, \citenamefont {Kaminishi},\ and\ \citenamefont
  {Ueda}}]{mori2018thermalization}%
  \BibitemOpen
  \bibfield  {author} {\bibinfo {author} {\bibfnamefont {Takashi}\ \bibnamefont
  {Mori}}, \bibinfo {author} {\bibfnamefont {Tatsuhiko~N}\ \bibnamefont
  {Ikeda}}, \bibinfo {author} {\bibfnamefont {Eriko}\ \bibnamefont
  {Kaminishi}}, \ and\ \bibinfo {author} {\bibfnamefont {Masahito}\
  \bibnamefont {Ueda}},\ }\bibfield  {title} {\enquote {\bibinfo {title}
  {Thermalization and prethermalization in isolated quantum systems: a
  theoretical overview},}\ }\href {\doibase 10.1088/1361-6455/aabcdf}
  {\bibfield  {journal} {\bibinfo  {journal} {Journal of Physics B: Atomic,
  Molecular and Optical Physics}\ }\textbf {\bibinfo {volume} {51}},\ \bibinfo
  {pages} {112001} (\bibinfo {year} {2018})}\BibitemShut {NoStop}%
\bibitem [{\citenamefont {Bukov}\ \emph {et~al.}(2016)\citenamefont {Bukov},
  \citenamefont {Heyl}, \citenamefont {Huse},\ and\ \citenamefont
  {Polkovnikov}}]{bukov2016heating}%
  \BibitemOpen
  \bibfield  {author} {\bibinfo {author} {\bibfnamefont {Marin}\ \bibnamefont
  {Bukov}}, \bibinfo {author} {\bibfnamefont {Markus}\ \bibnamefont {Heyl}},
  \bibinfo {author} {\bibfnamefont {David~A.}\ \bibnamefont {Huse}}, \ and\
  \bibinfo {author} {\bibfnamefont {Anatoli}\ \bibnamefont {Polkovnikov}},\
  }\bibfield  {title} {\enquote {\bibinfo {title} {Heating and many-body
  resonances in a periodically driven two-band system},}\ }\href {\doibase
  10.1103/PhysRevB.93.155132} {\bibfield  {journal} {\bibinfo  {journal} {Phys.
  Rev. B}\ }\textbf {\bibinfo {volume} {93}},\ \bibinfo {pages} {155132}
  (\bibinfo {year} {2016})}\BibitemShut {NoStop}%
\bibitem [{\citenamefont {Abanin}\ \emph {et~al.}(2017)\citenamefont {Abanin},
  \citenamefont {De~Roeck}, \citenamefont {Ho},\ and\ \citenamefont
  {Huveneers}}]{abanin2017effective}%
  \BibitemOpen
  \bibfield  {author} {\bibinfo {author} {\bibfnamefont {Dmitry~A.}\
  \bibnamefont {Abanin}}, \bibinfo {author} {\bibfnamefont {Wojciech}\
  \bibnamefont {De~Roeck}}, \bibinfo {author} {\bibfnamefont {Wen~Wei}\
  \bibnamefont {Ho}}, \ and\ \bibinfo {author} {\bibfnamefont
  {Fran\ifmmode~\mbox{\c{c}}\else\c{c}\fi{}ois}\ \bibnamefont {Huveneers}},\
  }\bibfield  {title} {\enquote {\bibinfo {title} {{Effective Hamiltonians,
  prethermalization, and slow energy absorption in periodically driven
  many-body systems}},}\ }\href {\doibase 10.1103/PhysRevB.95.014112}
  {\bibfield  {journal} {\bibinfo  {journal} {Phys. Rev. B}\ }\textbf {\bibinfo
  {volume} {95}},\ \bibinfo {pages} {014112} (\bibinfo {year}
  {2017})}\BibitemShut {NoStop}%
\bibitem [{\citenamefont {Amundsen}\ \emph {et~al.}(2024)\citenamefont
  {Amundsen}, \citenamefont {Brataas},\ and\ \citenamefont
  {Linder}}]{amundsen2024rkky}%
  \BibitemOpen
  \bibfield  {author} {\bibinfo {author} {\bibfnamefont {Morten}\ \bibnamefont
  {Amundsen}}, \bibinfo {author} {\bibfnamefont {Arne}\ \bibnamefont
  {Brataas}}, \ and\ \bibinfo {author} {\bibfnamefont {Jacob}\ \bibnamefont
  {Linder}},\ }\bibfield  {title} {\enquote {\bibinfo {title} {{RKKY
  interaction in Rashba altermagnets}},}\ }\href {\doibase
  10.1103/PhysRevB.110.054427} {\bibfield  {journal} {\bibinfo  {journal}
  {Phys. Rev. B}\ }\textbf {\bibinfo {volume} {110}},\ \bibinfo {pages}
  {054427} (\bibinfo {year} {2024})}\BibitemShut {NoStop}%
\bibitem [{\citenamefont {Rao}\ \emph {et~al.}(2024)\citenamefont {Rao},
  \citenamefont {Mook},\ and\ \citenamefont {Knolle}}]{rao2024tunable}%
  \BibitemOpen
  \bibfield  {author} {\bibinfo {author} {\bibfnamefont {Peng}\ \bibnamefont
  {Rao}}, \bibinfo {author} {\bibfnamefont {Alexander}\ \bibnamefont {Mook}}, \
  and\ \bibinfo {author} {\bibfnamefont {Johannes}\ \bibnamefont {Knolle}},\
  }\bibfield  {title} {\enquote {\bibinfo {title} {Tunable band topology and
  optical conductivity in altermagnets},}\ }\href {\doibase
  10.1103/PhysRevB.110.024425} {\bibfield  {journal} {\bibinfo  {journal}
  {Phys. Rev. B}\ }\textbf {\bibinfo {volume} {110}},\ \bibinfo {pages}
  {024425} (\bibinfo {year} {2024})}\BibitemShut {NoStop}%
\bibitem [{\citenamefont {Duan}\ \emph {et~al.}(2026)\citenamefont {Duan},
  \citenamefont {Fang}, \citenamefont {Deng}, \citenamefont {Wang},\ and\
  \citenamefont {Yang}}]{duan2026neel}%
  \BibitemOpen
  \bibfield  {author} {\bibinfo {author} {\bibfnamefont {Hou-Jian}\
  \bibnamefont {Duan}}, \bibinfo {author} {\bibfnamefont {Miao-Sheng}\
  \bibnamefont {Fang}}, \bibinfo {author} {\bibfnamefont {Ming-Xun}\
  \bibnamefont {Deng}}, \bibinfo {author} {\bibfnamefont {Ruiqiang}\
  \bibnamefont {Wang}}, \ and\ \bibinfo {author} {\bibfnamefont {Mou}\
  \bibnamefont {Yang}},\ }\bibfield  {title} {\enquote {\bibinfo {title}
  {{N\'eel vector and Rashba spin-orbit coupling effects on RKKY interaction in
  two-dimensional $d$-wave altermagnets}},}\ }\href {\doibase
  10.1103/gr5n-174c} {\bibfield  {journal} {\bibinfo  {journal} {Phys. Rev. B}\
  }\textbf {\bibinfo {volume} {113}},\ \bibinfo {pages} {075104} (\bibinfo
  {year} {2026})}\BibitemShut {NoStop}%
\bibitem [{\citenamefont {Bergman}\ and\ \citenamefont
  {Oganesyan}(2010)}]{bergman2010theory}%
  \BibitemOpen
  \bibfield  {author} {\bibinfo {author} {\bibfnamefont {Doron~L.}\
  \bibnamefont {Bergman}}\ and\ \bibinfo {author} {\bibfnamefont {Vadim}\
  \bibnamefont {Oganesyan}},\ }\bibfield  {title} {\enquote {\bibinfo {title}
  {{Theory of Dissipationless Nernst Effects}},}\ }\href {\doibase
  10.1103/PhysRevLett.104.066601} {\bibfield  {journal} {\bibinfo  {journal}
  {Phys. Rev. Lett.}\ }\textbf {\bibinfo {volume} {104}},\ \bibinfo {pages}
  {066601} (\bibinfo {year} {2010})}\BibitemShut {NoStop}%
\end{thebibliography}%

\end{document}